\documentclass{aa}  

\usepackage{graphicx}
\usepackage{txfonts}
\usepackage[utf8]{inputenc}
\usepackage{xcolor}

\usepackage{microtype}

\begin{document}

   \title{Spatially resolved stellar-to-total dynamical mass relation:}

\titlerunning{Spatially resolved stellar-to-total dynamical mass relation}
\authorrunning{L. Scholz-D\'iaz et al.}

    \subtitle{ Radial variations, gradients and profiles of galaxy stellar populations}

   \author{L. Scholz-D\'iaz
          \inst{1}\thanks{laura.scholzdiaz@inaf.it},
          A. R. Gallazzi
          \inst{1},
          S. Zibetti,
          \inst{1,2}
          \and
          D. Mattolini
          \inst{1,3}
          }

   \institute{INAF-Osservatorio Astrofisico di Arcetri, Largo Enrico Fermi 5, 50125 Firenze (FI), Italy
   \and Physics and Astronomy department of the University of Florence, Via G. Sansone, 50019 Sesto Fiorentino (FI), Italy
              \and Physics department of the University of Trento, Via Sommarive 14, 38123 Povo (TN), Italy
              }

   \date{Received September 15, 1996; accepted March 16, 1997}

  \abstract
   {In our standard cosmological model, galaxy assembly is inherently linked to the hierarchical growth of dark matter halos, which provide the gravitational framework in which highly complex baryonic processes unfold. Although galaxy evolution is governed by the interplay between baryonic physics and halo assembly, the extent to which halo properties shape observed galaxy properties remains unclear. With current observational challenges in measuring halo properties, the stellar-to-total dynamical mass relation is introduced as an alternative observationally-based plane, sensitive to the dark matter content within galaxies.}
   {We investigate how spatially-resolved stellar population properties vary across the stellar-to-total dynamical mass relation.}
   {We analyze optical integral-field spectrocopic data from the CALIFA survey coupled with photometry for a sample of 265 galaxies to derive their spatially resolved ages and metallicities through a Bayesian fitting framework fed with an extensive library of model spectra based on stochastic star formation and metallicity histories and dust attenuation. We study these properties in terms of both stellar and total dynamical mass, derived in a completely independent manner. Total masses correspond to enclosed masses within an aperture of three effective radii obtained through detailed Jeans dynamical modeling of the galaxies' stellar kinematics.}
   {We find that galaxy ages and metallicities measured at different radial annuli depend both on stellar and total mass, typically showing an anti-correlation with total mass after accounting for the strong correlation with stellar mass. Yet, age and [M/H] show a distinct behavior with galactocentric distance. While the dependence of age on total mass becomes more prominent in the outskirts, the one of [M/H] is significant in the inner regions. This behavior is reflected in the galaxies' stellar population profiles, which appear to be connected to the morphological type of galaxies. In particular, intermediate-mass ($M_{\star}\sim \rm10^{10.5-11}  \rm M_{\odot})$ early-types have higher stellar-to-total mass ratios and flatter age profiles with overall old ages, and steep negative [M/H] profiles, whereas later-types have lower stellar-to-total mass ratios, negative age profiles with younger ages and less steep negative [M/H] profiles. Consistently, the scatter of the stellar-total-dynamical mass relation is connected to differences in the stellar population gradients, although more strongly for age. Specifically, galaxies have flatter age gradients and steeper negative [M/H] ones with increasing stellar mass, at fixed total mass. Conversely, galaxies of a given stellar mass exhibit age gradients that become more negative with increasing total mass, while the metallicity gradients become less steep. }
   {Our findings reveal that, at fixed stellar mass, total dynamical mass is linked to systematic variations in the stellar populations and their radial gradients, suggesting a  relevant role of dark matter halos in shaping galaxy properties. These trends can be interpreted as the imprint of different halo assembly histories across the stellar-to-total dynamical mass relation, where earlier-formed halos host more evolved systems. }


   \maketitle

%
\section{Introduction}
\label{sec:intro}

In our standard cosmological model, galaxy assembly is primarily driven by halo growth \citep[e.g.,][]{1978MNRAS.183..341W,1984Natur.311..517B}, which is broadly well-understood thanks to large-scale dark matter-only and gravity-only 
cosmological numerical simulations \citep[e.g.,][]{2005MNRAS.364.1105S}. 
These halos set the first-order conditions in which highly non-linear baryonic physics operate (e.g., gas cooling, star formation, chemical enrichment, supernovae explosions and black hole growth, etc.). The diversity of observed galaxies arises from the complex interplay between these baryonic processes and halo assembly throughout cosmic evolution. Yet, despite its crucial role for understanding galaxy formation, this interplay remains a challenge for both theoretical models and observations \citep[e.g., see][review]{2018ARA&A..56..435W}.

A standard approach to study the complex interplay between galaxies and halos is through the link between their stellar ($M_{\star}$) and halo masses ($M_h$). To some extent stellar mass traces the overall hierarchical assembly of their host halos \citep[e.g.,][]{2010ApJ...725.2312O,2016MNRAS.458.2371R,2020MNRAS.497...81D,2024NatAs...8.1310A}, given the relatively tight well-known stellar-to-halo mass relation (SHMR)  for group/halo central galaxies\footnote{Observationally, a central galaxy is typically defined as the most massive o luminous one in a group or cluster, while in simulations is the one at the center of potential of its host dark matter halo} \citep[e.g.,][]{2006MNRAS.371..537W,2007ApJ...667..760Z,2012ApJ...758...50L,2012ApJ...744..159L,2013ApJ...770...57B,2013MNRAS.428.3121M,2018ARA&A..56..435W}. Yet, the SHMR is not a one-to-one relation, as theoretical models of galaxy formation predict that the SHMR has an intrinsic scatter \citep[e.g.,][]{2017MNRAS.465.2381M,2018MNRAS.480.3978A,2018ARA&A..56..435W,2021NatAs...5.1069C,2023MNRAS.518.5670C}, implying that galaxies of similar stellar masses can reside in halos of different mass. Interestingly, this scatter also encodes information about the efficiency of galaxy formation, as galaxies with higher stellar masses have been more efficient in forming stars than less massive ones, for a given halo mass. Thus, it is reasonable to consider that this scatter could be tied to the past evolutionary histories of galaxies.

To  understand their growth and evolution, galaxies have been extensively studied observationally through the baryonic term of the SHMR, i.e., stellar mass.  The latter is generally regarded as a primary metric that encapsulates key information about galaxy formation, given the well-known scaling relations between stellar mass and different stellar and gaseous properties of galaxies, such as age, SFR, stellar, gas-phase metallicity, gas mass, or structural properties such as size and concentration, in the Local Universe \citep[e.g.,][]{2003MNRAS.343..978S,2004ApJ...613..898T,2004MNRAS.351.1151B,2005MNRAS.362...41G,2005ApJ...621..673T,2010MNRAS.408.2115M,2011MNRAS.415...32S,2017ApJS..233...22S, 2015ApJ...801L..29R,2015MNRAS.448.3484M,2016MNRAS.463.2799I} and at higher redshifts  \citep[e.g.,][]{2004ApJ...604..521T,2007A&A...468...33E,2008A&A...488..463M,2014ApJ...788...72G,2014ApJ...795..104W,2014ApJ...788...28V,2023ApJ...948..140B,2024A&A...690A.150B,2025arXiv251111805G}. Specifically, more massive galaxies generally tend to be older and more metal-rich, have lower levels of star formation (SF), earlier-type morphologies, and have formed the bulk of their stars faster and at earlier cosmic times than their less massive counterparts.  However, regarding halo properties, no consensus has been established on whether they can influence observed galaxy properties beyond the effect of stellar mass. 

Stellar population properties, powerful fossil indicators of the past star formation and chemical enrichment histories of galaxies, have been typically studied as a function of halo properties to characterize the evolution of centrals and satellites \citep[e.g.,][]{2010MNRAS.407..937P,2014MNRAS.445.1977L,2021MNRAS.502.4457G,2021MNRAS.500.4469T}. There is growing observational evidence in the Local Universe indicating that halo mass can influence the stellar population properties of centrals of a given $M_{\star}$ or velocity dispersion ($\sigma$) \citep[e.g.,][]{2022MNRAS.511.4900S,2023MNRAS.518.6325S,2022ApJ...933...88O,2024ApJ...974...29O,2024MNRAS.527.3542L,2024MNRAS.530.4082Z}. Specifically, despite their strong correlation with $M_{\star}$ or $\sigma$, these physical properties show a secondary dependence on halo mass. However, the main limitation of these studies is that they rely on indirect halo mass estimations from group/cluster catalogs \citep[e.g.,][]{2007ApJ...671..153Y,2014A&A...566A...1T,2021ApJ...923..154T}, as direct halo mass measurements such as those obtained through cold HI gas rotation curves \citep[e.g.,][]{2019A&A...626A..56P} are not yet available for statistically large galaxy samples. Furthermore, alternative halo mass estimation methods such as the ones based on weak-lensing \citep[e.g.,][]{2006MNRAS.368..715M,2013ApJ...778...93T,2015MNRAS.447..298H,2016MNRAS.457.3200M,2020MNRAS.499.2896T} or satellite kinematics \citep[e.g.,][]{2011MNRAS.410..210M,2013MNRAS.428.2407W,2019MNRAS.482.4824L} have different, but significant limitations, as mass estimates are only available statistically for average galaxy populations.

With current challenges for measuring halo properties hindering the study of the co-evolution of galaxies and halos, \citet[][]{2024NatAs...8..648S} (hereafter SD24) introduced an alternative observationally-based proxy 
of halo mass. Total dynamical masses within 3 effective radii ($R_e$) (i.e., total enclosed mass within a projected elliptical aperture of 3$R_e$) \textcolor{black}{are measured through detailed Jeans dynamical modeling\footnote{Dynamical modeling of IFS data typically fits the stellar kinematics by either solving the Jeans equations of stellar dynamics \citep{1922MNRAS..82..122J,2008MNRAS.390...71C,2020MNRAS.494.4819C} or by using the Schwarzschild numerical orbital-superposition method \citep{1979ApJ...232..236S,2008MNRAS.385..647V,2022A&A...667A..51T}. See section 3.4 of \citet{2016ARA&A..54..597C} for a review.} of stellar kinematic maps obtained from high-quality integral-field spectroscopic (IFS) data of nearby galaxies from surveys like  CALIFA \citep{2012A&A...538A...8S,2014A&A...569A...1W}. } These dynamical mass estimates are sensitive to the total mass enclosed within a given aperture (e.g., stars, gas, dust, dark matter), which are typically dominated by the stellar and dark matter components. Thus, dynamical modeling techniques generally need to incorporate a dark matter mass component to reproduce the stellar motions.

When looking into galaxy properties across the stellar-to-total dynamical mass relation (STDMR), SD24 find that galaxy ages, metallicities ([M/H]), stellar apparent angular momentum ($\lambda_{Re}$), star formation rates (SFRs) and morphology are connected to the scatter about the relation, depending both on stellar and total mass. Moreover, all these baryonic properties mirror the behavior of galaxy ages, [M/H] and star formation histories (SFHs) across the SHMR \citep{2022MNRAS.511.4900S,2023MNRAS.518.6325S}. At fixed halo/total mass, higher stellar mass galaxies are older, more metal-rich, and dispersion dominated, have lower SFRs, earlier-type morphologies, and have formed the bulk of their stars earlier on and on shorter time-scales than less massive galaxies. Reversely, there is a secondary dependence on halo/total mass with galaxies lower total/halo mass at fixed $M_{\star}$ being younger, more metal-poor, rotationally supported, having higher SFRs, later-type morphologies and more extended SFHs than galaxies having higher total/halo masses. 

 To gain insights into the processes that have given rise to the observed trends, in this work we investigate spatially resolved stellar populations of galaxies across the STDMR. Being fossil records of galaxy star formation and chemical enrichment histories, the different physical processes that shape stellar populations within galaxies (e.g., accretion, feedback, and environmental processes) can also leave distinct signatures in their spatial distribution. As an example, gas-rich minor mergers or accretion of pristine gas in galaxy outskirts can fuel star formation and lead to younger and more metal-poor stellar populations formed after the dilution of the ISM. On the other hand, steep central metallicity gradients within similarly old stellar populations within massive galaxies may reflect very efficient star formation  at early times coupled with a high retention of metals in deep gravitational potential wells.

Gradients and profiles of galaxy stellar populations have been typically studied separately for early-types (ETGs) and late-types (LTGs) \citep[e.g., see][for a review]{2020ARA&A..58...99S}. ETGs typically exhibit steep negative metallicity gradients within their inner regions that flatten in their outskirts, with relatively flat age profiles, although some studies report mild positive or negative age gradients \citep[e.g.,][]{Mehlert:2003,Sanchez-Blazquez:2007,Reda:2007,Brough:2007,Spolaor:2009,Koleva:2011, kuntschner+10,2015MNRAS.448.3484M,Gonzalez-Delgado:2015aa,Goddard:2017b,2017A&A...608A..27G,martin-navarro+18,Li_Mao_Cappellari:2018,Parikh:2019,DominguezSanchez:2019,2019ApJ...880..111O,2020MNRAS.491.3562Z}. In the case of late-types, they tend to show negative gradients in both age and [M/H] \citep[e.g.,][]{2014A&A...570A...6S,Gonzalez-Delgado:2015aa,Goddard:2017b,2017A&A...604A...4R}. Furthermore, spatially-resolved SFHs of nearby galaxies indicate that the bulk of the stellar populations within inner regions of relatively massive galaxies ($M_{\star}$>$\rm10^{9.5}M_{\odot}$) are formed earlier than the ones in their outskirts \citep[e.g.,][]{2016MNRAS.463.2799I,2017A&A...608A..27G,2020ARA&A..58...99S}. This so-called local `downsizing' is not only connected to stellar mass --more massive galaxies assemble their stars faster than less massive ones also on local scales--, but also to morphology. For a given stellar mass, the inner regions of ETGs also form earlier and faster than the ones of LTGs and build up higher surface mass densities \citep[e.g.,][]{2020ARA&A..58...99S}, given their different central surface stellar mass densities. In fact, local stellar mass density is found to strongly affect local stellar population properties of galaxies \citep[e.g.,][]{2021MNRAS.508.4844N,2022MNRAS.512.1415Z}.

 In this work, we follow up on SD24 and investigate radial variations, gradients and profiles of stellar population properties across the STDMR for CALIFA, and how they depend both on stellar and total enclosed mass within 3$R_{e}$. 

The paper is organized as follows: Section \ref{sec:sample} introduces the CALIFA survey, our galaxy sample, and the ancillary global galaxy properties used in this study. Section \ref{sec:stelpopsanalysis} describes the spatially-resolved stellar population analysis. In Section \ref{sec:STDMR:radialregions} we show stellar population properties across the STDMR at different radial annuli. In Section \ref{sec:STDMR:gradients} we quantify these radial variations by means of stellar population gradients. Sections \ref{sec:total_mass} and \ref{sec:projectin_STDMR_mtot} show scaling relations with stellar and total mass, respectively, for the inner and outer parts of galaxies. In Section \ref{sec:profiles} we show stellar population profiles in a narrow stellar mass bin. We discuss our results in Section \ref{sec:disc} and summarize them in Section \ref{sec:conc}.
 
\section{Our CALIFA sample and dataset}
\label{sec:sample}

\subsection{The CALIFA survey, galaxy sample and data}
\label{sample:anci}

Our galaxy sample is based on 300 galaxies from the integral-field spectroscopic survey CALIFA \citep{2012A&A...538A...8S}, which are presented in \citet{2017A&A...597A..48F}. These galaxies are drawn from the CALIFA mother sample \citep{2014A&A...569A...1W}, which was diameter-selected (45''< r-band angular isophotal diamater < 80'') from the Sloan Digital Sky Survey (SDSS) DR7 in the redshift range $0.005 \leq z \leq 0.03$. 

Galaxies were observed with the Postdam Multi-Aperture Spectrograph, PMAS \citep{Roth:2005aa} on the PPAK mode \citep{Verheijen:2004aa,Kelz:2006aa} mounted on the 3.5 meter telescope of the Calar Alto Observatory. The hexagonal field of view (FoV) of $\rm 74\times64$  arcsec is covered by a bundle of 331 science fibers. An effective covering factor of nearly 100\% of the FoV is achieved by adopting a three pointing dithering scheme \citep{2012A&A...538A...8S}.

This subset of galaxies was observed with a high resolution configuration (R$\sim$1650 at $\sim$4500 \AA) over the spectral range 3650-4840~\AA \ (`V1200' setup), with the aim to derive high-quality stellar kinematics, and also with a lower resolution configuration (R$\sim$850 at $\sim$5000 \AA) with a wider spectral coverage of 3745-7300~\AA \ (`V500' setup). These galaxies are representative of the full sample, with morphologies ranging from ellipticals to late-type spirals. We note that the sample is strongly biased towards central galaxies\footnote{Based on the galaxies in common with the group and cluster catalog from \citet{2007ApJ...671..153Y}}, with 95\% being in isolation. We refer the reader to \citet{2017A&A...597A..48F} for more details on this galaxy subset and to their Table 1 for their basic properties, such as $z$, name or CALIFA ID.

We note that the 300 CALIFA galaxies were carefully selected for having high quality and regular kinematic data, with  good spatial sampling and no signs of interactions or perturbations in their stellar kinematic maps. This good quality requirement is essential to perform Jeans dynamical modeling and derive highly robust total enclosed/dynamical masses \citep[e.g.,][]{2013MNRAS.432.1862C,2023MNRAS.522.6326Z}. Yet, we note that it may impose selection biases, generally excluding irregular galaxies, merging or interacting systems and close galaxy pairs. We note that in this work we explore how morphology is connected to variations in the stellar population profiles within the STDMR (section \ref{sec:morphology}), dissecting some of these biases.

While total dynamical masses were obtained through detailed dynamical modeling  (see section \ref{sample:anci}, Lyubenova in prep.) of high-quality stellar kinematics by analyzing datacubes from the V1200 setup \citep{2017A&A...597A..48F}, the stellar population analysis makes use of the so-called ‘Main Sample’ COMBO cubes released as part of the CALIFA third and final public data release \citep{2016A&A...594A..36S}, which are available for 396 galaxies \citep{2017MNRAS.468.1902Z}. These cubes combine the datasets from the V500 and the V1200 setups, having an unvignetted spectral coverage of 3700-7140~\AA,\ with a spatial sampling of 1 arcsec per spaxel  (effective spatial resolution of $\rm \sim$2.57 arcsec FWHM). This spectral range includes key absorption features  sensitive to age and metallicity needed for the stellar population analysis  \citep[$\rm D4000_n$, H$\rm \beta$, Mgb, $\rm Mg_2$, Fe5015, Fe5270, Fe5335; ][]{1983ApJ...273..105B,1994ApJS...94..687W,1999ApJ...527...54B}.  These COMBO cubes typically reach a signal-to-noise ratio (SNR) of 3 per spaxel and spectral resolution element at $\rm \sim$23.4 mag/$\rm arcsec^{2}$ in the r-band.

Out of the original 300 galaxies observed with the `V1200' grating, there are 265 galaxies with COMBO cubes available (i.e., observed with both the `V1200' and `V500' setups), which constitute our final galaxy sample.

\subsection{Global properties}
\label{sample:anci}

The following global properties are drawn from previous studies:\\
(i) Stellar masses are obtained from Sunrise spectral energy distribution fits by assuming a \citet{2003PASP..115..763C} stellar initial mass function (IMF) \citep{2014A&A...569A...1W}. We note that the mass-to-light ratios from which stellar masses are derived  are estimated independently from the spatially-resolved stellar population properties (see section~\ref{sec:stelpopsanalysis}).
\\
 (ii) Total dynamical masses have been derived through Jeans dynamical modeling (Lyubenova et al. in prep.), and are presented in SD24. Briefly, the total dynamical mass of a galaxy corresponds to the total enclosed mass within an aperture of 3$R_e$, which is modeled with a stellar and a dark matter components. These total masses are determined by constructing axisymmetric Jeans dynamical models that fit the high-quality stellar kinematics of our CALIFA galaxies \citep[velocity and velocity dispersion fields determined by][]{2017A&A...597A..48F}. These dynamical models are based on a solutions of the Jeans equations of stellar dynamics as implemented by \citet{2008MNRAS.390...71C}. The reader is referred to SD24 for more details on their determination, while the dynamical models are fully described in a forthcoming paper (Lyubenova et al. \textit{in prep}). \\
  (iii) The morphological classification of the CALIFA mother sample results from averaging the classifications by five members of the CALIFA collaboration, based on visual inspection of gri color-composite SDSS images, and it is drawn from \citet{2014A&A...569A...1W}. Galaxies were classified into ellipticals, spirals or irregulars, with ellipticals being subdivided into 0-7 and spirals into 0, 0a, a, ab, b, bc, c, cd, d, m. For this work, we used the broader classification of \citet{2019A&A...632A..59F} and SD24, who group galaxies into the following Hubble-Types: ellipticals (E), lenticulars (S0), Sa, Sb, Sc, Sd, and Ir, according to the previous classification. Our final sample of 265 galaxies consists of 16\% of ellipticals, 13\% of lenticulars, 18\% of Sa, 36\% of Sb, 13\% of Sc, 5\% of Sd, with only one irregular galaxy.

\section{Spatially-resolved stellar population analysis}
\label{sec:stelpopsanalysis}

The spatially resolved stellar population analysis of CALIFA that we employed for this work was performed by \citet{2017MNRAS.468.1902Z}. Here, we provide an overview of the method, highlighting its main characteristics, but the reader is referred to \citet{2017MNRAS.468.1902Z} for a full description. 

\subsection{Spatially-resolved spectrophotometric maps}

First, to achieve the required SNR (in individual spaxels) required for the stellar population analysis described below, \citet{2017MNRAS.468.1902Z} performed an adaptive smoothing (preserving the original data sampling), which follows closely the ones described in \citet{2009MNRAS.400.1181Z,2009arXiv0911.4956Z}. The SNR is evaluated in a featureless region of the continuum at each spaxel. If the target SNR is reached, then the routine stops. Otherwise, the spectrum is iteratively averaged over an expanding set of neighboring pixels until a target SNR of 20 is reached or a maximum smoothing radius of 5 pixels  (i.e., 5 arcsecs), balancing improved SNR and spatial resolution. The spectrum of a spaxel is effectively replaced, spectral pixel by spectral pixel, with the median of the surrounding spaxels.  Moreover, this smoothing scheme preserves much better the spatial information compared to other  binnig schemes widely used in IFS studies \citep[e.g., Voronoi binning, ][]{2003MNRAS.342..345C}, yet at the expense of the statistical independence of the spaxels. We also note that CALIFA datacubes have already a instrinsic non-negligible correlation between adjacent spaxels, due to the dithering scheme and the spatial resolution. Yet, this does not crucially affect our analysis, as to achieve our objectives we do not require such high spatial resolution. Moreover, to ensure a consistent selection of the galaxy regions examined across the whole sample, the analysis is limited to spaxels with an r-band surface brightness of $\rm \mu_r \leq$ 22.5 mag arcsec$^{-2}$. We note that in the central regions of the galaxies (typically within $\rm\sim1R_e$) no smoothing is effectively required due to the high SNR. For illustration, in Appendix \ref{ap:smoothing} we show  the result of applying this adaptative smoothing to one of our CALIFA galaxies.

We note that the high SNR of 20 per spaxel required for our stellar population analysis  translates into our 
sample being 82\% complete up to $1.5 \, R_e$, but we start running into incompleteness at larger radii. However, the main results obtained in this work are also observed at $r \, $>$\, 1.5 \, R_e$ despite the lower statistics. In contrast, note that dynamical mass measurements extend up to larger radii, as the analysis does not require the high SNR criteria to be satisfied. 

Finally, ellipticities ($\epsilon$), position angles (PA) and effective semi-major axes, defined as the semi-major axis of the elliptical aperture enclosing half of the total flux ($R_e$), are obtained from the growth curve analysis described in 
\citet{2014A&A...569A...1W}.

\subsection{Stellar population analysis}
The stellar population analysis employed in this work is a Bayesian method fully described in detail in \citet{2017MNRAS.468.1902Z} and is based on the original work of \citet{2005MNRAS.362...41G}. This analysis allows to derive the following stellar population properties: Mean light-weighted and mass-weighted ages and metallicities, and stellar masses of each spaxel within the galaxies. Throughout this study, we employ light-weighted r-band ages and metallicities, as they are less sensitive than mass-weighted properties to model assumptions on the galaxy's SFHs. By construction, this is a general feature in spectral energy distribution (SED) and spectral fitting techniques. On the other hand, we note that light-weighted mean properties are more weighted toward the younger, brighter stellar generations, with respect to mass-weighted mean ones \citep[e..g.,][]{2017MNRAS.468.1902Z}. The degree of deviation between light- and mass-weighted properties depends on the amount of younger stars and on the total SFH duration. The effect is more evident in the ages and to a much lesser degree in stellar metallicity \citep[e.g.,][]{2009MNRAS.395..608T,2017MNRAS.468.1902Z,2020MNRAS.491.5406T,2021MNRAS.500.4469T,2025A&A...703A...5M}.

In this Bayesian approach, a selected set of stellar absorption features and photometric data is compared to those predicted by a large library of model spectra constructed by convolving simple stellar population models (SSP) with randomly generated SFHs, metallicity histories and dust attenuation parameters. 

The model library is described in \citet{2017MNRAS.468.1902Z} \citep[see also appendix A of ][]{2025A&A...703A...5M} and comprises 500,000 models obtained by convolving SSP models with Monte Carlo SFHs. The SSP models were built with the 2016 version of the \citet{2003MNRAS.344.1000B} stellar population synthesis code using the MILES empirical spectral libraries \citep{2006MNRAS.371..703S,2011A&A...532A..95F}, adopting a \citet{2003PASP..115..763C} initial mass function and \citet{1994A&AS..106..275B} isochrones. The SFHs are described by a parametric `á la Sandage' (or delayed gaussian) form \citep{1986A&A...161...89S,2002ApJ...576..135G}, which allows both for a rising and declining phase. On top of that continuum SFH, the modeling allows for star formation bursts in random number, age and intensity. The models also include a parametrized chemical enrichment history, as an evolving metallicity along the SFH, instead of assuming a constant value. With this paramerization the metallicity gradually increases with the formed stellar mass (more or less rapidly), although the inclusion of random bursts of a certain metallicity also allows for stochasticity. The resulting spectra are attenuated following a \cite{2000ApJ...539..718C} model that has two separated components of dust: from birth clouds and from the diffuse interstellar medium, in a randomly generated variable amount. 

The spectral and photometric diagnostics used are the spectral indices $\rm D4000_n$ \citep{1999ApJ...527...54B}, $\rm H\beta$ \citep{1994ApJS...94..687W} and $\rm H\delta_A+H\gamma_A$ \citep{1997ApJS..111..377W} (mainly sensitive to age), $\rm [Mg_2Fe]$ \citep{2003MNRAS.344.1000B} and $\rm [MgFe]'$ \citep{2003MNRAS.339..897T} (mostly metal-sensitive indices and relatively insensitive to the abundance of $\rm \alpha$-elements with respect to the one of iron-peak elements) and the photometric fluxes in the five SDSS bands \textit{ugriz}. 

The posterior probability distribution function of a chosen physical parameter is determined by marginalizing over all the other parameters and weighting the models by the likelihood of the fit. We note that this Bayesian approach allows us to obtain uncertainties that account for both observational errors and degeneracies in the assumed prior distributions for SFHs, metallicity histories, and dust.

\section{Stellar-to-total dynamical mass relation}
\label{sec:STDMR_all}

\begin{figure*}
\centering
\includegraphics[width=\hsize]{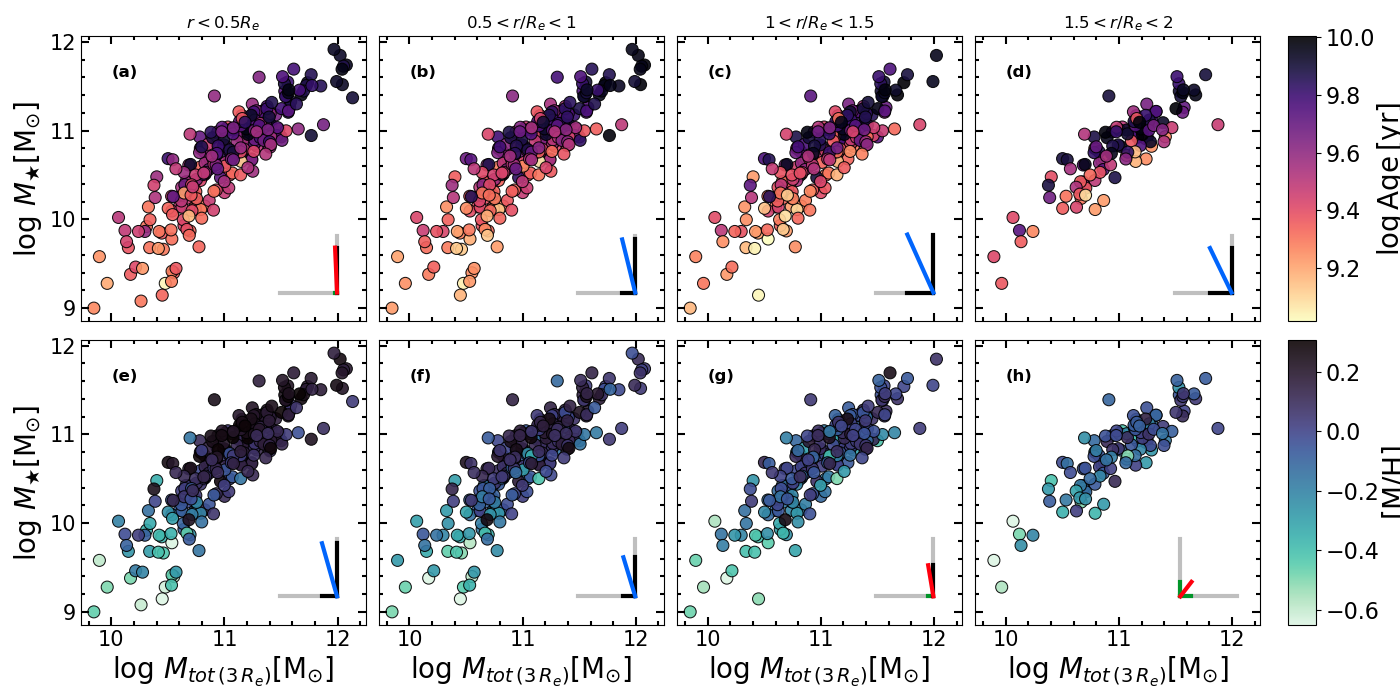}
  \caption{Stellar-to-total dynamical mass relation for our CALIFA galaxies in terms of stellar population properties measured at different annul. Each column corresponds to a different annulus with increasing galactocentric distance from left to right (see text). Galaxies are shown as circles colored-coded by median ages (upper rows) and metallicities (bottom rows) measured within their corresponding annuli. Partial correlation coefficient strengths are shown in the bottom right corner (solid black or green lines) between the stellar population parameters and $M_{\star}$ (vertical) and $M_{tot}$ (horizontal). Black (green) lines correspond to statistically (not) significant partial correlation coefficients. Grey solid lines have a length which corresponds to a correlation coefficient of 0.6 for reference. The direction of maximal increase of the stellar population parameters (see text) is indicated as a blue solid line when both partial correlation coefficients are significant, while red if at least one of the two is not significant}.
     \label{fig:stel_pops_annuli}
\end{figure*}

\renewcommand{\arraystretch}{1.3}
\begin{table*}
\centering
\caption{Partial correlation coefficients for age and metallicity measured different radial annuli ($r/R_e$)}
\setlength{\tabcolsep}{4pt}
\begin{tabular}{|l||cc|cc|}
\hline
& \multicolumn{2}{c||}{Age partial correlations} & \multicolumn{2}{c|}{Metallicity partial correlations} \\
\cline{2-5}
 & $\rho_{Age-M_*}$ & $\rho_{Age-M_{tot}}$  & $\rho_{Met-M_*}$ & $\rho_{Met-M_{tot}}$ \\
 
\hline
$r/R_e < 0.5$ & $0.474^{+0.086}_{-0.104}$\textsuperscript{***}  & $-0.021^{-0.119}_{+0.079}$\textsuperscript{ns}  & $0.560^{+0.080}_{-0.090}$\textsuperscript{***} & $-0.159^{-0.111}_{+0.119}$\textsuperscript{*} \\

 $0.5 \leq r/R_e < 1$ & $0.560^{+0.079}_{-0.090}$\textsuperscript{***} & $-0.138^{-0.122}_{+0.128}$\textsuperscript{*}  & $0.411^{+0.099}_{-0.110}$\textsuperscript{***}  & $-0.125^{-0.115}_{+0.125}$\textsuperscript{*} \\
 
$1 \leq r/R_e < 1.5$ & $0.611^{+0.076}_{-0.094}$\textsuperscript{***} & $-0.275^{-0.115}_{+0.125}$\textsuperscript{***}  & $0.326^{+0.114}_{-0.126}$\textsuperscript{***}  & $-0.055^{-0.135}_{+0.135}$\textsuperscript{ns}  \\

$1.5 \leq r/R_e < 2$ & $0.450^{+0.150}_{-0.140}$\textsuperscript{***} & $-0.227^{-0.173}_{+0.187}$\textsuperscript{*}  & $0.152^{+0.178}_{-0.112}$\textsuperscript{ns}  & $0.118^{+0.182}_{-0.048}$\textsuperscript{ns}  \\

\hline
\end{tabular}
\tablefoot{Partial correlation coefficients of age or metallicity and stellar and total mass. Quoted uncertainties correspond to the 95\% confidence intervals and significance levels are indicated according to the associated p-values: $^{***}p<0.001$, $^{**}p<0.01$, $^{*}p<0.05$, $^{ns}$not significant. }
\label{table:1}
\end{table*}

While SD24 focused on integrated ages and [M/H] (averaged within 1$R_e$) across the STDMR, here we take advantage of CALIFA's 2D spatial information. In this section, we investigate how stellar population properties measured in different radial regions of the galaxies vary across the STDMR. Moreover, we also quantify the radial variations of ages and metallicities
across the STDMR by means of radial gradients.

\subsection{Ages and [M/H] at different radial annuli}
\label{sec:STDMR:radialregions}

First, we inspect how stellar populations measured at different wide radial annuli within the galaxies behave across the stellar-to-total dynamical mass relation. For each galaxy, we compute the median ages and [M/H] within elliptical annuli centered on the nucleus of the galaxy. For that, we use the semi-major axis of the ellipse, which we denote by $r$ for simplicity. With $r$ normalized by the half-light semi-major axis $R_e$, we define four different regions that move from the inside to the outside as follows:

\vspace{2pt}
\noindent (i)  $r/R_e \leq 0.5 $ \\
 (ii)  $ 0.5 \leq r/R_e < 1$\\
 (iii)  $1 \leq r/R_e < 1.5$\\
 (iv)  $ 1.5 \leq r/R_e < 2$
\vspace{2pt}

Note that we require that each annulus has at least 70\% spaxel coverage (i.e., spaxels with stellar population measurements available) in order to compute median properties. We caution the reader that the outermost annulus has lower statistics because of our surface brightness limit criteria for performing our stellar population analysis. This translates into our sample being reduced to 42\% in the outermost annulus\footnote{In another words, 42\% of our original sample has at least 70\% of spaxels with stellar population measurements available in the outermost annulus.}. Thus, our reference study is based on results up to 1.5$R_e$, but we highlight that the main results described below are also seen in the outermost annulus, despite the lower statistics.

Fig. \ref{fig:stel_pops_annuli} shows the STDMR for all galaxies satisfying the completeness criteria color-coded with median ages (upper row) and metallicities (bottom row). Each column corresponds to a different radial annulus (with increasing galactocentric distance from left to right), where stellar population properties are measured.

To quantify the dependence of the stellar population parameters on stellar/total mass we perform a partial correlation analysis following \citet{2022MNRAS.511.4900S,2024NatAs...8..648S}. As stellar and total mass also correlate with each other, this method allows to truly assess the dependencies by removing intercorrelations between the data \citep[e.g.,][]{2017MNRAS.471.2687B,2020MNRAS.499..230B,2020MNRAS.492...96B}. 
We note that partial correlation analyses inherently do not incorporate error uncertainties. Following \citet{2022ApJ...926..117S}, we consider that a correlation is significant if its corresponding confidence intervals excludes zero, and the \textit{p-value} is below 0.05. Partial correlation coefficients were computed using Spearman rank correlations between the variables of interest as described in  \citet{2022MNRAS.511.4900S,2024NatAs...8..648S}. We computed partial correlation coefficients ($\rho$), their 95\% confidence intervals (CIs) and their corresponding \textit{p-values}. The partial correlation coefficients along with their CI and significance are summarized in Table \ref{table:1}. Although the partial correlation coefficients do not capture the full two-dimensional information across the STDMR, they allow us to evaluate the statistical significance of the global trends with stellar/total mass.

In the bottom right corner of each panel of Fig. \ref{fig:stel_pops_annuli}, we show partial correlation coefficient strengths (solid lines) between the stellar population parameters and stellar mass (vertical line) and total dynamical mass (horizontal line) (see caption for full details). To guide the eye, the direction of the maximal increase of the stellar population properties is also indicated as a blue/red line, whose slope was computed using the partial correlation coefficients following SD24.

 \vspace{-10pt}
\paragraph{ \emph{Age} --} In the upper panels, (b), (c) and (d), of Fig. \ref{fig:stel_pops_annuli} we observe that age depends both on stellar and total mass. By eye, at fixed $M_{tot}$, more massive galaxies are older, while , at fixed $M_{\star}$, galaxies with higher total masses are younger. Visually, this is especially clear for the intermediate stellar mass regime ($\rm \sim 10^{ 10}-10^{11} M_{\odot}$), in agreement with SD24. More quantitavely, the partial correlation coefficient strengths indicate that age correlates more strongly with stellar mass (vertical black line), but there is still also a secondary anti-correlation with total dynamical mass (horizontal black line) (Table \ref{table:1}). This can also be seen by looking at the direction of maximal increase of age across the relation, which is neither vertical nor horizontal, indicating a dependence on both parameters. However, in the central region (panel a), although we still see a strong dependence with stellar mass, the dependence on total mass is negligible and not statistically significant.  

By looking at the different panels as a whole, we see that the correlation coefficient strength with total dynamical mass increases from the inner to the outer regions (from left to right). This is suggestive of an increasing dependence of stellar population age on total dynamical mass with increasing galactocentric distance.

\vspace{-10pt}
\paragraph{\emph{[M/H]} --} Panel (e) of Fig. \ref{fig:stel_pops_annuli} shows that the metallicity measured in the inner region the galaxies behaves very similarly to the integrated properties within 1$R_e$ seen in SD24, with [M/H] depending both on stellar mass and total mass. Visually, we observe how more massive galaxies are more metal-rich at fixed $M_{tot}$, while the trend at fixed $M_{\star}$ are less clear than in the case of Age. By quantifying the dependences on both parameters with partial correlations (Table \ref{table:1}), we observe that [M/H] mainly depends on stellar mass, but still has a secondary weak dependence on total mass that is statistically significant. Moving outward from the center, panel (f) shows an analogous behavior, although with slightly weaker correlation strengths.  These trends continue to weaken with increasing galactocentric distance, as panel (g) shows even weaker coefficient strengths, and the correlation with total mass is not significant. At the largest galactocentric distance, panel (h) show very small correlation coefficients for both stellar and total mass, which are not statistically significant.

In this sense,  metallicity is much less coupled with age when comparing the strengths of the partial correlation coefficients at different annuli. The correlation's strength of [M/H] with stellar mass decreases with galactocentric distance (from left to right). While [M/H] depends on total mass in the inner region, the correlation strength substantially decreases with galactocentric distance, being not statistically significant in the outer regions. 

Additionally, by looking at the different panels we clearly observe a difference in average metallicity from the center to the outskirts, with higher metallicities in the central regions, consistent with the steep metallicity gradients seen in literature  (see sections \ref{sec:intro} and \ref{sec:disc} for more details). 

Finally, as a model-independent alternative to the stellar populations obtained with our Bayesian fitting scheme, in Appendix \ref{ap:break} we also show the STDMR color-coded with the  $\rm D4000_n$ break (mainly sensitive to age) for different radial annuli. We find that $\rm D4000_n$  shows analogous trends as the ones found for age. Similarly as for age, partial correlations also indicate that $\rm D4000_n$ has a main dependence on stellar mass and a secondary one on total mass. This remarkably good agreement further supports the robustness of our stellar population measurements.

\subsection{Gradients across the STDMR}
\label{sec:STDMR:gradients}

\begin{figure*}
\centering
\includegraphics[width=0.85\hsize]{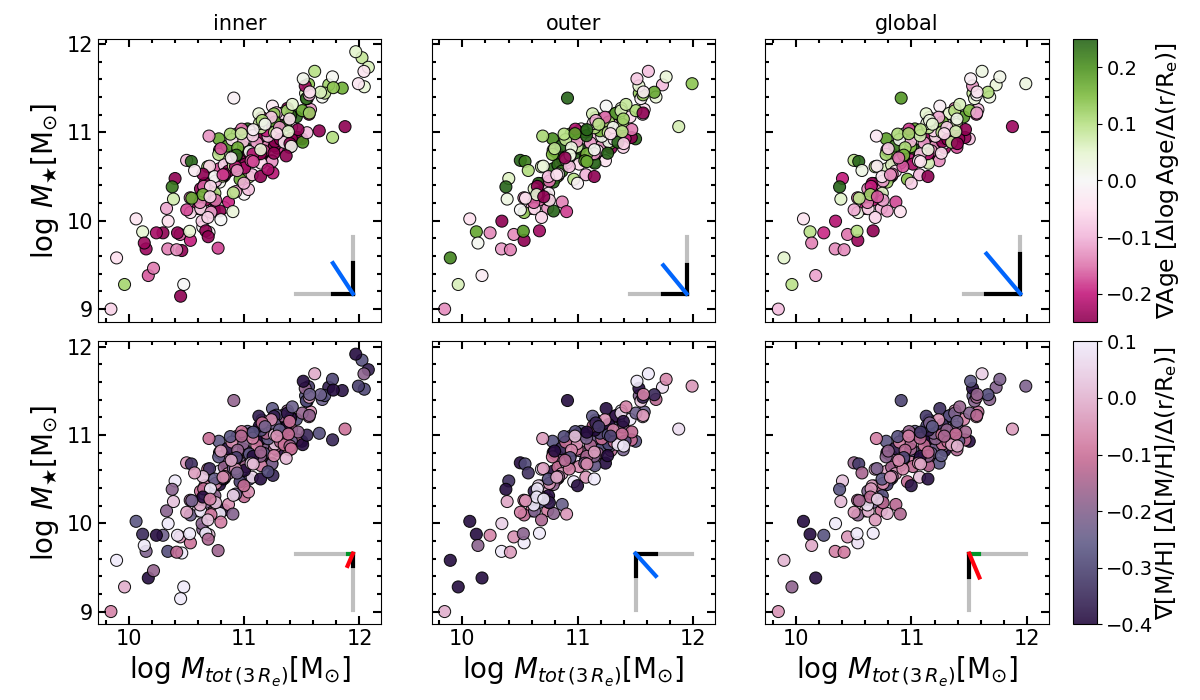}
  \caption{ Stellar population gradients across the stellar-to-total dynamical mass relation. Galaxies are shown as circles colored-coded by age gradient, $\nabla \log \rm Age$, (upper panels), and metallicity gradient, $\nabla \rm [M/H]$, (lower panels). Inner gradient (left panels), outer gradient (middle panels) and the global gradient (right panels). Partial correlation coefficient strengths are shown in the bottom right corner (solid black or green lines)  between the $\nabla \log \rm Age$ / $\nabla \rm [M/H]$ and $M_{\star}$ (vertical) and $M_{tot}$ (horizontal). Black (green) lines correspond to statistically (not) significant partial correlation coefficients. Grey solid lines have a length which corresponds to a correlation coefficient of 0.6 for reference. The direction of maximal increase of the stellar population parameters (see text) is indicated as a blue solid line when both partial correlation coefficients are significant, while red if at least one of the two is not significant.}
     \label{fig:gradients}
\end{figure*}

\begin{table*}
\centering
\caption{Partial correlation coefficients for the inner, outer and global age and metallicity gradients}
\setlength{\tabcolsep}{4pt}
\begin{tabular}{|l||cc|cc|}
\hline
& \multicolumn{2}{c||}{$\nabla$Age partial correlations} & \multicolumn{2}{c|}{$\nabla$[M/H] partial correlations} \\
\cline{2-5}
 & $\rho_{\nabla Age-M_*}$ & $\rho_{\nabla Age-M_{tot}}$  & $\rho_{\nabla Met-M_*}$ & $\rho_{\nabla Met-M_{tot}}$ \\
 
\hline
inner & $0.324^{-0.114}_{+0.107}$\textsuperscript{***}  & $-0.212^{+0.122}_{-0.118}$\textsuperscript{***}  &  $-0.130^{+0.130}_{-0.112}$\textsuperscript{*} &  $0.056^{+0.126}_{-0.124}$\textsuperscript{ns} \\

outer &  $0.301^{-0.141}_{+0.130}$\textsuperscript{***} &  $-0.246^{+0.146}_{-0.134}$\textsuperscript{***}  &  $-0.236^{+0.146}_{-0.134}$\textsuperscript{**}  & $0.214^{-0.144}_{+0.136}$\textsuperscript{**} \\
 
global &  $0.423^{-0.133}_{+0.117}$\textsuperscript{***} &  $-0.358^{+0.138}_{-0.122}$\textsuperscript{***}  &  $-0.253^{+0.143}_{-0.138}$\textsuperscript{***}  &  $0.106^{-0.147}_{0.143}$\textsuperscript{ns}  \\

\hline
\end{tabular}
\tablefoot{Partial correlation coefficients of age or metallicity gradients and stellar and total mass. Quoted uncertainties correspond to the 95\% confidence intervals and significance levels are indicated according to the associated p-values: $^{***}p<0.001$, $^{**}p<0.01$, $^{*}p<0.05$, $^{ns}$not significant. }
\label{table:2}
\end{table*}

To quantify radial variations of age and metallicity across the STDMR, we follow \citet{2020MNRAS.491.3562Z} and compute radial gradients as the ratio of finite differences between stellar population properties measured in two distinct thin radial annuli. These radial gradients are defined as:

\vspace{-15pt}

\begin{equation}
\nabla \log \rm{Age}= \frac{\Delta \log \rm{Age}}{\Delta (r/R_e)} \ ; \ \nabla \rm{[M/H]}  = \frac{ \Delta \rm{[M/H]}} {\Delta (r/R_e)}
\end{equation}

We employ three radial regions to compute inner, outer and global gradients. For that, analogously as in section \ref{sec:STDMR:radialregions}, stellar population properties are measured within thin radial annuli of $0.1 \, R_e$ width defined as follows:

\vspace{3.5pt}
\noindent (i)  $0.2 \, R_e$ ($0.15 \leq r/R_e < 0.25$)\footnote{We selected $0.2 \, R_e$ as inner region, given that it is more robust than the very central region both from a statistical and an observational point of view ($0.2 \, R_e$ contains more spaxels, and it is less sensitive to the residual PSF mismatches between photometry and IFS).}\\
(ii) $1 \, R_e$ ($ 0.95 \leq r/R_e < 1.05$)\\
(iii) $1.5 \, R_e$ ($ 1.45 \leq r/R_e < 1.55$)
\vspace{3.5pt}

Then, we estimate the inner gradient as the ratio of finite differences measured between $1 \, R_e$ and $0.2 \, R_e$, the outer one between $1.5 \, R_e$ and $1 \, R_e$, and the global one between $1.5 \, R_e$ and $0.2 \, R_e$.

We show the inner, outer and global age and metallicity gradients as a function of stellar mass in Appendix \ref{ap:gradients}. Consistently with the literature (see section \ref{sec:intro}), we generally find negative [M/H] gradients. Inner and global negative [M/H] gradients tend to become steeper with increasing stellar mass, while the outer one shows more scatter with stellar mass. In contrast, age gradients tend to be negative in the lower-mass regime, and transition to positive ones for massive galaxies, resulting in a large scatter in the intermediate mass regime.\\

Fig. \ref{fig:gradients} shows the STDMR color-coded with stellar population properties gradients: $\nabla \log \rm Age$ (upper panels) and $\nabla$[M/H] (bottom panels). We show the inner gradient measured between $1 \, R_e$ and $0.2 \,R_e$ (left panels), the outer one measured between $1.5 \, R_e$ and $1 \, R_e$ (middle panels), and the global one measured between $1.5 \,R_e$ and $0.2 \, R_e$ (right panels). Similarly to Fig. \ref{fig:stel_pops_annuli}, in the bottom right corner of each panel, we plot partial correlation coefficient strengths (solid  lines) between the stellar population parameters and stellar mass (vertical line) and total dynamical mass (horizontal line) (see caption for full details). The direction of the maximal increase of the stellar population properties is also indicated as a blue/red line, whose slope was computed using the partial correlation coefficients following SD24. As in Table \ref{table:1}, the partial correlation coefficients along with their CI intervals and significance
are summarized in Table \ref{table:2}.

\vspace{-10pt}

\paragraph{\emph{Age} --} Fig. \ref{fig:gradients} shows that galaxies in the upper part of STDMR tend to have flat or positive global gradients, while the ones in the lower part of the relation have mild negative gradients (right upper panel). Although with more scatter, we see a similar behavior for the inner and outer gradients (left and middle upper panels). Additionally, looking at the partial correlation coefficients, we see that the direction of increase of $\nabla \rm Age$ across the STDMR is qualitatively very similar to the one seen for age in the outer parts of the galaxies (Fig. \ref{fig:stel_pops_annuli}), with $\nabla \rm Age$ depending both on stellar and total mass. 

\vspace{-10pt}

\paragraph{\emph{[M/H]} --} In contrast to age, in the case of metallicity galaxies generally display steep negative gradients, in agreement with the literature (see section \ref{sec:intro}). When looking into outer and the global gradient visually (middle and right lower panel), we see that galaxies in the lower STDMR tend to have milder gradients than galaxies in the upper STDMR, which display steeper gradients. This is visible in the outer and global gradient, while the inner gradient does not show a clear trend (left lower panel), similarly to age. More quantitatively, the partial correlation coefficients show a weak dependence of the outer and global gradients on stellar and total mass, whereas it is negligible for the inner one.

\section{Stellar population scaling relations with stellar mass: Above and below the STDMR}
\label{sec:total_mass}
\label{sec:projectin_STDMR}

\begin{figure}
\centering
\includegraphics[width=0.7\hsize]{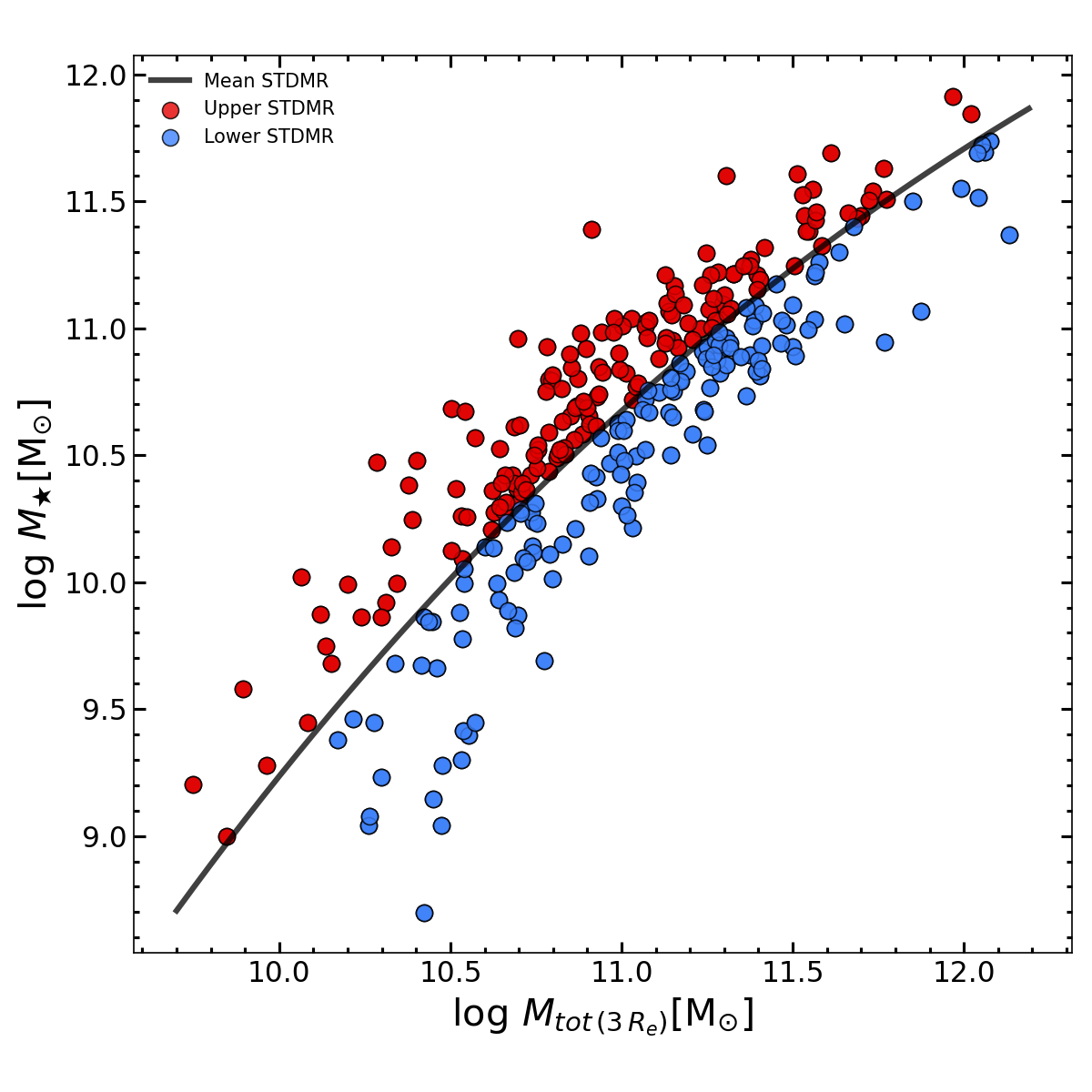}
  \vspace{-10pt}
  \caption{ Mean stellar-to-total dynamical mass relation for CALIFA. Galaxies are divided according to the position of the mean STDMR (black solid line). Galaxies above the mean relation are shown with red, and while the ones below with blue circles.  }
     \label{fig:meanTDMR_class}
\end{figure}

\begin{figure*}
\centering
\includegraphics[width=0.75\hsize]{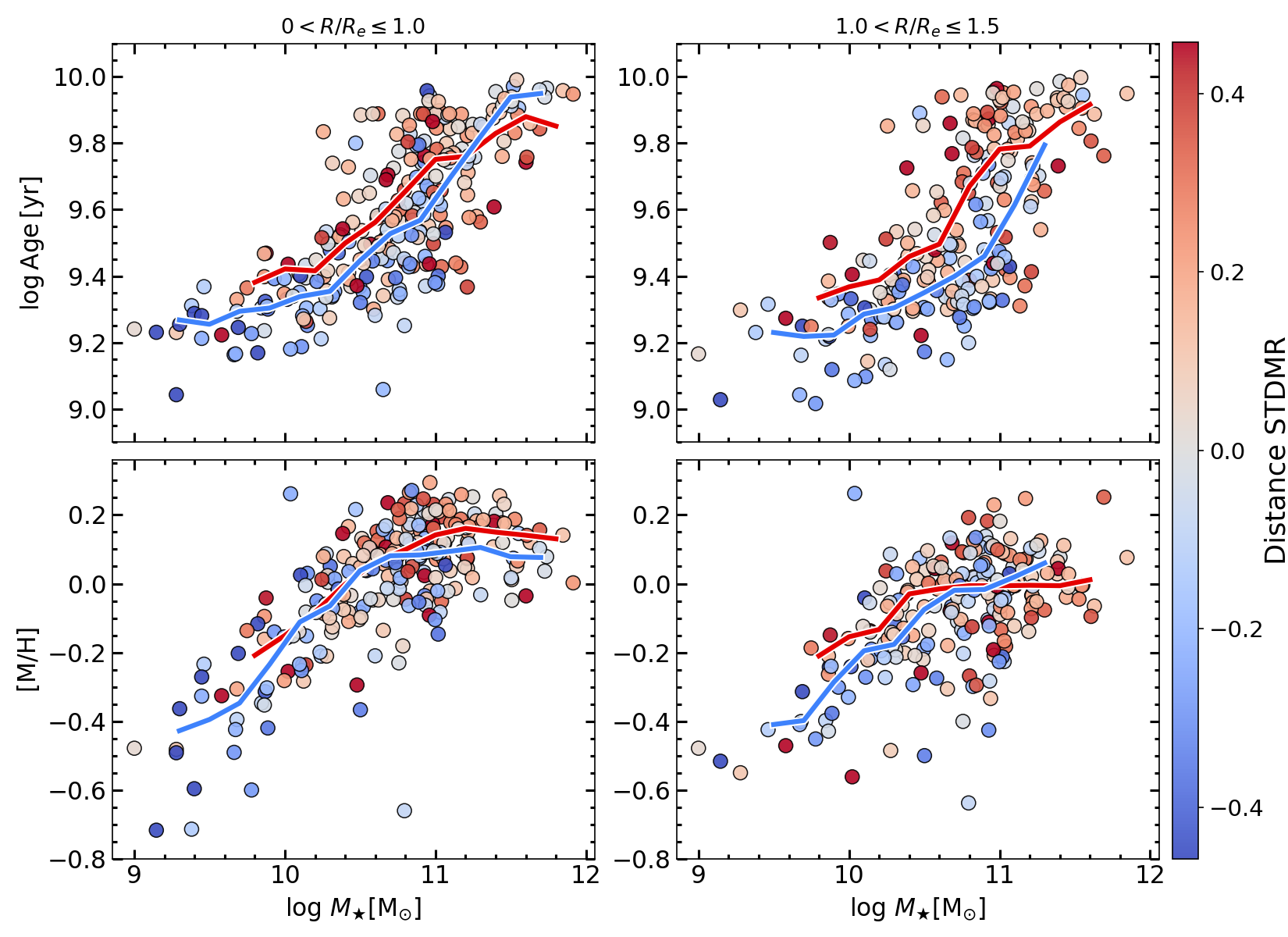}
  \caption{Scaling relations between age/[M/H] and stellar mass for galaxies with different positions with respect to the mean STDMR. Scaling relations with age are shown in the upper panels and the ones with [M/H] in the bottom ones. Ages ([M/H]) are measured within the inner ($r/R_e \leq 1 $) and within outer regions  ($ 1 <r/R_e \leq 1.5 $) of the galaxies (left and right panels respectively). For individual galaxies, median ages ([M/H]) (computed within the corresponding annuli) are plotted as a function of stellar mass color-coded with their vertical offset with respect to the mean STDMR (in the STDMR plane). Median ages ([M/H]) of the galaxies above and below the STDMR (see Fig. \ref{fig:meanTDMR_class}) are indicated with solid lines, red for galaxies above the mean STDMR and blue for the ones below. }
     \label{fig:age_mstar}
\end{figure*}

In section \ref{sec:STDMR:radialregions}, we observed a complex behavior of ages and [M/H] across the STDMR, showing dependencies both on stellar and total mass that vary with radii. The strength of the dependence of age on total mass varies at different radial annuli within the galaxies, while for metallicity it is mainly present in the inner regions, as indicated by the partial correlation analysis. However, the partial correlation analysis does not indicate how the effect of total mass varies with stellar mass. In this section, we aim to further investigate this effect of total mass and its dependence with radii. For that, we investigate the stellar mass regime in which this dependence is more significant both in the inner and outer parts of galaxies.

To this goal, we divide the galaxies according to their position with respect to the mean STDMR (Fig. \ref{fig:meanTDMR_class}). The mean STDMR (<STDMR>) is computed through symbolic regression, as described in detail in Appendix \ref{sec:symbreg}. In short, symbol regression algorithms find interpretable symbolic equations by using mathematical formulas to approximate the relation between input and output variables. Here, we looked for an expression of the form $\log M_{\star}=f(\log M_{tot})$:

\vspace{-10pt}
\begin{equation}
    \frac{\log M_\star}{\log10^{11}} = - \frac{\log M_{\text{tot}}}{\log10^{11}} - \frac{2.22 \times \log10^{11}}{\log M_{\text{tot}}} + 4.19 
\end{equation}
This equation is shown as a black solid line in Fig. \ref{fig:meanTDMR_class}. Then, we divide the galaxies into two groups: above and below the relation (red and blue circles in Fig. \ref{fig:meanTDMR_class}, respectively). By construction, note that galaxies above the relation have on average lower total masses than the ones below the relation  (at fixed stellar mass), or in other words a higher stellar-to-total mass ratio. 

Then, we investigate the scaling relations between the stellar population properties and stellar mass for galaxies above and below the STDMR. Stellar population properties are measured within the inner and outer regions of the galaxies to explore how these scaling relations vary with galactocentric distance.

Fig. \ref{fig:age_mstar} show the scaling relations with stellar mass of age (upper panels) and M/H (bottom panels), for galaxies with different positions with respect to the mean STDMR. Stellar populations are measured within the inner ($r/R_e \leq 1 $) and outer regions ($ 1 <r/R_e \leq 1.5 $) of the galaxies (left and right panels, respectively), as in section \ref{sec:STDMR:radialregions}. Each panel, shows median ages/[M/H] as a function of stellar mass, where individual galaxies are indicated with circles color-coded with their vertical offset with respect to the mean STDMR. We perform a running median using a moving stellar mass window of 0.5 dex, with an overlapping fraction of 0.6, imposing a minimum of 5 galaxies per bin. Median ages/[M/H] of the galaxy distributions are indicated with solid lines, red for galaxies above the mean STDMR and blue for the ones below.

\vspace{-10pt}
\paragraph{\emph{Age} --} At first glance, the upper panels of Fig. \ref{fig:age_mstar} show that stellar populations of galaxies above the relation tend to have older ages than galaxies below. The difference between the two galaxy groups is evident mostly at $M_\star < \rm 10^{11} M_{\odot}$ and stronger for the outer galaxy regions. More in particular, we also observe that the scaling relation's normalization for galaxies below the STDMR decreases in their outskirts, having overall younger stellar populations.

The dispersion in the $\rm Age-M_{\star}$ relation is highest at intermediate masses ($\rm \sim10^{10.5}-10^{11}M_{\odot}$), showing a bimodality that appears to be connected to the distance from the STDMR. Galaxies with positive deviation from the STDMR (i.e. lower at fixed $M_{\star}$) tend to populate an old sequence, whereas galaxies with a negative offset tend to populate a young sequence. The separation between the two sequences is larger when looking at the age of the galaxies' outskirts, consistent with the stronger dependence on $M_{tot}$ observed in \ref{fig:stel_pops_annuli}. Nevertheless, we notice that galaxies above the mean STDMR do not sharply separate from those below the STDMR. On the other hand, looking in more detail into galaxies above the relation, there is a broadening of their distribution for intermediate stellar masses: some galaxies above the STDMR have younger ages,similar to those galaxies below the STDMR. This suggests that just separating the galaxies into above and below the STDMR does not fully capture the behavior of stellar population properties across the STDMR. In fact, we observe in Fig. \ref{fig:stel_pops_annuli} that the direction of increase of stellar population parameters deviates from being perpendicular to the mean STDMR.

\vspace{-10pt}
\paragraph{\emph{[M/H]} --} The bottom panels of Fig. \ref{fig:age_mstar} show that the [M/H] of galaxies above and below the mean STDMR follow similar scaling relations with $M_{\star}$. Yet, galaxies below the relation show a lower normalization (i.e.,  overall lower metallicities), although with a difference between the two groups less pronounced than for age (at fixed $M_{\star}$). This difference is more noticeable within the inner regions  for intermediate mass galaxies ($ M \gtrsim 10^{10.5} M_{\odot}$), while this distinction is less clear at larger radii. In fact, we note that the differences in metallicity are mainly found in the central regions  (r<$0.5 \, R_e$) according to Fig. \ref{fig:stel_pops_annuli}. Thus, averaging the stellar populations within $1 \, R_e$ could explain the weakening of the trends with total mass within the inner regions of intermediate and high mass galaxies.

\section{Stellar populations scaling relations with total mass: Above and below the STDMR}
\label{sec:projectin_STDMR_mtot}
\label{sec:project_tot_STDMR}

\begin{figure*}
\centering
\includegraphics[width=0.75\hsize]{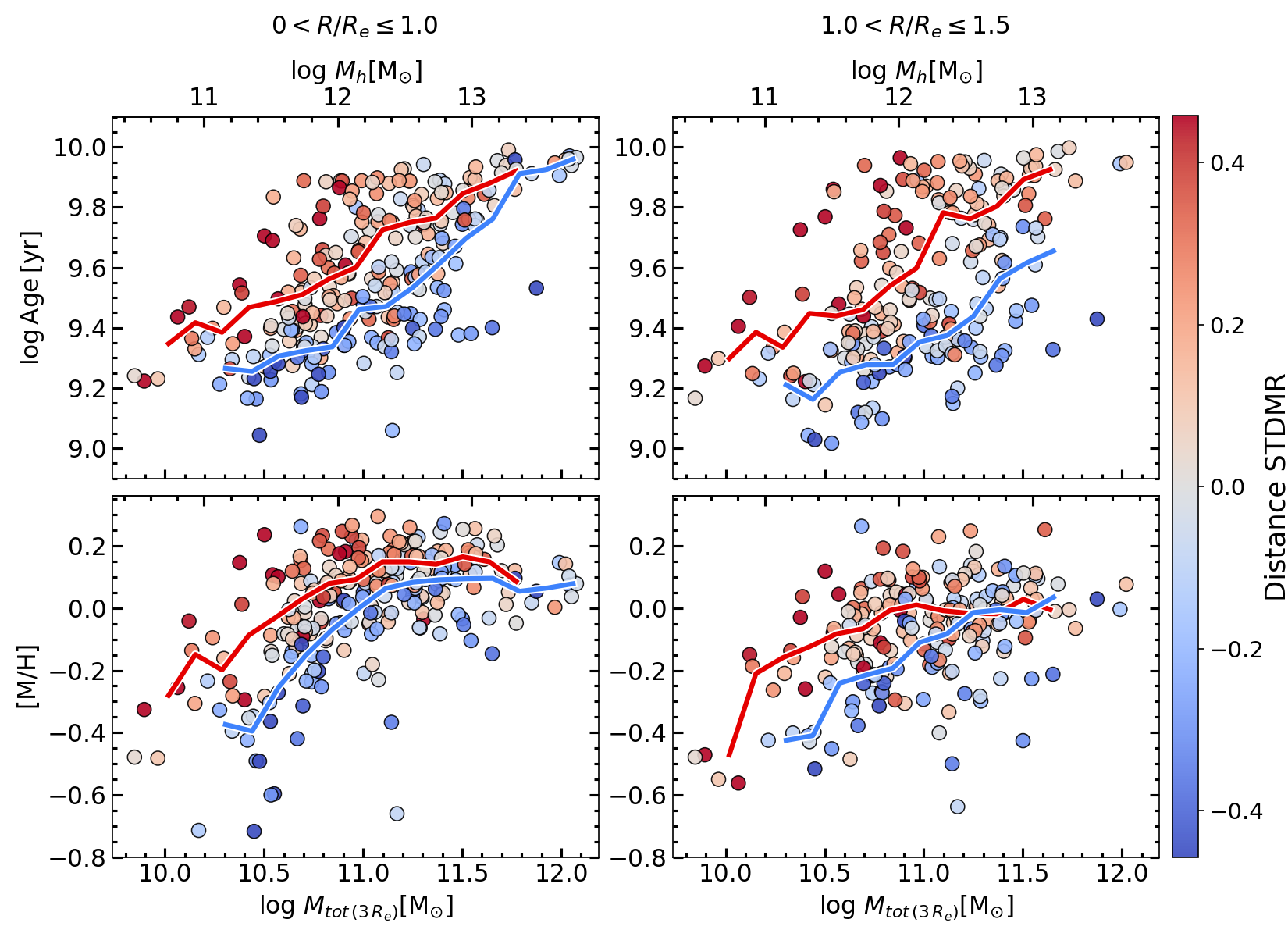}
  \caption{Scaling relations between age/[M/H] and total mass for galaxies at different radial annuli, with different positions with respect to the mean STDMR. Scaling relations with age are shown in the upper panels, and the ones with [M/H] in the bottom ones.  Ages ([M/H]) are measured within the inner ($r/R_e \leq 1 $) and within outer regions  ($ 1 <r/R_e \leq 1.5 $) of the galaxies (left and right panels respectively). For individual galaxies, median ages ([M/H]) (computed within the corresponding annuli) are plotted as a function of total mass color-coded with their vertical offset with respect to the mean STDMR (in the STDMR plane). Median ages ([M/H]) of the galaxies above and below the STDMR are indicated with solid lines, red for galaxies above the mean STDMR and blue for the ones below (see Fig. \ref{fig:meanTDMR_class}). To guide the eye, a halo mass conversion is reported in the upper axis (see text).} 
     \label{fig:age_mtot}
\end{figure*}

To investigate how the stellar populations behave across the STDMR at fixed $M_{tot}$, we perform a similar analysis to the one shown in section \ref{sec:total_mass}. Note that we also show this study to facilitate the comparison with different works studying the SHMR, given that the trends at fixed $M_{\star}$ cannot be directly extrapolated or interpreted at fixed $M_{h}$ due to the inversion problem see SD24 for a detailed discussion). 

We explore scaling relations of stellar population properties with total mass for galaxies above and below the STDMR. This division according to the position of the galaxies with respect to the mean STDMR corresponds to the one described in section \ref{sec:total_mass}. We measured stellar population properties within the inner and outer regions of the galaxies to study their dependence with radii, analogously to section \ref{sec:projectin_STDMR}.

In Fig. \ref{fig:age_mtot} we show the scaling relations of age (upper panels) and [M/H] (bottom panels) with total mass, for galaxies with different positions with respect to the mean STDMR. Stellar populations are measured within the inner ($r/R_e \leq 1 $) and outer regions ($ 1 <r/R_e \leq 1.5 $) of the galaxies (left and right panels, respectively). Each panel shows median ages/[M/H] as a function of total mass, where individual galaxies are indicated with circles color-coded with their vertical offset ($\Delta \log M_{\star}$) with respect to the mean STDMR. We perform a running median using a moving stellar mass window of 0.3 dex, with an overlapping fraction of 0.55, imposing a minimum of 5 galaxies per bin. Median ages/[M/H] of the galaxy distributions are indicated with solid lines, red for galaxies above the mean STDMR and blue for the ones below. The correspondence between total dynamical mass and dark matter halo mass is indicated in the upper axis. To convert total masses into halo masses, we follow SD24 and use the linear fit of the correlation between CALIFA total dynamical masses and the halo masses from the \citet{2007ApJ...671..153Y} group and cluster catalog (SD24), which is based on the subsample of galaxies in common.

\vspace{-10pt}
\paragraph{\emph{Age} --} Upper panels of Fig. \ref{fig:age_mtot} clearly shows how galaxies in the upper STDMR tend to have older ages compared to the ones below the STDMR both in the inner and outer regions of the galaxies. We observe this difference across the whole total mass range, except in the low and very massive regime where the statistics is poor. Moreover, galaxies in the lower STDMR tend to have younger ages in their outer regions, lowering  the normalization of the age-$\rm M_{tot}$ relation and increasing the difference between the two groups. 

On the other hand, similarly as in Fig. \ref{fig:age_mstar}, we also observe a large scatter in the relation for both the inner and outer regions of galaxies above the STDMR (left and right panels). Galaxies with positive distances with respect to the mean STDMR can be both very old and similarly young as those with negative distances. This is observed especially in the intermediate mass regime ($M_{tot}\sim 10^{10.7-11.5} M_{\odot}$).

\vspace{-10pt}
\paragraph{\emph{[M/H]} --} In the bottom panels of Fig. \ref{fig:age_mtot} we see clearly that galaxies in the upper STDMR tend to have higher metallicities compared to galaxies below the STDMR at all radii. In this case, the difference in [M/H] between galaxies from the two groups is more noticeable than the one observed in Fig. \ref{fig:age_mstar} when studying the [M/H]-$\rm M_{\star}$ relation. Looking at the median relation, we observe a steep increase at lower masses followed by a flattening at higher total masses. Comparing the two radial regions, we observe that galaxies in the inner regions tend to have higher metallicities than in their outskirts at all total masses. The offset between the two groups is quite similar in both regions, except at high mass where the differences are smaller or even vanishing at $r$>$ 1 \, R_e$.

\section{Stellar population profiles}
\label{sec:profiles}

In section \ref{sec:STDMR:radialregions} we found that the strength of the dependence of stellar populations on total mass varies according to their spatial location within the galaxies, with the behavior of age and [M/H] being decoupled. In section \ref{sec:projectin_STDMR} we also found that intermediate mass ($\rm \sim 10^{10.5}-10^{11} M_{\odot}$) galaxies above and below the STDMR seem to show an age bimodality (at fixed stellar mass), which is more noticeable in their outskirts. Thus, in section \ref{sec:profilessovebelow}, we further investigate how the dependence on total mass varies with galactocentric distance by exploring age and [M/H] profiles of galaxies across the STDMR in this intermediate mass regime.

Furthermore, we expect that the bimodality observed in the age profiles of galaxies with different total masses is connected to morphology. SD24 shows how the scatter of the STDMR is correlates with the morphological type of the galaxies, with earlier-types (ETGs) being more massive than late-types (LTGs) at fixed total mass. At the same time, it has been extensively discussed in the literature that ETGs and LTGs in the Local Universe display different age and [M/H] gradients and profiles (see more details in Section \ref{sec:intro}). Hence,  we also explore how morphology is connected to the stellar population profiles shown in section \ref{sec:morphology}.

\subsection{Above and below the STDMR}
\label{sec:profilessovebelow}

\begin{figure*}
\centering
\includegraphics[width=0.9\hsize]{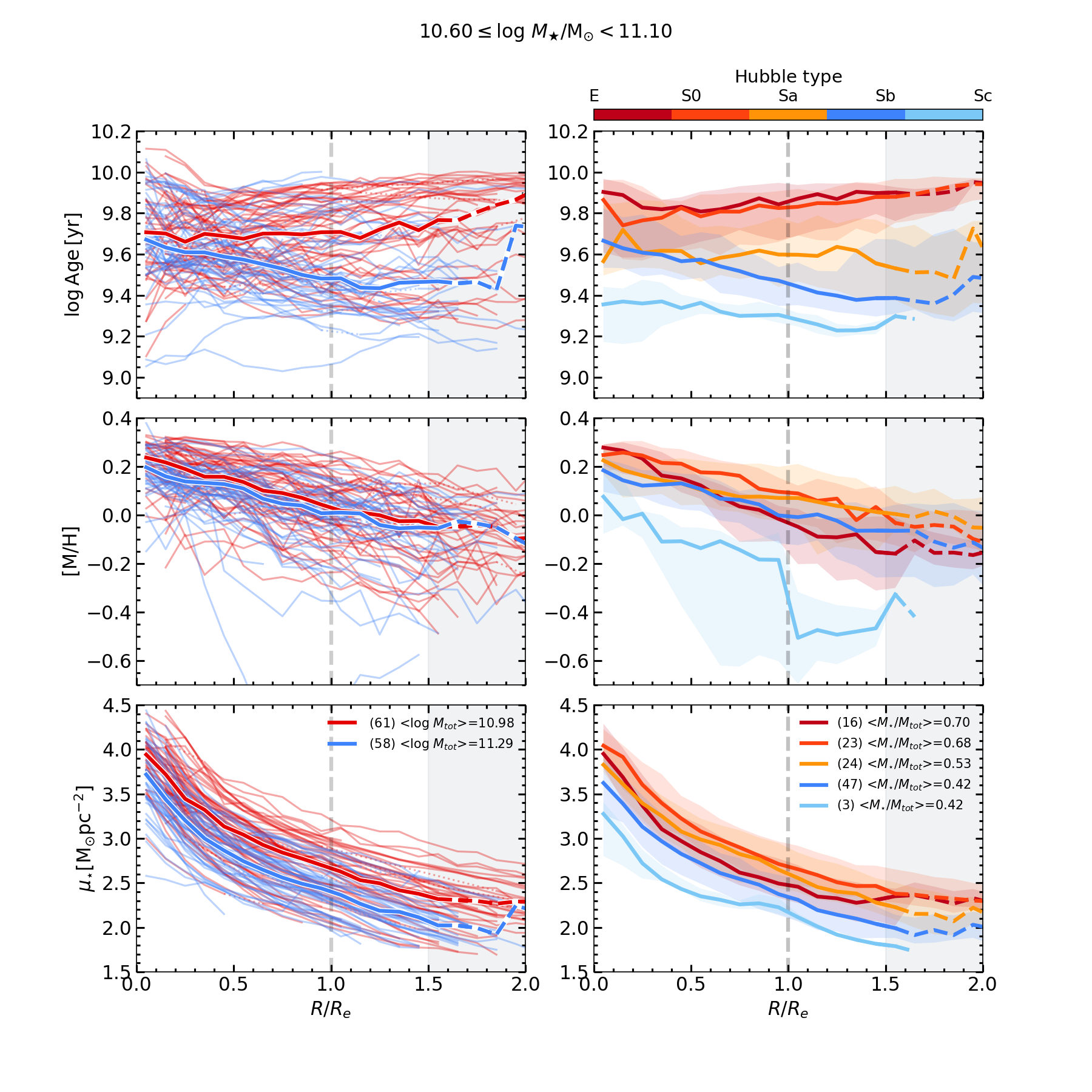}
  \vspace{-25pt}
  \caption{Dependence on total mass and morphology of stellar population profiles in narrow stellar mass bin $ \log M_{\star} \rm /M_{\odot} \ \epsilon $  $\rm (10.6, 11.1]$. Age (upper panels), [M/H] (middle panels), $\mu _{\star}$ (bottom panels) profiles are plotted as a function of galactocentric distance. \textit{Left column:} Their dependence on total mass: Profiles of individual galaxies are indicated with dashed lines color-coded according to their position with respect to the mean STDMR (red for galaxies above the mean relation, and blue the ones below). Median profiles of galaxies above and below the mean STDMR are shown as a red and blue solid lines, respectively. By construction, galaxies above the relation (red) have lower total masses than galaxies below (blue). The caption indicates the median total mass and number of galaxies for both galaxy subsamples. \textit{Right column:} Their dependence on morphology: Median profiles of different morphological types are shown with different colors and  shaded regions indicate the 16th and 84th percentiles of the corresponding distributions. The caption indicates the median stellar-to-total mass of each galaxy subsample. For reference, $1 \, R_e$ is indicated with a vertical grey dashed line in each panel.}
     \label{fig:age_met_profiles}
\end{figure*}

In order to isolate the effect of total mass, or in other words, its correlation with stellar population properties, we first select galaxies in a narrow intermediate stellar mass bin where the dynamic range in total mass is larger and the effect of total mass appears more prominent (see section \ref{sec:projectin_STDMR}): $ \log M_{\star} \rm /M_{\odot} \ \epsilon $ $\rm (10.6, 11.1]$. Then, we divide the galaxies according to the position with respect to the mean STDMR (i.e., above and below the relation), analogously to section \ref{sec:projectin_STDMR}. By construction, galaxies above the relation have on average lower total masses than the ones below.  We also checked that the stellar mass distributions of galaxies above and below the relation are not biased and have a similar median stellar mass.

We show the dependence on total mass of age, [M/H]  and stellar mass surface density ($\mu_{\star}$) profiles in the left column of Fig. \ref{fig:age_met_profiles} for this narrow intermediate stellar mass bin. We show age, [M/H] and $\mu_{\star}$ profiles in the upper, middle and bottom panels, respectively. Each panel shows azimuthally median-averaged ages/[M/H]/$\mu_{\star}$ profiles as a function of galactocentric distance (the elliptical semi-major axis normalized by the half-light semi-major axis $R_e$). To construct the age/[M/H]/$\mu_{\star}$ profile of a galaxy, we compute the median age/[M/H]/$\mu_{\star}$ of the spaxels contained in thin elliptical annuli ($0.1 \, R_e$ width) centered on the galaxy’s nucleus using the elliptical semi-major axis, analogously to section \ref{sec:STDMR:radialregions}. The profiles of individual galaxies are indicated with lines color-coded according to their position with respect to the mean STDMR (red for galaxies above the mean relation, and blue for the ones below). Median profiles of galaxies above and below the mean STDMR are shown as thick red and blue solid lines, respectively. For reference, $1 \, R_e$ is indicated with a vertical grey dashed line. We also checked that we recover similar median profiles if we exclude galaxies that do not extend beyond $1.5 \, R_e$, in order to have the median profile computed with the same galaxies at all radii.

\vspace{-10pt}

\paragraph{\emph{Age} --} The upper left panel of Fig.  \ref{fig:age_met_profiles} shows that there is a clear difference between the age profiles of galaxies that have different total masses at fixed $M_{\star}$. Median profiles of galaxies above the STDMR (i.e., the ones with lower total masses) tend to have flat age profiles, that even increase at the galaxies' outskirts. In contrast, age profiles of galaxies below the STDMR (i.e., higher total masses) tend to decrease with galactocentric distance. In this sense, the difference between the two groups increases with galactocentric distance, as already hinted at Fig. \ref{fig:stel_pops_annuli}. 

Similarly to what we observed in Fig. \ref{fig:age_mstar}, we also observe a bimodality in the shape of the age profiles, which is not fully captured by splitting galaxies into upper and lower STDMR. In fact, we see that although the majority of galaxies in the upper STDMR have approximately flat profiles, there are some with negative slopes, akin the ones in the lower STDMR. 

\vspace{-10pt}

\paragraph{\emph{[M/H]} --} In the middle left panel of Fig. \ref{fig:age_met_profiles} we observe that galaxies above and below the STDMR relation have negative steep [M/H] profiles. [M/H] profiles in the upper STDMR seem to be steeper than the ones of lower STDMR galaxies, with the former also having higher [M/H] normalizations in the inner regions. Thus, this translates into the difference in metallicity being mainly observed in the inner regions. As the profiles for lower STDMR galaxies are less steep, the difference in metallicity between the two groups eventually disappears at larger radii ($r> 1 \, R_e$).

\vspace{-10pt}
\paragraph{\emph{$\mu _{\star}$} --} The bottom left panel of Fig. \ref{fig:age_met_profiles} indicates that $\mu _{\star}$ profiles of galaxies above the STDMR have on average a higher normalization than the ones of galaxies below the STDMR, by 0.3 dex ($\sim$ a factor of two) on average. As the stellar mass distributions of the two galaxy samples are very similar, we checked that this offset in normalization is due to a combination of differences in $R_e$ and profile shape, with upper STDMR having smaller $R_e$.

Our findings indicate that not only integrated properties of intermediate mass galaxies are connected to the scatter of the STDMR, but also their stellar population profiles, especially in the case of age. Finally, we also checked that when selecting a narrow total mass bin, e.g., $\log M_{tot} \ \rm  \epsilon \ (10.9,11.4)$, that covers an intermediate stellar mass regime, we recover very similar stellar population profiles to the ones shown in the left column of Fig. \ref{fig:age_met_profiles} for galaxies above and below the STDMR.

\subsection{Morphology}
\label{sec:morphology}

To explore the connection between stellar population profiles and morphology, we now divide galaxies within a narrow stellar mass bin, as in section \ref{sec:profilessovebelow}, according to their morphological type and compute their median stellar population profiles. In the the right panels of Fig. \ref{fig:age_met_profiles}, we show the median profiles of age (upper), [M/H] (middle), $\mu_{\star}$ (lower) for galaxies with different morphological types through different colors. Shaded colored areas indicate the 16th -- 84th percentile range of the corresponding distributions. Median $M_{\star}/M_{tot}$ ratios of galaxies with different morphology are indicated in the legend. We note that to ensure sufficient statistics, the stellar mass selection window cannot be too narrow. As a result, a limitation of this analysis is the different stellar mass distributions among different morphologies, which result from the width of this selection window. In this stellar mass bin, earlier-type galaxies have slightly higher median stellar masses than later-types, which also enhance the difference of their $M_{\star}/M_{tot}$ ratios. Yet, the difference in stellar mass between ellipticals and Sa or Sb is lower than 0.15 dex. We checked that when using narrower stellar mass bins we found stellar population profiles in agreement with the ones presented in this work.

\paragraph{\emph{Age} --} The shape bimodality observed previously when comparing the age profiles of galaxies with different total masses, is seen again here, but now related to morphology. We clearly observe how earlier-types (Ellipticals and S0) are generally characterized by having older ages in the inner regions, flat age profiles, and lower total masses. On the contrary, spiral galaxies have on average slightly negative age profiles in the inner regions that flatten in their outskirts, with later-types having a lower normalization (i.e., overall younger ages). We find relatively flat age profiles for Sc galaxies, the low statistics prevent us to draw any conclusions.

\paragraph{\emph{[M/H]} --}  To first-order, we observe that all morphological types tend to have negative metallicity profiles, yet few difference arise when comparing their steepness and normalization. Ellipticals have the steepest negative profiles with higher metallicities in the inner regions, slightly flattening in the outer regions. S0s behave in a similar manner to late-types, with less steep negative profiles than ellipticals. The normalization decreases moving to later-types (i.e., overall more metal-poor), similarly to age. Yet, the steep [M/H] profiles of ellipticals translate, on average, into more metal-poor stellar populations at $r>0.5 \, R_e$ compared to S0s and spirals (except for Sc). In fact, in Fig. \ref{fig:age_mstar} we observed that the metallicity of the outer regions of the galaxies is on average flat with $M_{\star}$ for galaxies above the STDMR, and lower than the metallicity of galaxies below the STDMR at high masses. 

\paragraph{\emph{$\mu _{\star}$} --}  All morphological types have profiles that decrease monotonically with $r$, declining more rapidly in their inner parts ($r\lesssim0.5 \, R_e$), and flattening in their outskirts. Yet, the main differences between profiles of different morphologies are found in their steepness. We observe a steeper decline for ellipticals than for S0 and spirals in the inner regions. The normalization in the inner regions is higher and very similar for ellipticals, S0 and Sa, while later-type spirals have lower $\mu_{\star }$ values. Note that the slightly different median stellar masses can also lead to differences in normalization.\\

Finally, this connection with morphology is also seen when looking at the stellar population gradients and stellar mass (Appendix \ref{ap:gradients}). We observe that age gradients of ETGs and LTGs exhibit parallel sequences as a function of stellar mass. For both morphological types, age gradients decrease with increasing stellar mass. Yet, they have a different normalization and different $M_{\star}$/$M_{tot}$ ratios. ETGs tend to  have generally flat or positive age gradients and higher $M_{\star}$/$M_{tot}$ values, while LTGs negative gradients and lower $M_{\star}$/$M_{tot}$ ratios. In this sense, these two sequences coexist in the intermediate mass regime ($  M_{\star}\rm \sim10^{10.5-11}M_{\odot}$),  resulting in ETGs with lower $M_{tot}$ having flat or mild positive gradients and LTGs with higher $M_{tot}$ showing negative age gradients, at fixed $M_{\star}$. In the case of metallicity, although both ETGs and LTGs show negative [M/H] gradients, ETGs tend to show slightly steeper ones and higher $M_{\star}$/$M_{tot}$ ratios at fixed $M_{\star}$ ($ M_{\star}\rm \sim10^{10.5-11}M_{\odot}$).

\section{Discussion}
\label{sec:disc}

Our analysis of stellar populations across the stellar-to-total dynamical mass relation reveals a connection between the scatter of this relation and spatially-resolved stellar populations. In this section, we discuss our results in terms of previous literature as well as different scenarios that could explain our findings.

\subsection{Connection with the SHMR} 

Currently, there is no consensus on how galaxy properties are connected to the scatter of the SHMR (see SD24 for a detailed discussion). Under the assumption that total dynamical mass could act as a proxy of halo mass (see SD24), our findings are in line with works indicating that in the intermediate halo mass regime ($\sim10^{12-13.5} \rm M_{\odot}$), more massive galaxies at fixed halo mass tend to be older and passive. This is also in agreement with SD24 despite the different stellar population analysis and models involved. Going beyond the global properties, we also discuss our findings in terms of the spatially-resolved SHMR.

Previous works studying spatially-resolved stellar population properties across the SHMR are based on MaNGA \citep{2015ApJ...798....7B} galaxies. Note that we compare our results with the ones of central galaxies because our sample is highly biased towards central and isolated galaxies by construction (see SD24 and section \ref{sec:sample}). 

For intermediate-mass centrals, \citet{2022ApJ...933...88O} finds negative [Fe/H] gradients, with galaxies in the upper SHMR (i.e., higher $M_{\star}$ at fixed $M_h$, or lower $M_h$ at fixed stellar mass) having slighlty higher [Fe/H] values, similarly to what we find for metallicity. This dependence on halo mass can be traced up to 1.5$R_e$, in contrast to our results, as we only observe it within the inner regions (r<0.5$R_e$). Yet, we highlight the broad agreement, given the different galaxy sample (they focus on passive galaxies only), different spectral fitting approach and models, and the use of halo mass instead of total mass. In the case of age, our results do not fully agree in the mass regime in common. Yet, this is not surprising as that scatter of the STDMR correlates with morphology and SFR (SD24). Hence, selecting only passive or early-type galaxies would naturally dilute the differences we find in age. The lack of star-foming galaxies in the \citet{2022ApJ...933...88O} sample could also be the reason why the dependence on metallicity is preserved in the outskirts of the galaxies, as LTGs would have less steep negative [M/H] profiles, leading to similar [M/H] at larger radii. 

On the other hand, \citet{2024MNRAS.530.4082Z} find that the centers (r<0.5$R_e$) of more massive galaxies  formed the bulk of their mass earlier on than less massive ones at fixed halo mass, in agreement with our results. This is traced by both their Mg/Fe abundances, and the formation time of half of their present-day stellar mass, $t_{50}$. Additionally, although their sample is restricted to S0s and later-types, they also find that the scatter of the SHMR also correlates with morphology as well. Similarly to the results of SD24 across the STDMR, they find later-types at the bottom part of the SHMR (at fixed halo mass).

\subsection{Galaxy bimodality and connection with morphology}
\label{disc:origin}

When looking at galaxies of the Local Universe, there is a bimodality on several properties related to their stellar populations, dynamical state and level of star formation \citep[e.g.,][]{2001AJ....122.1861S,2003MNRAS.341...54K,2004ApJ...600..681B,2007ApJS..173..267S}. Young, late-type, star-forming galaxies dominate the low-mass regime, while old, early-type, and passive galaxies are more prominent at the high-mass end of the galaxy population. More relevant for this work, ETGs and LTGs also exhibit different stellar population gradients and profiles (see section \ref{sec:intro}). These two galaxy populations co-exist together in the intermediate-mass regime ($\sim10^{10.5-11} \rm M_{\odot}$) \citep[e.g.,][]{2025A&A...703A...5M}. This range is quite interesting as galaxies of $\rm M_{\star}\sim10^{10.5}M_{\odot}$ typically probe the peak of galaxy formation efficiency in terms of the SHMR. On the other hand, despite having similar stellar masses at present-day, the physical drivers that result in all the different properties of ETGs and LTGs today remain under debate.

In agreement with the literature, in this work we also observe a bimodality in the scaling relations between age and stellar mass, which is enhanced when only the ages of the outer regions of the galaxies are considered (for the $\sim10^{10.5-11} \rm M_{\odot}$ stellar mass regime). Most interestingly, we find that this bimodality is connected to the galaxies' offset with respect to the mean STDMR, and hence, to their total masses at fixed stellar mass: older galaxies tend to populate the upper STDMR while younger galaxies occupy the bottom part of the relation. Consistently with the fact that these age differences are enhanced in the stellar populations of the galaxies' outskirts, we also observed this bimodality in their age profiles. Galaxies in the upper part of the STDMR (i.e., with lower total masses) have relatively flat age profiles, while galaxies in the bottom part of the relation (i.e., higher total masses) have negative ones.

In the literature, relatively flat age profiles have been typically associated to ETGs, while LTGs tend to exhibit negative ones (see section \ref{sec:intro}). Thus, we also explored the connection between stellar population profiles and morphology \citep[see also e.g., ][]{Gonzalez-Delgado:2015aa,Parikh:2019}, as well as how this relates to the galaxies' total masses. For that, we study stellar population profiles as a function of morphology in a narrow stellar mass bin  (right panels of Fig. \ref{fig:age_met_profiles}). We find that ETGs display steep negative metallicity gradients within 1$R_e$ that flatten in their outskirts. We observe that ETGs have broadly flat age profiles, although ellipticals have valleys around $r\sim0.5R_e$ \citep[see also ][]{2020MNRAS.491.3562Z}, and mildly increasing age profiles for S0s. On the contrary, spirals tend to show negative gradients in both age and [M/H]. In the case of Sc galaxies, they are relatively flat, although our statistics are poor. Furthermore, in addition to different morphologies and stellar population profiles, we find that these galaxies have different total masses. Specifically, ETGs have higher stellar-to-total mass ratios than LTGs at fixed stellar mass.

Notably, this connection between morphology and stellar-to-total mass ratios is likely driving the variations of stellar population properties with radius across the STDMR (Fig. \ref{fig:stel_pops_annuli}). The morphological type of the galaxies correlates with the scatter about the STDMR, showing a primary dependence on stellar mass and a secondary one on total mass (SD24). This complex behavior across the STDMR results in ETGs occupying the upper part of the relation (i.e., higher stellar-total mass ratios), and LTGs being in the bottom one (i.e., lower stellar-total mass ratios). Thus, the distinct stellar population profiles of ETGs and LTGs (Fig. \ref{fig:age_met_profiles}) naturally explain why the dependence on total mass is diluted in the galaxies' outskirts for metallicity, while it is enhanced for age. The difference between relatively flat or slightly increasing age profiles of ETGs and negative ones for LTGs naturally gives rise to a greater age difference in the outskirts of the galaxies. In the case of S0s and later-types they do not have so steep negative metallicity profiles as ellipticals in the inner regions, leading to similar [M/H] values at $r>0.5R_e$.

SD24 also shows that the stellar apparent angular momentum and SFRs behave in a similar manner across the STDMR plane (their Figure 2), correlating with both stellar and total mass. These properties are sensitive to physical processes that are associated to different time-scales during the formation of a galaxy (i.e., SFR traces the last 10-100 Myr, whereas age the bulk of star formation in the history of a galaxy; morphology and angular momentum are sensitive to the dynamical state of a galaxy and set earlier on its evolution). Yet, all of them show a similar primary dependence on stellar mass, and a secondary one on total mass. SD24 argues that the scatter about the STDMR maps different evolutionary pathways of galaxies. At fixed total mass, galaxies with higher stellar masses tend to exhibit features of a more advanced evolution: they are older, more metal-rich and more dispersion dominated, lower SFRs, have early-type morphologies (their Figs. 1 and 2), and, as reported here, relatively flat age profiles and steep negative metallicity gradients. On the contrary, less massive galaxies seem to be less evolved, as they are younger, more metal-poor and more rotationally supported galaxies, have higher SFRs, late-type morphologies, as found in this work, negative age and less steep negative metallicity gradients. An analogous argument can be done at fixed stellar mass, but reversed. 

\subsection{Formation scenarios and physical origin}

\subsubsection{Insights from stellar population gradients}

To form the old, metal-rich cores of ETGs \citep[e.g.,][]{1973ApJ...179..731F,1989PhDT.......149P,2005ApJ...621..673T,2015MNRAS.448.3484M}, star formation must have been extremely efficient at $z\gtrsim3$. Deep central potentials would help retain metals, while rapid chemical enrichment and star formation would lead to steep negative metallicity gradients and flat age gradients. Yet, despite the gas-rich environment at $z\sim2-3$, star formation needs to be halted early on to preserve the properties imprinted in the early Universe. In fact, the size growth of ETGs since $z\sim2$ 
\citep[e.g.,][]{2007MNRAS.382..109T,2014ApJ...788...28V} is generally attributed to dry mergers with low-mass satellites \citep[e.g.,][]{2009ApJ...697.1290B,2012ApJ...744...63O,Huang_2013}. In this sense, ETG's stellar populations are generally understood as a result of in-situ growth in their cores and ex-situ growth from accretion of low-mass satellites in their outskirts. In fact, Fig. \ref{fig:age_met_profiles} shows some evidence of increased stellar mass surface densities in the outskirts of massive ETGs, consistent with the ex-situ growth scenario in their outer parts.

In contrast, late-type galaxies generally show ongoing star formation and younger populations, especially in their outer regions  \citep[e.g.,][]{2000MNRAS.312..497B,Gonzalez-Delgado:2015aa,2017A&A...608A..27G}, producing negative age gradients. While outflows and pristine gas inflows cause dilution of metals in their outskirts, deep central potential wells are more efficient in retaining metals in the inner regions, leading to negative metallicity gradients. In this way, the key difference between the formation scenarios of ETGs and LTGs is that the bulk of the LTGs' population typically does not show signs of early quenching, in contrast to ETGs, whose later growth is largely driven by accretion in their outskirts. Furthermore, recent studies indicate that the stellar mass of the outer envelopes of galaxies can trace their halo masses \citep{2020MNRAS.492.3685H,2022MNRAS.515.4722H}, suggesting halo properties' information could be encoded in the stellar mass distributions of massive galaxies. 

Additionally, spatially resolved SFHs also show the cores of massive ETGs form earlier and faster than those of LTGs of the same mass, driven by the higher central stellar mass surface densities of ETGs. In fact, local stellar ages and metallicities correlate strongly with the local stellar mass density \citep[e.g.,][]{2021MNRAS.508.4844N,2022MNRAS.512.1415Z} and the local velocity dispersion \citep[e.g.,][]{2025MNRAS.540.1069F}, both tracers of the local gravitational potential.

In this sense, the older, more chemically-enriched and more concentrated stellar populations that we find in ETGs' cores together with their steeper metallicity and stellar surface density gradients are consistent with an earlier stellar mass assembly with higher efficiency (higher SFRs) of their inner regions. The higher gas densities in their cores that lead to higher star formation efficiencies would naturally be a consequence of the shape of the galaxies' local gravitational potential at the time.

\subsubsection{Different evolutionary stages across the STDMR?}

In the intermediate-mass regime, galaxies in the upper part of the STDMR tend to show overall flat age profiles and steep negative metallicity gradients, compared to galaxies in the bottom part of the STDMR, which have both negative age and metallicity profiles. However, note that just separating between the upper and lower STDMR does not fully capture the behavior of stellar population properties across the relation. In Fig. \ref{fig:stel_pops_annuli}  we clearly observe how the direction of maximal increase of age (in the outskirts) and metallicity (in the inner regions) is not fully perpendicular to the mean STDMR (i.e., it is not a 45 degree angle). In fact, when looking at individual age profiles in the left column of Fig. \ref{fig:age_met_profiles}, some galaxies in the upper STDMR have similar negative age profiles as galaxies below the STDMR. Actually, morphology better disentangles the different profiles given that the direction of maximal increase of morphology across the STDMR is notparallel, neither perpendicular (SD24), instead, it is quite similar to the one observed for stellar population properties. 

In this work, by definition, galaxies above the STDMR have lower total masses (within 3$R_e$) than galaxies below the STDMR for a given $M_{\star}$. These total masses are tracers of the underlying global gravitational potential. As galaxies start to transition away from being dominated by the stellar component at r>$R_e$ \citep[e.g.,][]{2005ApJ...623L...5F}, SD24 discusses how these total dynamical masses could be tracing the mass of host dark matter halos of the galaxies (see section \ref{sec:intro}). In the context of the SHMR, this complex behavior of galaxy properties across the STDMR may be related to the underlying properties their host halos. Many theoretical models predict that SHMR's scatter is connected to the formation time or concentration of the halos \citep[e.g.,][]{2017MNRAS.470.3720T,2017MNRAS.465.2381M,2018ApJ...853...84Z,2018MNRAS.480.3978A}. For given halo mass, higher stellar mass galaxies are hosted by earlier-formed or more concentrated halos. In fact, the $M_{\star}/M_h$ ratio has been used as a proxy of halo formation time at fixed $M_h$ \citep[e.g,][]{2016MNRAS.455..499L,2017MNRAS.470.3720T}. In some sense, this is a reflection that higher density regions of the large-scale structure collapse at earlier cosmic times in the $\rm \Lambda$CDM cosmology. Additionally, by applying machine learning algorithms to SDSS galaxies and semi-analytical models, recent findings indicate that stellar populations' ages and the $M_{\star}/M_h$ ratios of central galaxies are the most significant features for inferring halo formation times \citep{2024ApJ...972..108L}.

Thus, galaxies in the upper STDMR  (i.e., with higher stellar masses at fixed $M_{tot}$) are possibly hosted by halos that collapsed earlier. Focusing on the majority of galaxies in the upper STDMR (with flat age profiles), they could have experienced a rapid intense formation in the early Universe driven by an earlier formation of their host halos. These galaxies have overall old and metal-rich stellar population in their cores, together with steep metallicity gradients indicating that they could have had a high star formation efficiency at early times, i.e., more efficiently retained their gas during this early star formation phase. This is likely driven by higher gas densities due to deeper central gravitational potential resulting in higher surface stellar mass densities and an earlier stellar mass assembly. Thus, these deeper central potential wells could be a reflection of more concentrated halos. In fact, at fixed $M_h$, older and more concentrated halos are predicted to bind the baryonic content (i.e., stellar and gaseous components) of centrals more strongly \citep[e.g.,][]{2002ApJ...568...52W}.

\begin{figure*}
\centering
\includegraphics[width=\hsize]{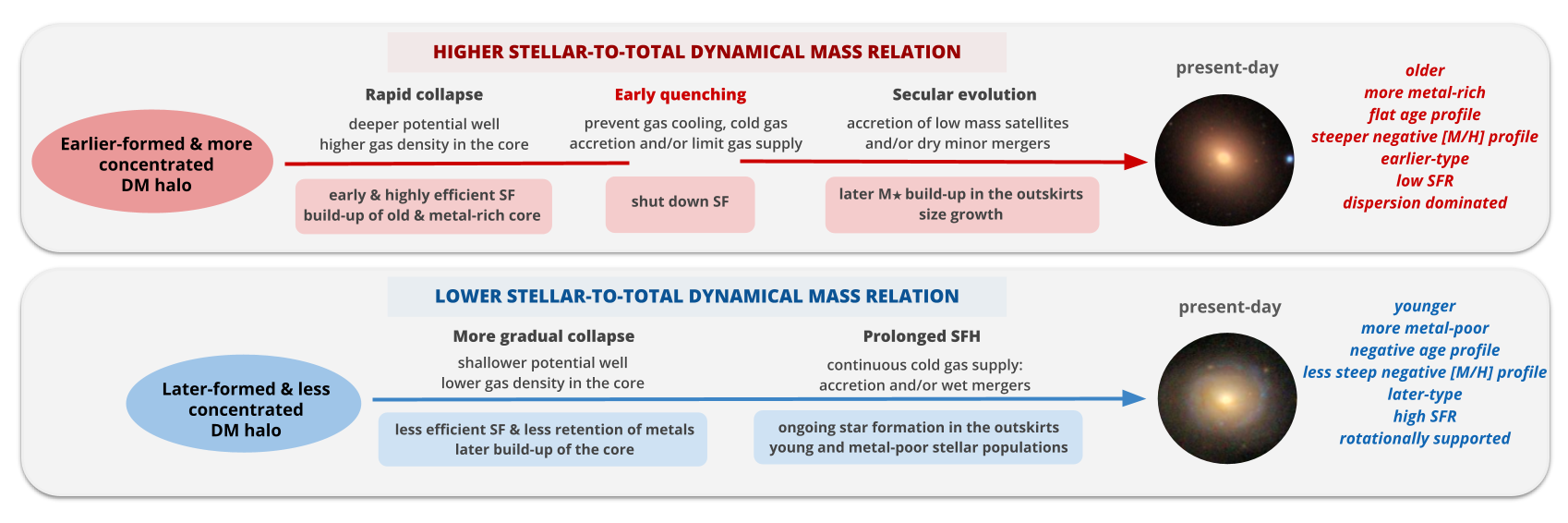}
  \vspace{-5pt}
  \caption{Illustration of the proposed evolutionary scenarios corresponding to galaxies in the upper STDMR (upper row) and the lower STDMR (lower row). The sequence evolves from left to right, ending with the different galaxy properties at present-day (on the right).
  }
     \label{fig:illustration}
\end{figure*}

Moreover, the overall old stellar population of these upper STDMR galaxies indicate that they shut down their star formation early on, and kept it shut off during their later evolution. Their flat age profiles and flattening of the metallicity profiles at $r>R_e$ is consistent with their growth in the outer parts being driven by accretion of low-mass quenched satellites and dry minor mergers. In this scenario, the earlier formation of the halos could be connected to an earlier onset of quenching in these galaxies. A schematic illustration of these scenario is presented in Figure \ref{fig:illustration}.

An earlier onset of quenching could be the result of a combination between the earlier halo assembly coupled to an earlier onset of black hole feedback. In fact, state-of-the art zoom-in hydrodynamical simulations \citep[based on the EAGLE galaxy formation model; ][]{2015MNRAS.446..521S} predict that halo assembly time can influence the evolution of the stellar and gas content of galaxies \citep{2021MNRAS.501..236D}. By \textit{genetically} modifying the assembly history of a halo (of a given mass) --essentially shifting it earlier or later in time by changing the halo's initial conditions--, its hosted galaxy ends up with different properties at present-day. In these simulations, while an earlier-formed halo would host a passive galaxy, which has lowered its rotational support, has a lower gas fraction, and has suffered strong morphological transformation from disc-like to spheroidal, a later-formed halo would host a galaxy that remains star-forming, disc-like and gas-rich. A combination of halo assembly and black hole feedback is responsible for these differences in the simulations, with earlier-formed halos fostering the growth of more massive central black holes, that inject more energy into the halos, facilitating galaxy quenching.

In the light of these simulations, our findings could be naturally explained with different galaxy evolutionary stages reflecting different halo assembly histories at fixed halo mass. Earlier-formed halos would host more massive galaxies that assembled the bulk of their inner stellar component quickly and efficiently, early on and quenched also earlier, thus ending up with older and more-metal rich stellar populations at present-day, very steep metallicity gradients, and low SFRs. Moreover, this early quenching on more massive galaxies at fixed total/halo mass, would naturally lead to a more accentuated age difference in the galaxies' outskirts, where galaxies in the upper STDMR no longer form new stars, in contrast to the more extended SFHs of lower STDMR galaxies. Moreover, the imprint of the rapid chemical enrichment of these more massive galaxies would be preserved in the inner regions, although the accretion of pristine gas and wet metal-poor minor mergers would dilute the ISM, resulting in younger and more metal-poor stellar populations in their outskirts.

Note that in these simulations, in addition to the galaxies' present SFR, an earlier halo formation affects as well their morphology and level of rotational support. As we mentioned before, in addition to morphology, the scatter of STDMR is connected to the stellar apparent angular momentum and present-day SFRs (SD24). Galaxies in the upper STDMR show earlier-type morphologies, lower SFRs and are dispersion dominated in comparison to galaxies in the bottom part of the STDMR. Thus, an earlier halo formation could also be connected to a morphological and dynamical transformation, explaining that galaxies in the upper STDMR are more dispersion-dominated with typically early-type morphologies.

However, we caution the reader, as this connection between halo assembly and black hole feedback in numerical simulations is heavily sensitive to the back hole feedback model employed. Ill-defined numerical implementations and prescriptions of baryonic physics, such as black hole feedback, can drastically change the predictions of how galaxy properties are connected to the scatter of the SHMR \citep[e.g.,][]{2024ApJ...976..148P,2024MNRAS.531.2262P,2025ApJ...987....4L}. Additionally, the connection between halo formation time and the scatter of the SHMR is model dependent \citet[see section 5.2.2. of ][]{2022ApJ...933...88O}. Semi-empirical models \citep{2019MNRAS.488.3143B} show that the correlation between  halo assembly time and the scatter of the SHMR is strongest at the high-mass end, being driven by accreted stellar populations \citep{2020MNRAS.493..337B}. While in large-scale hydrodynamical simulations the stochasticity of late mergers can instead weaken this connection for high halo masses \citep{2017MNRAS.465.2381M}.

Hence, further observational tests and constraints on secondary halo properties are crucial to bring light to this scenario. Although observations of secondary halo properties such as halo formation time are not generally available for large galaxy samples, recent works have estimated halo formation times for SDSS galaxies using semi-analytical models and machine learning techniques \citep{2024ApJ...972..108L}. On the other hand, an observational alternative to test this scenario consists on using the large-scale environment of galaxies as a tracer of the formation time of dark matter halos. In the framework of the standard cosmological model, the well-known correlation between halo concentration and formation time reflects that  higher density regions in the large-scale structure collapse at earlier cosmic times \citep[e.g.,][]{1999MNRAS.302..111L}. In fact, \citet[]{2024ApJ...974...29O} use the number density of galaxies on large-scales (10 Mpc) as a proxy of halo formation time. At fixed $M_h$, they find that more massive central SDSS galaxies are older and also have higher galaxy number densities than less massive centrals (for $M_h>10^{12} \rm M_{\odot}$).
According to an analytical model based on dark matter-only simulations, they find that stellar mass and large-scale density variations at fixed halo mass are consistent with differences in halo formation time.

Finally, we note that although the assumption that total enclosed mass within 3$R_e$ acts as a proxy for halo mass requires further observational testing, recent findings point to its plausibility (see SD24 for a detailed discussion). \citet{2025A&A...699A.301A} shows that `total' masses in the zoom-in hydro-dynamical simulations AURIGA \citep{2017MNRAS.467..179G}, HESTIA \citep{2020MNRAS.498.2968L} and NIHAO \citep{2015MNRAS.454...83W} correlate with $M_{200}$ (at fixed  $M_{\star}$). In this case, total masses are computed within $R_1$ \citep{2020MNRAS.493...87T} for $M_{\star} \gtrsim 10^{9} \rm M_{\odot}$. $R_1$ is a physically motivated definition of galaxy size based on a stellar mass density threshold, defined in order to probe the whole optical extent of galaxies, being comparable to 3 half-mass radius for $M_{\star}\sim10^{9} \rm M_{\odot}$ galaxies, and extending to larger radii for larger stellar masses. Hence, we highlight the remarkable good agreement across different simulation suites. Moreover, \citet{2025ApJ...983...93M} recently found that in the FIREBox cosmological simulations \citep{2023MNRAS.522.3831F} halo mass is the most important predictor of the inner dark matter content of galaxies (within 1 half-mass radius).

\section{Summary and conclusions}
\label{sec:conc}

We have investigated how spatially-resolved stellar populations of CALIFA galaxies behave across the stellar-to-total dynamical mass relation. We explored how average ages and metallicities across the STDMR vary with galactocentric distance by measuring them in different radial annuli. Then, we projected the STDMR to look into scaling relations of these properties as a function of stellar and total mass (both in the inner and outer regions of the galaxies). We also looked at dependence of their stellar population profiles on total mass and morphology. Finally, we show stellar population gradients across the STDMR. 

(i) We observe that the scatter of the STDMR is connected to galaxy ages and metallicities measured at different radial annuli (Fig. \ref{fig:stel_pops_annuli}). These observables depend both on stellar and total mass, although the strength and significance of these correlations varies with galactocentric distance. While the dependence of age on total mass becomes more prominent in the outer regions of galaxies (r>1$R_e$), the one of metallicity is only significant in the inner regions (r<0.5$R_e$). 

(ii) By studying the dependence of age and metallicity separately on stellar mass and total mass, respectively, we find the following: (1) We see that galaxies with higher $M_{\star}$ are older and have more metal-rich stellar populations than the ones with lower $M_{\star}$, at fixed total mass (Fig. \ref{fig:age_mtot}); (2) At fixed stellar mass, galaxies have older and slightly more metal-rich stellar populations with decreasing total mass (Fig. \ref{fig:age_mstar}). These trends are more noticeable in the intermediate-mass regime ($ M_{\star} \rm /M_{\odot} \sim 10^{10.5}-10^{11}$).

(iii) Focusing on a narrow range of intermediate stellar mass, for which we can cover with good statistical representativeness the full range of total mass, we find that galaxies with different total masses have different stellar population profiles, especially in the case of age (left column of Fig. \ref{fig:age_met_profiles}). At fixed stellar mass, galaxies with lower total masses have generally relatively flat age profiles and steep negative metallicity profiles. Whereas galaxies with higher total masses have both negative age and [M/H] profiles, yet the [M/H] ones are less steep and have a slightly lower normalization in the inner regions.

(iv) In addition to different total masses, these galaxies have also different morphological types (right column of Fig. \ref{fig:age_met_profiles}). The steepness and normalization of the profiles is connected to morphology, with earlier-types having relatively flat age profiles with overall old stellar populations and quite steep negative metallicity. In contrast, late-types have less steep negative metallicity profiles and negative age gradients.

(v) These radial variations translate into a connection between the STDMR and age and [M/H] gradients (Fig. \ref{fig:gradients}). At fixed $M_{tot}$, global gradients show that galaxies with larger $M_{\star}$ have flatter or slightly positive age gradients and more steep negative [M/H] gradients. Whereas galaxies with smaller $M_{\star}$ have mild negative age gradients and less steep metallicity ones, with a slighly lower normalization.

(vi) Interestingly, we find that the different radial dependence on total mass of age and [M/H] can be explained by the connection between the scatter of the STDMR and morphology, given the different stellar population profiles and gradients of ETGs and LTGs (Fig. \ref{fig:age_met_profiles}). 

(vii) Under the assumption that total dynamical mass can trace the halo masses of the galaxies, we connect the STDMR and the stellar-to-halo mass relation. We interpret our findings in the context of halo evolution, with dark matter halos forming at different cosmic times naturally modulating the observed trends across the STDMR. At fixed halo/total mass, earlier-formed halos would host more `evolved' galaxies: with higher $M_{\star}$, older, quenched, earlier-type galaxies that assembled their stellar population earlier on, resulting in steep negative [M/H] gradients and relatively flat age ones, with overall old stellar populations. As opposed, later-formed halos, would host less `evolved' galaxies: with lower $M_{\star}$, younger, star-forming, later-type galaxies that have assembled the bulk of their stars with more prolonged SFHs, leading to less steep negative metallicity gradients and negative age profiles.\\

Our results indicate that, at given stellar mass, dark matter halos have a key role modulating a wide range of baryonic properties of galaxies while leaving an imprint in the spatial distribution of their stellar populations. To further test observationally the role of total dynamical mass as a tracer of halo mass, we need to derive total dynamical masses within larger apertures as well as to compare them with direct halo mass measurements, while also using larger statistics. This will be possible in the near future thanks to IFS instruments like WEAVE. Its large aperture will allow to spectroscopically cover the baryonic extent of nearby galaxies in order to derive their total enclosed masses as well as different stellar and gas properties. Moreover, upcoming surveys such as WEAVE-Apertif \citep{2020AAS...23545903H} not only will provide larger statistics, but also direct halo mass measurements through galaxy rotation curves of cold HI gas. This will allow to truly assess in detail the role of total dynamical mass as a proxy of halo mass. Moreover, to confirm our proposed scenario and better understand how secondary halo properties (e.g., halo formation time) affect galaxy formation, the large-scale environment of galaxies can be used as an observational window to trace halo evolution. Upcoming spectroscopic facilities such as 4MOST will be key to trace the detailed characterizations of large-scale environment at high-z in order to investigate baryonic galaxy properties at these earlier times both in terms of their host halos and their cosmic web environment.

\begin{acknowledgements}
   We would like to thank the referee for her/his constructive feedback, which has notably improved this paper. We thank Filippo Mannucci for useful discussions that helped shaping the analysis presented in this work. LSD thanks Eirini Angeloudi for her helpful suggestions on symbol regression algorithms and Francesco La Barbera for insightful discussions. LSD, ARG and SZ  acknowledge support from PRIN-MUR project “PROMETEUS” financed by the European Union - Next Generation EU, Mission 4 Component 1 CUP B53D23004750006. LSD acknowledges support through PID2021-123313NA-I00 of MICIN/AEI/10.13039/501100011033/FEDER, UE.  ARG acknowledges support from the INAF-Minigrant-2022 "LEGA-C" 1.05.12.04.01. SZ acknowledges support from the INAF-Minigrant-2023 "Enabling the study of galaxy evolution through unresolved stellar population analysis" 1.05.23.04.01. DM acknowledges support from the Italian national inter-university PhD programme in Space Science and Technology.\\

This research made use of Astropy, a community-developed core Python package for Astronomy \citep[][]{astropy:2013,astropy:2018}, and of the Numpy \citep[][]{2020Natur.585..357H}, Scipy \citep[][]{2020SciPy-NMeth}, Matplotlib \citep[][]{2007CSE.....9...90H}, Pandas \citep{mckinney-proc-scipy-2010}, Pingouin \citep{Vallat2018}, Seaborn \citep{Waskom2021}, PySR  \citep{cranmerInterpretableMachineLearning2023} and Scientific color maps \citep{crameri_2021_5501399} Python libraries.

\end{acknowledgements}

\bibliographystyle{aa}
\bibliography{references}

\begin{thebibliography}{184}
\expandafter\ifx\csname natexlab\endcsname\relax\def\natexlab#1{#1}\fi

\bibitem[{{Angeloudi} {et~al.}(2024){Angeloudi}, {Falc{\'o}n-Barroso},
  {Huertas-Company}, {Boecker}, {Sarmiento}, {Eisert}, \&
  {Pillepich}}]{2024NatAs...8.1310A}
{Angeloudi}, E., {Falc{\'o}n-Barroso}, J., {Huertas-Company}, M., {et~al.}
  2024, Nature Astronomy, 8, 1310

\bibitem[{{Arjona} \& {Nesseris}(2020)}]{2020PhRvD.101l3525A}
{Arjona}, R. \& {Nesseris}, S. 2020, \prd, 101, 123525

\bibitem[{{Arjona-G{\'a}lvez} {et~al.}(2025){Arjona-G{\'a}lvez},
  {Cardona-Barrero}, {Grand}, {Di Cintio}, {Dalla Vecchia}, {Benavides},
  {Macci{\`o}}, {Libeskind}, \& {Knebe}}]{2025A&A...699A.301A}
{Arjona-G{\'a}lvez}, E., {Cardona-Barrero}, S., {Grand}, R. J.~J., {et~al.}
  2025, \aap, 699, A301

\bibitem[{{Artale} {et~al.}(2018){Artale}, {Zehavi}, {Contreras}, \&
  {Norberg}}]{2018MNRAS.480.3978A}
{Artale}, M.~C., {Zehavi}, I., {Contreras}, S., \& {Norberg}, P. 2018, \mnras,
  480, 3978

\bibitem[{{Astropy Collaboration} {et~al.}(2018){Astropy Collaboration},
  {Price-Whelan}, {Sip{\H{o}}cz}, {G{\"u}nther}, {Lim}, {Crawford}, {Conseil},
  {Shupe}, {Craig}, {Dencheva}, {Ginsburg}, {Vand erPlas}, {Bradley},
  {P{\'e}rez-Su{\'a}rez}, {de Val-Borro}, {Aldcroft}, {Cruz}, {Robitaille},
  {Tollerud}, {Ardelean}, {Babej}, {Bach}, {Bachetti}, {Bakanov}, {Bamford},
  {Barentsen}, {Barmby}, {Baumbach}, {Berry}, {Biscani}, {Boquien}, {Bostroem},
  {Bouma}, {Brammer}, {Bray}, {Breytenbach}, {Buddelmeijer}, {Burke},
  {Calderone}, {Cano Rodr{\'\i}guez}, {Cara}, {Cardoso}, {Cheedella}, {Copin},
  {Corrales}, {Crichton}, {D'Avella}, {Deil}, {Depagne}, {Dietrich}, {Donath},
  {Droettboom}, {Earl}, {Erben}, {Fabbro}, {Ferreira}, {Finethy}, {Fox},
  {Garrison}, {Gibbons}, {Goldstein}, {Gommers}, {Greco}, {Greenfield},
  {Groener}, {Grollier}, {Hagen}, {Hirst}, {Homeier}, {Horton}, {Hosseinzadeh},
  {Hu}, {Hunkeler}, {Ivezi{\'c}}, {Jain}, {Jenness}, {Kanarek}, {Kendrew},
  {Kern}, {Kerzendorf}, {Khvalko}, {King}, {Kirkby}, {Kulkarni}, {Kumar},
  {Lee}, {Lenz}, {Littlefair}, {Ma}, {Macleod}, {Mastropietro}, {McCully},
  {Montagnac}, {Morris}, {Mueller}, {Mumford}, {Muna}, {Murphy}, {Nelson},
  {Nguyen}, {Ninan}, {N{\"o}the}, {Ogaz}, {Oh}, {Parejko}, {Parley}, {Pascual},
  {Patil}, {Patil}, {Plunkett}, {Prochaska}, {Rastogi}, {Reddy Janga},
  {Sabater}, {Sakurikar}, {Seifert}, {Sherbert}, {Sherwood-Taylor}, {Shih},
  {Sick}, {Silbiger}, {Singanamalla}, {Singer}, {Sladen}, {Sooley},
  {Sornarajah}, {Streicher}, {Teuben}, {Thomas}, {Tremblay}, {Turner},
  {Terr{\'o}n}, {van Kerkwijk}, {de la Vega}, {Watkins}, {Weaver}, {Whitmore},
  {Woillez}, {Zabalza}, \& {Astropy Contributors}}]{astropy:2018}
{Astropy Collaboration}, {Price-Whelan}, A.~M., {Sip{\H{o}}cz}, B.~M., {et~al.}
  2018, \aj, 156, 123

\bibitem[{{Astropy Collaboration} {et~al.}(2013){Astropy Collaboration},
  {Robitaille}, {Tollerud}, {Greenfield}, {Droettboom}, {Bray}, {Aldcroft},
  {Davis}, {Ginsburg}, {Price-Whelan}, {Kerzendorf}, {Conley}, {Crighton},
  {Barbary}, {Muna}, {Ferguson}, {Grollier}, {Parikh}, {Nair}, {Unther},
  {Deil}, {Woillez}, {Conseil}, {Kramer}, {Turner}, {Singer}, {Fox}, {Weaver},
  {Zabalza}, {Edwards}, {Azalee Bostroem}, {Burke}, {Casey}, {Crawford},
  {Dencheva}, {Ely}, {Jenness}, {Labrie}, {Lim}, {Pierfederici}, {Pontzen},
  {Ptak}, {Refsdal}, {Servillat}, \& {Streicher}}]{astropy:2013}
{Astropy Collaboration}, {Robitaille}, T.~P., {Tollerud}, E.~J., {et~al.} 2013,
  \aap, 558, A33

\bibitem[{{Bait} {et~al.}(2017){Bait}, {Barway}, \&
  {Wadadekar}}]{2017MNRAS.471.2687B}
{Bait}, O., {Barway}, S., \& {Wadadekar}, Y. 2017, \mnras, 471, 2687

\bibitem[{{Baldry} {et~al.}(2004){Baldry}, {Glazebrook}, {Brinkmann},
  {Ivezi{\'c}}, {Lupton}, {Nichol}, \& {Szalay}}]{2004ApJ...600..681B}
{Baldry}, I.~K., {Glazebrook}, K., {Brinkmann}, J., {et~al.} 2004, \apj, 600,
  681

\bibitem[{{Balogh} {et~al.}(1999){Balogh}, {Morris}, {Yee}, {Carlberg}, \&
  {Ellingson}}]{1999ApJ...527...54B}
{Balogh}, M.~L., {Morris}, S.~L., {Yee}, H.~K.~C., {Carlberg}, R.~G., \&
  {Ellingson}, E. 1999, \apj, 527, 54

\bibitem[{{Bartlett} {et~al.}(2022){Bartlett}, {Desmond}, \&
  {Ferreira}}]{2022arXiv221111461B}
{Bartlett}, D.~J., {Desmond}, H., \& {Ferreira}, P.~G. 2022, arXiv e-prints,
  arXiv:2211.11461

\bibitem[{{Bayron Orjuela-Quintana} {et~al.}(2022){Bayron Orjuela-Quintana},
  {Nesseris}, \& {Cardona}}]{2022arXiv221106393B}
{Bayron Orjuela-Quintana}, J., {Nesseris}, S., \& {Cardona}, W. 2022, arXiv
  e-prints, arXiv:2211.06393

\bibitem[{{Behroozi} {et~al.}(2019){Behroozi}, {Wechsler}, {Hearin}, \&
  {Conroy}}]{2019MNRAS.488.3143B}
{Behroozi}, P., {Wechsler}, R.~H., {Hearin}, A.~P., \& {Conroy}, C. 2019,
  \mnras, 488, 3143

\bibitem[{{Behroozi} {et~al.}(2013){Behroozi}, {Wechsler}, \&
  {Conroy}}]{2013ApJ...770...57B}
{Behroozi}, P.~S., {Wechsler}, R.~H., \& {Conroy}, C. 2013, \apj, 770, 57

\bibitem[{{Bell} \& {de Jong}(2000)}]{2000MNRAS.312..497B}
{Bell}, E.~F. \& {de Jong}, R.~S. 2000, \mnras, 312, 497

\bibitem[{{Bernal} {et~al.}(2021){Bernal}, {Caputo}, {Villaescusa-Navarro}, \&
  {Kamionkowski}}]{2021PhRvL.127m1102B}
{Bernal}, J.~L., {Caputo}, A., {Villaescusa-Navarro}, F., \& {Kamionkowski}, M.
  2021, \prl, 127, 131102

\bibitem[{{Bertelli} {et~al.}(1994){Bertelli}, {Bressan}, {Chiosi}, {Fagotto},
  \& {Nasi}}]{1994A&AS..106..275B}
{Bertelli}, G., {Bressan}, A., {Chiosi}, C., {Fagotto}, F., \& {Nasi}, E. 1994,
  \aaps, 106, 275

\bibitem[{{Bevacqua} {et~al.}(2024){Bevacqua}, {Saracco}, {Boecker}, {D'Ago},
  {De Lucia}, {De Propris}, {La Barbera}, {Pasquali}, {Spiniello}, \&
  {Tortora}}]{2024A&A...690A.150B}
{Bevacqua}, D., {Saracco}, P., {Boecker}, A., {et~al.} 2024, \aap, 690, A150

\bibitem[{{Beverage} {et~al.}(2023){Beverage}, {Kriek}, {Conroy}, {Sandford},
  {Bezanson}, {Franx}, {van der Wel}, \& {Weisz}}]{2023ApJ...948..140B}
{Beverage}, A.~G., {Kriek}, M., {Conroy}, C., {et~al.} 2023, \apj, 948, 140

\bibitem[{{Bezanson} {et~al.}(2009){Bezanson}, {van Dokkum}, {Tal},
  {Marchesini}, {Kriek}, {Franx}, \& {Coppi}}]{2009ApJ...697.1290B}
{Bezanson}, R., {van Dokkum}, P.~G., {Tal}, T., {et~al.} 2009, \apj, 697, 1290

\bibitem[{{Bluck} {et~al.}(2020{\natexlab{a}}){Bluck}, {Maiolino},
  {Piotrowska}, {Trussler}, {Ellison}, {S{\'a}nchez}, {Thorp}, {Teimoorinia},
  {Moreno}, \& {Conselice}}]{2020MNRAS.499..230B}
{Bluck}, A. F.~L., {Maiolino}, R., {Piotrowska}, J.~M., {et~al.}
  2020{\natexlab{a}}, \mnras, 499, 230

\bibitem[{{Bluck} {et~al.}(2020{\natexlab{b}}){Bluck}, {Maiolino},
  {S{\'a}nchez}, {Ellison}, {Thorp}, {Piotrowska}, {Teimoorinia}, \&
  {Bundy}}]{2020MNRAS.492...96B}
{Bluck}, A. F.~L., {Maiolino}, R., {S{\'a}nchez}, S.~F., {et~al.}
  2020{\natexlab{b}}, \mnras, 492, 96

\bibitem[{{Blumenthal} {et~al.}(1984){Blumenthal}, {Faber}, {Primack}, \&
  {Rees}}]{1984Natur.311..517B}
{Blumenthal}, G.~R., {Faber}, S.~M., {Primack}, J.~R., \& {Rees}, M.~J. 1984,
  \nat, 311, 517

\bibitem[{{Bradshaw} {et~al.}(2020){Bradshaw}, {Leauthaud}, {Hearin}, {Huang},
  \& {Behroozi}}]{2020MNRAS.493..337B}
{Bradshaw}, C., {Leauthaud}, A., {Hearin}, A., {Huang}, S., \& {Behroozi}, P.
  2020, \mnras, 493, 337

\bibitem[{{Brinchmann} {et~al.}(2004){Brinchmann}, {Charlot}, {White},
  {Tremonti}, {Kauffmann}, {Heckman}, \& {Brinkmann}}]{2004MNRAS.351.1151B}
{Brinchmann}, J., {Charlot}, S., {White}, S.~D.~M., {et~al.} 2004, \mnras, 351,
  1151

\bibitem[{{Brough} {et~al.}(2007){Brough}, {Proctor}, {Forbes}, {Couch},
  {Collins}, {Burke}, \& {Mann}}]{Brough:2007}
{Brough}, S., {Proctor}, R., {Forbes}, D.~A., {et~al.} 2007, \mnras, 378, 1507

\bibitem[{{Bruzual} \& {Charlot}(2003)}]{2003MNRAS.344.1000B}
{Bruzual}, G. \& {Charlot}, S. 2003, \mnras, 344, 1000

\bibitem[{{Bruzual A.}(1983)}]{1983ApJ...273..105B}
{Bruzual A.}, G. 1983, \apj, 273, 105

\bibitem[{{Bundy} {et~al.}(2015){Bundy}, {Bershady}, {Law}, {Yan}, {Drory},
  {MacDonald}, {Wake}, {Cherinka}, {S{\'a}nchez-Gallego}, {Weijmans}, {Thomas},
  {Tremonti}, {Masters}, {Coccato}, {Diamond-Stanic}, {Arag{\'o}n-Salamanca},
  {Avila-Reese}, {Badenes}, {Falc{\'o}n-Barroso}, {Belfiore}, {Bizyaev},
  {Blanc}, {Bland-Hawthorn}, {Blanton}, {Brownstein}, {Byler}, {Cappellari},
  {Conroy}, {Dutton}, {Emsellem}, {Etherington}, {Frinchaboy}, {Fu}, {Gunn},
  {Harding}, {Johnston}, {Kauffmann}, {Kinemuchi}, {Klaene}, {Knapen},
  {Leauthaud}, {Li}, {Lin}, {Maiolino}, {Malanushenko}, {Malanushenko}, {Mao},
  {Maraston}, {McDermid}, {Merrifield}, {Nichol}, {Oravetz}, {Pan}, {Parejko},
  {Sanchez}, {Schlegel}, {Simmons}, {Steele}, {Steinmetz}, {Thanjavur},
  {Thompson}, {Tinker}, {van den Bosch}, {Westfall}, {Wilkinson}, {Wright},
  {Xiao}, \& {Zhang}}]{2015ApJ...798....7B}
{Bundy}, K., {Bershady}, M.~A., {Law}, D.~R., {et~al.} 2015, \apj, 798, 7

\bibitem[{{Cappellari}(2008)}]{2008MNRAS.390...71C}
{Cappellari}, M. 2008, \mnras, 390, 71

\bibitem[{{Cappellari}(2016)}]{2016ARA&A..54..597C}
{Cappellari}, M. 2016, \araa, 54, 597

\bibitem[{{Cappellari}(2020)}]{2020MNRAS.494.4819C}
{Cappellari}, M. 2020, \mnras, 494, 4819

\bibitem[{{Cappellari} \& {Copin}(2003)}]{2003MNRAS.342..345C}
{Cappellari}, M. \& {Copin}, Y. 2003, \mnras, 342, 345

\bibitem[{{Cappellari} {et~al.}(2013){Cappellari}, {McDermid}, {Alatalo},
  {Blitz}, {Bois}, {Bournaud}, {Bureau}, {Crocker}, {Davies}, {Davis}, {de
  Zeeuw}, {Duc}, {Emsellem}, {Khochfar}, {Krajnovi{\'c}}, {Kuntschner},
  {Morganti}, {Naab}, {Oosterloo}, {Sarzi}, {Scott}, {Serra}, {Weijmans}, \&
  {Young}}]{2013MNRAS.432.1862C}
{Cappellari}, M., {McDermid}, R.~M., {Alatalo}, K., {et~al.} 2013, \mnras, 432,
  1862

\bibitem[{{Chabrier}(2003)}]{2003PASP..115..763C}
{Chabrier}, G. 2003, \pasp, 115, 763

\bibitem[{{Charlot} \& {Fall}(2000)}]{2000ApJ...539..718C}
{Charlot}, S. \& {Fall}, S.~M. 2000, \apj, 539, 718

\bibitem[{{Chittenden} \& {Tojeiro}(2023)}]{2023MNRAS.518.5670C}
{Chittenden}, H.~G. \& {Tojeiro}, R. 2023, \mnras, 518, 5670

\bibitem[{Crameri(2021)}]{crameri_2021_5501399}
Crameri, F. 2021, Scientific colour maps

\bibitem[{Cranmer(2023)}]{cranmerInterpretableMachineLearning2023}
Cranmer, M. 2023, Interpretable {Machine} {Learning} for {Science} with {PySR}
  and {SymbolicRegression}.jl, arXiv:2305.01582 [astro-ph, physics:physics]

\bibitem[{{Cranmer} {et~al.}(2020){Cranmer}, {Sanchez-Gonzalez}, {Battaglia},
  {Xu}, {Cranmer}, {Spergel}, \& {Ho}}]{2020arXiv200611287C}
{Cranmer}, M., {Sanchez-Gonzalez}, A., {Battaglia}, P., {et~al.} 2020, arXiv
  e-prints, arXiv:2006.11287

\bibitem[{{Cui} {et~al.}(2021){Cui}, {Dav{\'e}}, {Peacock},
  {Angl{\'e}s-Alc{\'a}zar}, \& {Yang}}]{2021NatAs...5.1069C}
{Cui}, W., {Dav{\'e}}, R., {Peacock}, J.~A., {Angl{\'e}s-Alc{\'a}zar}, D., \&
  {Yang}, X. 2021, Nature Astronomy, 5, 1069

\bibitem[{{Davies} {et~al.}(2021){Davies}, {Crain}, \&
  {Pontzen}}]{2021MNRAS.501..236D}
{Davies}, J.~J., {Crain}, R.~A., \& {Pontzen}, A. 2021, \mnras, 501, 236

\bibitem[{{Davison} {et~al.}(2020){Davison}, {Norris}, {Pfeffer}, {Davies}, \&
  {Crain}}]{2020MNRAS.497...81D}
{Davison}, T.~A., {Norris}, M.~A., {Pfeffer}, J.~L., {Davies}, J.~J., \&
  {Crain}, R.~A. 2020, \mnras, 497, 81

\bibitem[{{Delgado} {et~al.}(2022){Delgado}, {Wadekar}, {Hadzhiyska}, {Bose},
  {Hernquist}, \& {Ho}}]{2022MNRAS.515.2733D}
{Delgado}, A.~M., {Wadekar}, D., {Hadzhiyska}, B., {et~al.} 2022, \mnras, 515,
  2733

\bibitem[{{Dom{\'\i}nguez S{\'a}nchez} {et~al.}(2019){Dom{\'\i}nguez
  S{\'a}nchez}, {Bernardi}, {Brownstein}, {Drory}, \&
  {Sheth}}]{DominguezSanchez:2019}
{Dom{\'\i}nguez S{\'a}nchez}, H., {Bernardi}, M., {Brownstein}, J.~R., {Drory},
  N., \& {Sheth}, R.~K. 2019, \mnras, 489, 5612

\bibitem[{{Elbaz} {et~al.}(2007){Elbaz}, {Daddi}, {Le Borgne}, {Dickinson},
  {Alexander}, {Chary}, {Starck}, {Brandt}, {Kitzbichler}, {MacDonald},
  {Nonino}, {Popesso}, {Stern}, \& {Vanzella}}]{2007A&A...468...33E}
{Elbaz}, D., {Daddi}, E., {Le Borgne}, D., {et~al.} 2007, \aap, 468, 33

\bibitem[{{Faber}(1973)}]{1973ApJ...179..731F}
{Faber}, S.~M. 1973, \apj, 179, 731

\bibitem[{{Falc{\'o}n-Barroso} {et~al.}(2017){Falc{\'o}n-Barroso}, {Lyubenova},
  {van de Ven}, {Mendez-Abreu}, {Aguerri}, {Garc{\'\i}a-Lorenzo},
  {Bekerait{\'e}}, {S{\'a}nchez}, {Husemann}, {Garc{\'\i}a-Benito}, {Mast},
  {Walcher}, {Zibetti}, {Barrera-Ballesteros}, {Galbany},
  {S{\'a}nchez-Bl{\'a}zquez}, {Singh}, {van den Bosch}, {Wild}, {Zhu},
  {Bland-Hawthorn}, {Cid Fernandes}, {de Lorenzo-C{\'a}ceres}, {Gallazzi},
  {Gonz{\'a}lez Delgado}, {Marino}, {M{\'a}rquez}, {P{\'e}rez}, {P{\'e}rez},
  {Roth}, {Rosales-Ortega}, {Ruiz-Lara}, {Wisotzki}, {Ziegler}, \& {CALIFA
  Collaboration}}]{2017A&A...597A..48F}
{Falc{\'o}n-Barroso}, J., {Lyubenova}, M., {van de Ven}, G., {et~al.} 2017,
  \aap, 597, A48

\bibitem[{{Falc{\'o}n-Barroso} {et~al.}(2011){Falc{\'o}n-Barroso},
  {S{\'a}nchez-Bl{\'a}zquez}, {Vazdekis}, {Ricciardelli}, {Cardiel}, {Cenarro},
  {Gorgas}, \& {Peletier}}]{2011A&A...532A..95F}
{Falc{\'o}n-Barroso}, J., {S{\'a}nchez-Bl{\'a}zquez}, P., {Vazdekis}, A.,
  {et~al.} 2011, \aap, 532, A95

\bibitem[{{Falc{\'o}n-Barroso} {et~al.}(2019){Falc{\'o}n-Barroso}, {van de
  Ven}, {Lyubenova}, {Mendez-Abreu}, {Aguerri}, {Garc{\'\i}a-Lorenzo},
  {Bekerait{\'e}}, {S{\'a}nchez}, {Husemann}, {Garc{\'\i}a-Benito},
  {Gonz{\'a}lez Delgado}, {Mast}, {Walcher}, {Zibetti}, {Zhu},
  {Barrera-Ballesteros}, {Galbany}, {S{\'a}nchez-Bl{\'a}zquez}, {Singh}, {van
  den Bosch}, {Wild}, {Bland-Hawthorn}, {Cid Fernandes}, {de
  Lorenzo-C{\'a}ceres}, {Gallazzi}, {Marino}, {M{\'a}rquez}, {Peletier},
  {P{\'e}rez}, {P{\'e}rez}, {Roth}, {Rosales-Ortega}, {Ruiz-Lara}, {Wisotzki},
  \& {Ziegler}}]{2019A&A...632A..59F}
{Falc{\'o}n-Barroso}, J., {van de Ven}, G., {Lyubenova}, M., {et~al.} 2019,
  \aap, 632, A59

\bibitem[{{Feldmann} {et~al.}(2023){Feldmann}, {Quataert},
  {Faucher-Gigu{\`e}re}, {Hopkins}, {{\c{C}}atmabacak}, {Kere{\v{s}}},
  {Bassini}, {Bernardini}, {Bullock}, {Cenci}, {Gensior}, {Liang}, {Moreno}, \&
  {Wetzel}}]{2023MNRAS.522.3831F}
{Feldmann}, R., {Quataert}, E., {Faucher-Gigu{\`e}re}, C.-A., {et~al.} 2023,
  \mnras, 522, 3831

\bibitem[{{Ferreras} {et~al.}(2005){Ferreras}, {Saha}, \&
  {Williams}}]{2005ApJ...623L...5F}
{Ferreras}, I., {Saha}, P., \& {Williams}, L. L.~R. 2005, \apjl, 623, L5

\bibitem[{{Ferreras} {et~al.}(2025){Ferreras}, {Trevisan}, {Lahav}, {de
  Carvalho}, \& {Silk}}]{2025MNRAS.540.1069F}
{Ferreras}, I., {Trevisan}, M., {Lahav}, O., {de Carvalho}, R.~R., \& {Silk},
  J. 2025, \mnras, 540, 1069

\bibitem[{{Gallazzi} {et~al.}(2014){Gallazzi}, {Bell}, {Zibetti}, {Brinchmann},
  \& {Kelson}}]{2014ApJ...788...72G}
{Gallazzi}, A., {Bell}, E.~F., {Zibetti}, S., {Brinchmann}, J., \& {Kelson},
  D.~D. 2014, \apj, 788, 72

\bibitem[{{Gallazzi} {et~al.}(2005){Gallazzi}, {Charlot}, {Brinchmann},
  {White}, \& {Tremonti}}]{2005MNRAS.362...41G}
{Gallazzi}, A., {Charlot}, S., {Brinchmann}, J., {White}, S. D.~M., \&
  {Tremonti}, C.~A. 2005, \mnras, 362, 41

\bibitem[{{Gallazzi} {et~al.}(2021){Gallazzi}, {Pasquali}, {Zibetti}, \&
  {Barbera}}]{2021MNRAS.502.4457G}
{Gallazzi}, A.~R., {Pasquali}, A., {Zibetti}, S., \& {Barbera}, F.~L. 2021,
  \mnras, 502, 4457

\bibitem[{{Gallazzi} {et~al.}(2025){Gallazzi}, {Zibetti}, {van der Wel},
  {Nersesian}, {Kaushal}, {Bezanson}, {Mattolini}, {Bell}, {Scholz-Diaz},
  {Leja}, {D'Eugenio}, {Wu}, {Pacifici}, \& {Maseda}}]{2025arXiv251111805G}
{Gallazzi}, A.~R., {Zibetti}, S., {van der Wel}, A., {et~al.} 2025, arXiv
  e-prints, arXiv:2511.11805, accepted for publication in A\&A

\bibitem[{{Garc{\'\i}a-Benito} {et~al.}(2017){Garc{\'\i}a-Benito},
  {Gonz{\'a}lez Delgado}, {P{\'e}rez}, {Cid Fernandes}, {Cortijo-Ferrero},
  {L{\'o}pez Fern{\'a}ndez}, {de Amorim}, {Lacerda}, {Vale Asari}, \&
  {S{\'a}nchez}}]{2017A&A...608A..27G}
{Garc{\'\i}a-Benito}, R., {Gonz{\'a}lez Delgado}, R.~M., {P{\'e}rez}, E.,
  {et~al.} 2017, \aap, 608, A27

\bibitem[{{Gavazzi} {et~al.}(2002){Gavazzi}, {Bonfanti}, {Sanvito}, {Boselli},
  \& {Scodeggio}}]{2002ApJ...576..135G}
{Gavazzi}, G., {Bonfanti}, C., {Sanvito}, G., {Boselli}, A., \& {Scodeggio}, M.
  2002, \apj, 576, 135

\bibitem[{{Goddard} {et~al.}(2017){Goddard}, {Thomas}, {Maraston}, {Westfall},
  {Etherington}, {Riffel}, {Mallmann}, {Zheng}, {Argudo-Fern{\'a}ndez}, {Lian},
  {Bershady}, {Bundy}, {Drory}, {Law}, {Yan}, {Wake}, {Weijmans}, {Bizyaev},
  {Brownstein}, {Lane}, {Maiolino}, {Masters}, {Merrifield}, {Nitschelm},
  {Pan}, {Roman-Lopes}, {Storchi-Bergmann}, \& {Schneider}}]{Goddard:2017b}
{Goddard}, D., {Thomas}, D., {Maraston}, C., {et~al.} 2017, \mnras, 466, 4731

\bibitem[{{Gonz{\'a}lez Delgado} {et~al.}(2015){Gonz{\'a}lez Delgado},
  {Garc{\'{\i}}a-Benito}, {P{\'e}rez}, {Cid Fernandes}, {de Amorim},
  {Cortijo-Ferrero}, {Lacerda}, {L{\'o}pez Fern{\'a}ndez}, {Vale-Asari},
  {S{\'a}nchez}, {Moll{\'a}}, {Ruiz-Lara}, {S{\'a}nchez-Bl{\'a}zquez},
  {Walcher}, {Alves}, {Aguerri}, {Bekerait{\'e}}, {Bland-Hawthorn}, {Galbany},
  {Gallazzi}, {Husemann}, {Iglesias-P{\'a}ramo}, {Kalinova},
  {L{\'o}pez-S{\'a}nchez}, {Marino}, {M{\'a}rquez}, {Masegosa}, {Mast},
  {M{\'e}ndez-Abreu}, {Mendoza}, {del Olmo}, {P{\'e}rez}, {Quirrenbach}, \&
  {Zibetti}}]{Gonzalez-Delgado:2015aa}
{Gonz{\'a}lez Delgado}, R.~M., {Garc{\'{\i}}a-Benito}, R., {P{\'e}rez}, E.,
  {et~al.} 2015, \aap, 581, A103

\bibitem[{{Graham} {et~al.}(2012){Graham}, {Djorgovski}, {Mahabal}, {Donalek},
  {Drake}, \& {Longo}}]{2012arXiv1208.2480G}
{Graham}, M.~J., {Djorgovski}, S.~G., {Mahabal}, A., {et~al.} 2012, arXiv
  e-prints, arXiv:1208.2480

\bibitem[{{Grand} {et~al.}(2017){Grand}, {G{\'o}mez}, {Marinacci}, {Pakmor},
  {Springel}, {Campbell}, {Frenk}, {Jenkins}, \& {White}}]{2017MNRAS.467..179G}
{Grand}, R. J.~J., {G{\'o}mez}, F.~A., {Marinacci}, F., {et~al.} 2017, \mnras,
  467, 179

\bibitem[{{Harris} {et~al.}(2020){Harris}, {Millman}, {van der Walt},
  {Gommers}, {Virtanen}, {Cournapeau}, {Wieser}, {Taylor}, {Berg}, {Smith},
  {Kern}, {Picus}, {Hoyer}, {van Kerkwijk}, {Brett}, {Haldane}, {del R{\'\i}o},
  {Wiebe}, {Peterson}, {G{\'e}rard-Marchant}, {Sheppard}, {Reddy}, {Weckesser},
  {Abbasi}, {Gohlke}, \& {Oliphant}}]{2020Natur.585..357H}
{Harris}, C.~R., {Millman}, K.~J., {van der Walt}, S.~J., {et~al.} 2020, \nat,
  585, 357

\bibitem[{{Hess} {et~al.}(2020){Hess}, {Falcon-Barroso}, {Ascasibar},
  {Perez-Martin}, {Serra}, {Weijmans}, \& {Weave-Apertif
  Team}}]{2020AAS...23545903H}
{Hess}, K.~M., {Falcon-Barroso}, J., {Ascasibar}, Y., {et~al.} 2020, in
  American Astronomical Society Meeting Abstracts, Vol. 235, American
  Astronomical Society Meeting Abstracts \#235, 459.03

\bibitem[{Huang {et~al.}(2013)Huang, Ho, Peng, Li, \& Barth}]{Huang_2013}
Huang, S., Ho, L.~C., Peng, C.~Y., Li, Z.-Y., \& Barth, A.~J. 2013, The
  Astrophysical Journal Letters, 768, L28

\bibitem[{{Huang} {et~al.}(2022){Huang}, {Leauthaud}, {Bradshaw}, {Hearin},
  {Behroozi}, {Lange}, {Greene}, {DeRose}, {Speagle}, \&
  {Xhakaj}}]{2022MNRAS.515.4722H}
{Huang}, S., {Leauthaud}, A., {Bradshaw}, C., {et~al.} 2022, \mnras, 515, 4722

\bibitem[{{Huang} {et~al.}(2020){Huang}, {Leauthaud}, {Hearin}, {Behroozi},
  {Bradshaw}, {Ardila}, {Speagle}, {Tenneti}, {Bundy}, {Greene}, {Sif{\'o}n},
  \& {Bahcall}}]{2020MNRAS.492.3685H}
{Huang}, S., {Leauthaud}, A., {Hearin}, A., {et~al.} 2020, \mnras, 492, 3685

\bibitem[{{Hudson} {et~al.}(2015){Hudson}, {Gillis}, {Coupon}, {Hildebrandt},
  {Erben}, {Heymans}, {Hoekstra}, {Kitching}, {Mellier}, {Miller}, {Van
  Waerbeke}, {Bonnett}, {Fu}, {Kuijken}, {Rowe}, {Schrabback}, {Semboloni},
  {van Uitert}, \& {Velander}}]{2015MNRAS.447..298H}
{Hudson}, M.~J., {Gillis}, B.~R., {Coupon}, J., {et~al.} 2015, \mnras, 447, 298

\bibitem[{{Hunter}(2007)}]{2007CSE.....9...90H}
{Hunter}, J.~D. 2007, Computing in Science and Engineering, 9, 90

\bibitem[{{Ibarra-Medel} {et~al.}(2016){Ibarra-Medel}, {S{\'a}nchez},
  {Avila-Reese}, {Hern{\'a}ndez-Toledo}, {Gonz{\'a}lez}, {Drory}, {Bundy},
  {Bizyaev}, {Cano-D{\'\i}az}, {Malanushenko}, {Pan}, {Roman-Lopes}, \&
  {Thomas}}]{2016MNRAS.463.2799I}
{Ibarra-Medel}, H.~J., {S{\'a}nchez}, S.~F., {Avila-Reese}, V., {et~al.} 2016,
  \mnras, 463, 2799

\bibitem[{{Jeans}(1922)}]{1922MNRAS..82..122J}
{Jeans}, J.~H. 1922, \mnras, 82, 122

\bibitem[{{Kauffmann} {et~al.}(2003){Kauffmann}, {Heckman}, {White}, {Charlot},
  {Tremonti}, {Peng}, {Seibert}, {Brinkmann}, {Nichol}, {SubbaRao}, \&
  {York}}]{2003MNRAS.341...54K}
{Kauffmann}, G., {Heckman}, T.~M., {White}, S. D.~M., {et~al.} 2003, \mnras,
  341, 54

\bibitem[{{Kelz} {et~al.}(2006){Kelz}, {Verheijen}, {Roth}, {Bauer}, {Becker},
  {Paschke}, {Popow}, {S{\'a}nchez}, \& {Laux}}]{Kelz:2006aa}
{Kelz}, A., {Verheijen}, M.~A.~W., {Roth}, M.~M., {et~al.} 2006, \pasp, 118,
  129

\bibitem[{{Koleva} {et~al.}(2011){Koleva}, {Prugniel}, {De Rijcke}, \&
  {Zeilinger}}]{Koleva:2011}
{Koleva}, M., {Prugniel}, P., {De Rijcke}, S., \& {Zeilinger}, W.~W. 2011,
  \mnras, 417, 1643

\bibitem[{{Kuntschner} {et~al.}(2010){Kuntschner}, {Emsellem}, {Bacon},
  {Cappellari}, {Davies}, {de Zeeuw}, {Falc{\'o}n-Barroso}, {Krajnovi{\'c}},
  {McDermid}, {Peletier}, {Sarzi}, {Shapiro}, {van den Bosch}, \& {van de
  Ven}}]{kuntschner+10}
{Kuntschner}, H., {Emsellem}, E., {Bacon}, R., {et~al.} 2010, \mnras, 408, 97

\bibitem[{{La Barbera} {et~al.}(2014){La Barbera}, {Pasquali}, {Ferreras},
  {Gallazzi}, {de Carvalho}, \& {de la Rosa}}]{2014MNRAS.445.1977L}
{La Barbera}, F., {Pasquali}, A., {Ferreras}, I., {et~al.} 2014, \mnras, 445,
  1977

\bibitem[{{Lange} {et~al.}(2019){Lange}, {van den Bosch}, {Zentner}, {Wang}, \&
  {Villarreal}}]{2019MNRAS.482.4824L}
{Lange}, J.~U., {van den Bosch}, F.~C., {Zentner}, A.~R., {Wang}, K., \&
  {Villarreal}, A.~S. 2019, \mnras, 482, 4824

\bibitem[{{Leauthaud} {et~al.}(2012){Leauthaud}, {Tinker}, {Bundy}, {Behroozi},
  {Massey}, {Rhodes}, {George}, {Kneib}, {Benson}, {Wechsler}, {Busha},
  {Capak}, {Cort{\^e}s}, {Ilbert}, {Koekemoer}, {Le F{\`e}vre}, {Lilly},
  {McCracken}, {Salvato}, {Schrabback}, {Scoville}, {Smith}, \&
  {Taylor}}]{2012ApJ...744..159L}
{Leauthaud}, A., {Tinker}, J., {Bundy}, K., {et~al.} 2012, \apj, 744, 159

\bibitem[{{Lemson} \& {Kauffmann}(1999)}]{1999MNRAS.302..111L}
{Lemson}, G. \& {Kauffmann}, G. 1999, \mnras, 302, 111

\bibitem[{{Li} {et~al.}(2012){Li}, {Jing}, {Mao}, {Han}, {Peng}, {Yang}, {Mo},
  \& {van den Bosch}}]{2012ApJ...758...50L}
{Li}, C., {Jing}, Y.~P., {Mao}, S., {et~al.} 2012, \apj, 758, 50

\bibitem[{{Li} {et~al.}(2018){Li}, {Mao}, {Cappellari}, {Ge}, {Long}, {Li},
  {Mo}, {Li}, {Zheng}, {Bundy}, {Thomas}, {Brownstein}, {Roman Lopes}, {Law},
  \& {Drory}}]{Li_Mao_Cappellari:2018}
{Li}, H., {Mao}, S., {Cappellari}, M., {et~al.} 2018, \mnras, 476, 1765

\bibitem[{{Libeskind} {et~al.}(2020){Libeskind}, {Carlesi}, {Grand},
  {Khalatyan}, {Knebe}, {Pakmor}, {Pilipenko}, {Pawlowski}, {Sparre}, {Tempel},
  {Wang}, {Courtois}, {Gottl{\"o}ber}, {Hoffman}, {Minchev}, {Pfrommer},
  {Sorce}, {Springel}, {Steinmetz}, {Tully}, {Vogelsberger}, \&
  {Yepes}}]{2020MNRAS.498.2968L}
{Libeskind}, N.~I., {Carlesi}, E., {Grand}, R. J.~J., {et~al.} 2020, \mnras,
  498, 2968

\bibitem[{{Lim} {et~al.}(2016){Lim}, {Mo}, {Wang}, \&
  {Yang}}]{2016MNRAS.455..499L}
{Lim}, S.~H., {Mo}, H.~J., {Wang}, H., \& {Yang}, X. 2016, \mnras, 455, 499

\bibitem[{{Liu} {et~al.}(2025){Liu}, {Xu}, {Zhang}, {Wang}, \&
  {Liu}}]{2025ApJ...987....4L}
{Liu}, Z., {Xu}, K., {Zhang}, J., {Wang}, W., \& {Liu}, C. 2025, \apj, 987, 4

\bibitem[{{Lorenzoni} {et~al.}(2024){Lorenzoni}, {Rembold}, \& {de
  Carvalho}}]{2024MNRAS.527.3542L}
{Lorenzoni}, V., {Rembold}, S.~B., \& {de Carvalho}, R.~R. 2024, \mnras, 527,
  3542

\bibitem[{{Lyu} {et~al.}(2024){Lyu}, {Peng}, {Jing}, {Yang}, {Ho}, {Renzini},
  {Zhao}, {Mannucci}, {Mo}, {Wang}, {Wang}, {Xu}, {Dou}, {Gallazzi}, {Gu},
  {Maiolino}, {Wang}, \& {Yuan}}]{2024ApJ...972..108L}
{Lyu}, C., {Peng}, Y., {Jing}, Y., {et~al.} 2024, \apj, 972, 108

\bibitem[{{Maiolino} {et~al.}(2008){Maiolino}, {Nagao}, {Grazian}, {Cocchia},
  {Marconi}, {Mannucci}, {Cimatti}, {Pipino}, {Ballero}, {Calura}, {Chiappini},
  {Fontana}, {Granato}, {Matteucci}, {Pastorini}, {Pentericci}, {Risaliti},
  {Salvati}, \& {Silva}}]{2008A&A...488..463M}
{Maiolino}, R., {Nagao}, T., {Grazian}, A., {et~al.} 2008, \aap, 488, 463

\bibitem[{{Mandelbaum} {et~al.}(2006){Mandelbaum}, {Seljak}, {Kauffmann},
  {Hirata}, \& {Brinkmann}}]{2006MNRAS.368..715M}
{Mandelbaum}, R., {Seljak}, U., {Kauffmann}, G., {Hirata}, C.~M., \&
  {Brinkmann}, J. 2006, \mnras, 368, 715

\bibitem[{{Mandelbaum} {et~al.}(2016){Mandelbaum}, {Wang}, {Zu}, {White},
  {Henriques}, \& {More}}]{2016MNRAS.457.3200M}
{Mandelbaum}, R., {Wang}, W., {Zu}, Y., {et~al.} 2016, \mnras, 457, 3200

\bibitem[{{Mannucci} {et~al.}(2010){Mannucci}, {Cresci}, {Maiolino}, {Marconi},
  \& {Gnerucci}}]{2010MNRAS.408.2115M}
{Mannucci}, F., {Cresci}, G., {Maiolino}, R., {Marconi}, A., \& {Gnerucci}, A.
  2010, \mnras, 408, 2115

\bibitem[{{Mart{\'\i}n-Navarro} {et~al.}(2018){Mart{\'\i}n-Navarro},
  {Vazdekis}, {Falc{\'o}n-Barroso}, {La Barbera}, {Y{\i}ld{\i}r{\i}m}, \& {van
  de Ven}}]{martin-navarro+18}
{Mart{\'\i}n-Navarro}, I., {Vazdekis}, A., {Falc{\'o}n-Barroso}, J., {et~al.}
  2018, \mnras, 475, 3700

\bibitem[{{Matchev} {et~al.}(2022){Matchev}, {Matcheva}, \&
  {Roman}}]{2022ApJ...930...33M}
{Matchev}, K.~T., {Matcheva}, K., \& {Roman}, A. 2022, \apj, 930, 33

\bibitem[{{Matthee} {et~al.}(2017){Matthee}, {Schaye}, {Crain}, {Schaller},
  {Bower}, \& {Theuns}}]{2017MNRAS.465.2381M}
{Matthee}, J., {Schaye}, J., {Crain}, R.~A., {et~al.} 2017, \mnras, 465, 2381

\bibitem[{{Mattolini} {et~al.}(2025){Mattolini}, {Zibetti}, {Gallazzi},
  {Scholz-D{\'\i}az}, \& {Pratesi}}]{2025A&A...703A...5M}
{Mattolini}, D., {Zibetti}, S., {Gallazzi}, A.~R., {Scholz-D{\'\i}az}, L., \&
  {Pratesi}, J. 2025, \aap, 703, A5

\bibitem[{{McDermid} {et~al.}(2015){McDermid}, {Alatalo}, {Blitz}, {Bournaud},
  {Bureau}, {Cappellari}, {Crocker}, {Davies}, {Davis}, {de Zeeuw}, {Duc},
  {Emsellem}, {Khochfar}, {Krajnovi{\'c}}, {Kuntschner}, {Morganti}, {Naab},
  {Oosterloo}, {Sarzi}, {Scott}, {Serra}, {Weijmans}, \&
  {Young}}]{2015MNRAS.448.3484M}
{McDermid}, R.~M., {Alatalo}, K., {Blitz}, L., {et~al.} 2015, \mnras, 448, 3484

\bibitem[{{Mehlert} {et~al.}(2003){Mehlert}, {Thomas}, {Saglia}, {Bender}, \&
  {Wegner}}]{Mehlert:2003}
{Mehlert}, D., {Thomas}, D., {Saglia}, R.~P., {Bender}, R., \& {Wegner}, G.
  2003, \aap, 407, 423

\bibitem[{{Mercado} {et~al.}(2025){Mercado}, {Moreno}, {Feldmann}, {Zeender},
  {Benavides}, {Piotrowska}, {Klein}, {Wheeler}, {Necib}, {Bullock}, \&
  {Hopkins}}]{2025ApJ...983...93M}
{Mercado}, F.~J., {Moreno}, J., {Feldmann}, R., {et~al.} 2025, \apj, 983, 93

\bibitem[{{More} {et~al.}(2011){More}, {van den Bosch}, {Cacciato}, {Skibba},
  {Mo}, \& {Yang}}]{2011MNRAS.410..210M}
{More}, S., {van den Bosch}, F.~C., {Cacciato}, M., {et~al.} 2011, \mnras, 410,
  210

\bibitem[{{Moster} {et~al.}(2013){Moster}, {Naab}, \&
  {White}}]{2013MNRAS.428.3121M}
{Moster}, B.~P., {Naab}, T., \& {White}, S. D.~M. 2013, \mnras, 428, 3121

\bibitem[{{Neumann} {et~al.}(2021){Neumann}, {Thomas}, {Maraston}, {Goddard},
  {Lian}, {Hill}, {Dom{\'\i}nguez S{\'a}nchez}, {Bernardi},
  {Margalef-Bentabol}, {Barrera-Ballesteros}, {Bizyaev}, {Boardman}, {Drory},
  {Fern{\'a}ndez-Trincado}, \& {Lane}}]{2021MNRAS.508.4844N}
{Neumann}, J., {Thomas}, D., {Maraston}, C., {et~al.} 2021, \mnras, 508, 4844

\bibitem[{{Oser} {et~al.}(2012){Oser}, {Naab}, {Ostriker}, \&
  {Johansson}}]{2012ApJ...744...63O}
{Oser}, L., {Naab}, T., {Ostriker}, J.~P., \& {Johansson}, P.~H. 2012, \apj,
  744, 63

\bibitem[{{Oser} {et~al.}(2010){Oser}, {Ostriker}, {Naab}, {Johansson}, \&
  {Burkert}}]{2010ApJ...725.2312O}
{Oser}, L., {Ostriker}, J.~P., {Naab}, T., {Johansson}, P.~H., \& {Burkert}, A.
  2010, \apj, 725, 2312

\bibitem[{{Oyarz{\'u}n} {et~al.}(2019){Oyarz{\'u}n}, {Bundy}, {Westfall},
  {Belfiore}, {Thomas}, {Maraston}, {Lian}, {Arag{\'o}n-Salamanca}, {Zheng},
  {Gonzalez-Perez}, {Law}, {Drory}, \& {Andrews}}]{2019ApJ...880..111O}
{Oyarz{\'u}n}, G.~A., {Bundy}, K., {Westfall}, K.~B., {et~al.} 2019, \apj, 880,
  111

\bibitem[{{Oyarz{\'u}n} {et~al.}(2022){Oyarz{\'u}n}, {Bundy}, {Westfall},
  {Tinker}, {Belfiore}, {Argudo-Fern{\'a}ndez}, {Zheng}, {Conroy}, {Masters},
  {Wake}, {Law}, {McDermid}, {Arag{\'o}n-Salamanca}, {Parikh}, {Yan},
  {Bershady}, {S{\'a}nchez}, {Andrews}, {Fern{\'a}ndez-Trincado}, {Lane},
  {Bizyaev}, {Boardman}, {Lacerna}, {Brownstein}, {Drory}, \&
  {Zhang}}]{2022ApJ...933...88O}
{Oyarz{\'u}n}, G.~A., {Bundy}, K., {Westfall}, K.~B., {et~al.} 2022, \apj, 933,
  88

\bibitem[{{Oyarz{\'u}n} {et~al.}(2024){Oyarz{\'u}n}, {Tinker}, {Bundy},
  {Xhakaj}, \& {Wyithe}}]{2024ApJ...974...29O}
{Oyarz{\'u}n}, G.~A., {Tinker}, J.~L., {Bundy}, K., {Xhakaj}, E., \& {Wyithe},
  J. S.~B. 2024, \apj, 974, 29

\bibitem[{{Parikh} {et~al.}(2019){Parikh}, {Thomas}, {Maraston}, {Westfall},
  {Lian}, {Fraser-McKelvie}, {Andrews}, {Drory}, \&
  {Meneses-Goytia}}]{Parikh:2019}
{Parikh}, T., {Thomas}, D., {Maraston}, C., {et~al.} 2019, \mnras, 483, 3420

\bibitem[{{Pasquali} {et~al.}(2010){Pasquali}, {Gallazzi}, {Fontanot}, {van den
  Bosch}, {De Lucia}, {Mo}, \& {Yang}}]{2010MNRAS.407..937P}
{Pasquali}, A., {Gallazzi}, A., {Fontanot}, F., {et~al.} 2010, \mnras, 407, 937

\bibitem[{{Pei} {et~al.}(2024){Pei}, {Guo}, {Shao}, {He}, \&
  {Gu}}]{2024MNRAS.531.2262P}
{Pei}, W., {Guo}, Q., {Shao}, S., {He}, Y., \& {Gu}, Q. 2024, \mnras, 531, 2262

\bibitem[{{Peletier}(1989)}]{1989PhDT.......149P}
{Peletier}, R.~F. 1989, PhD thesis, -

\bibitem[{{Porras-Valverde} {et~al.}(2024){Porras-Valverde}, {Forbes},
  {Somerville}, {Stevens}, {Holley-Bockelmann}, {Berlind}, \&
  {Genel}}]{2024ApJ...976..148P}
{Porras-Valverde}, A.~J., {Forbes}, J.~C., {Somerville}, R.~S., {et~al.} 2024,
  \apj, 976, 148

\bibitem[{{Posti} {et~al.}(2019){Posti}, {Fraternali}, \&
  {Marasco}}]{2019A&A...626A..56P}
{Posti}, L., {Fraternali}, F., \& {Marasco}, A. 2019, \aap, 626, A56

\bibitem[{{Reda} {et~al.}(2007){Reda}, {Proctor}, {Forbes}, {Hau}, \&
  {Larsen}}]{Reda:2007}
{Reda}, F.~M., {Proctor}, R.~N., {Forbes}, D.~A., {Hau}, G. K.~T., \& {Larsen},
  S.~S. 2007, \mnras, 377, 1772

\bibitem[{{Renzini} \& {Peng}(2015)}]{2015ApJ...801L..29R}
{Renzini}, A. \& {Peng}, Y.-j. 2015, \apjl, 801, L29

\bibitem[{{Rodriguez-Gomez} {et~al.}(2016){Rodriguez-Gomez}, {Pillepich},
  {Sales}, {Genel}, {Vogelsberger}, {Zhu}, {Wellons}, {Nelson}, {Torrey},
  {Springel}, {Ma}, \& {Hernquist}}]{2016MNRAS.458.2371R}
{Rodriguez-Gomez}, V., {Pillepich}, A., {Sales}, L.~V., {et~al.} 2016, \mnras,
  458, 2371

\bibitem[{{Roth} {et~al.}(2005){Roth}, {Kelz}, {Fechner}, {Hahn}, {Bauer},
  {Becker}, {B{\"o}hm}, {Christensen}, {Dionies}, {Paschke}, {Popow}, {Wolter},
  {Schmoll}, {Laux}, \& {Altmann}}]{Roth:2005aa}
{Roth}, M.~M., {Kelz}, A., {Fechner}, T., {et~al.} 2005, \pasp, 117, 620

\bibitem[{{Ruiz-Lara} {et~al.}(2017){Ruiz-Lara}, {P{\'e}rez}, {Florido},
  {S{\'a}nchez-Bl{\'a}zquez}, {M{\'e}ndez-Abreu}, {S{\'a}nchez-Menguiano},
  {S{\'a}nchez}, {Lyubenova}, {Falc{\'o}n-Barroso}, {van de Ven}, {Marino}, {de
  Lorenzo-C{\'a}ceres}, {Catal{\'a}n-Torrecilla}, {Costantin},
  {Bland-Hawthorn}, {Galbany}, {Garc{\'\i}a-Benito}, {Husemann}, {Kehrig},
  {M{\'a}rquez}, {Mast}, {Walcher}, {Zibetti}, {Ziegler}, \& {CALIFA
  Team}}]{2017A&A...604A...4R}
{Ruiz-Lara}, T., {P{\'e}rez}, I., {Florido}, E., {et~al.} 2017, \aap, 604, A4

\bibitem[{{Saintonge} {et~al.}(2017){Saintonge}, {Catinella}, {Tacconi},
  {Kauffmann}, {Genzel}, {Cortese}, {Dav{\'e}}, {Fletcher},
  {Graci{\'a}-Carpio}, {Kramer}, {Heckman}, {Janowiecki}, {Lutz}, {Rosario},
  {Schiminovich}, {Schuster}, {Wang}, {Wuyts}, {Borthakur}, {Lamperti}, \&
  {Roberts-Borsani}}]{2017ApJS..233...22S}
{Saintonge}, A., {Catinella}, B., {Tacconi}, L.~J., {et~al.} 2017, \apjs, 233,
  22

\bibitem[{{Saintonge} {et~al.}(2011){Saintonge}, {Kauffmann}, {Kramer},
  {Tacconi}, {Buchbender}, {Catinella}, {Fabello}, {Graci{\'a}-Carpio}, {Wang},
  {Cortese}, {Fu}, {Genzel}, {Giovanelli}, {Guo}, {Haynes}, {Heckman},
  {Krumholz}, {Lemonias}, {Li}, {Moran}, {Rodriguez-Fernandez}, {Schiminovich},
  {Schuster}, \& {Sievers}}]{2011MNRAS.415...32S}
{Saintonge}, A., {Kauffmann}, G., {Kramer}, C., {et~al.} 2011, \mnras, 415, 32

\bibitem[{{Salim} {et~al.}(2007){Salim}, {Rich}, {Charlot}, {Brinchmann},
  {Johnson}, {Schiminovich}, {Seibert}, {Mallery}, {Heckman}, {Forster},
  {Friedman}, {Martin}, {Morrissey}, {Neff}, {Small}, {Wyder}, {Bianchi},
  {Donas}, {Lee}, {Madore}, {Milliard}, {Szalay}, {Welsh}, \&
  {Yi}}]{2007ApJS..173..267S}
{Salim}, S., {Rich}, R.~M., {Charlot}, S., {et~al.} 2007, \apjs, 173, 267

\bibitem[{{S{\'a}nchez}(2020)}]{2020ARA&A..58...99S}
{S{\'a}nchez}, S.~F. 2020, \araa, 58, 99

\bibitem[{{S{\'a}nchez} {et~al.}(2016){S{\'a}nchez}, {Garc{\'\i}a-Benito},
  {Zibetti}, {Walcher}, {Husemann}, {Mendoza}, {Galbany}, {Falc{\'o}n-Barroso},
  {Mast}, {Aceituno}, {Aguerri}, {Alves}, {Amorim}, {Ascasibar},
  {Barrado-Navascues}, {Barrera-Ballesteros}, {Bekerait{\`e}},
  {Bland-Hawthorn}, {Cano D{\'\i}az}, {Cid Fernandes}, {Cavichia}, {Cortijo},
  {Dannerbauer}, {Demleitner}, {D{\'\i}az}, {Dettmar}, {de
  Lorenzo-C{\'a}ceres}, {del Olmo}, {Galazzi}, {Garc{\'\i}a-Lorenzo}, {Gil de
  Paz}, {Gonz{\'a}lez Delgado}, {Holmes}, {Igl{\'e}sias-P{\'a}ramo}, {Kehrig},
  {Kelz}, {Kennicutt}, {Kleemann}, {Lacerda}, {L{\'o}pez Fern{\'a}ndez},
  {L{\'o}pez S{\'a}nchez}, {Lyubenova}, {Marino}, {M{\'a}rquez},
  {Mendez-Abreu}, {Moll{\'a}}, {Monreal-Ibero}, {Ortega Minakata},
  {Torres-Papaqui}, {P{\'e}rez}, {Rosales-Ortega}, {Roth},
  {S{\'a}nchez-Bl{\'a}zquez}, {Schilling}, {Spekkens}, {Vale Asari}, {van den
  Bosch}, {van de Ven}, {Vilchez}, {Wild}, {Wisotzki}, {Y{\i}ld{\i}r{\i}m}, \&
  {Ziegler}}]{2016A&A...594A..36S}
{S{\'a}nchez}, S.~F., {Garc{\'\i}a-Benito}, R., {Zibetti}, S., {et~al.} 2016,
  \aap, 594, A36

\bibitem[{{S{\'a}nchez} {et~al.}(2012){S{\'a}nchez}, {Kennicutt}, {Gil de Paz},
  {van de Ven}, {V{\'\i}lchez}, {Wisotzki}, {Walcher}, {Mast}, {Aguerri},
  {Albiol-P{\'e}rez}, {Alonso-Herrero}, {Alves}, {Bakos}, {Bart{\'a}kov{\'a}},
  {Bland-Hawthorn}, {Boselli}, {Bomans}, {Castillo-Morales}, {Cortijo-Ferrero},
  {de Lorenzo-C{\'a}ceres}, {Del Olmo}, {Dettmar}, {D{\'\i}az}, {Ellis},
  {Falc{\'o}n-Barroso}, {Flores}, {Gallazzi}, {Garc{\'\i}a-Lorenzo},
  {Gonz{\'a}lez Delgado}, {Gruel}, {Haines}, {Hao}, {Husemann},
  {Igl{\'e}sias-P{\'a}ramo}, {Jahnke}, {Johnson}, {Jungwiert}, {Kalinova},
  {Kehrig}, {Kupko}, {L{\'o}pez-S{\'a}nchez}, {Lyubenova}, {Marino},
  {M{\'a}rmol-Queralt{\'o}}, {M{\'a}rquez}, {Masegosa}, {Meidt},
  {Mendez-Abreu}, {Monreal-Ibero}, {Montijo}, {Mour{\~a}o}, {Palacios-Navarro},
  {Papaderos}, {Pasquali}, {Peletier}, {P{\'e}rez}, {P{\'e}rez}, {Quirrenbach},
  {Rela{\~n}o}, {Rosales-Ortega}, {Roth}, {Ruiz-Lara},
  {S{\'a}nchez-Bl{\'a}zquez}, {Sengupta}, {Singh}, {Stanishev}, {Trager},
  {Vazdekis}, {Viironen}, {Wild}, {Zibetti}, \&
  {Ziegler}}]{2012A&A...538A...8S}
{S{\'a}nchez}, S.~F., {Kennicutt}, R.~C., {Gil de Paz}, A., {et~al.} 2012,
  \aap, 538, A8

\bibitem[{{S{\'a}nchez-Bl{\'a}zquez} {et~al.}(2007){S{\'a}nchez-Bl{\'a}zquez},
  {Forbes}, {Strader}, {Brodie}, \& {Proctor}}]{Sanchez-Blazquez:2007}
{S{\'a}nchez-Bl{\'a}zquez}, P., {Forbes}, D.~A., {Strader}, J., {Brodie}, J.,
  \& {Proctor}, R. 2007, \mnras, 377, 759

\bibitem[{{S{\'a}nchez-Bl{\'a}zquez} {et~al.}(2006){S{\'a}nchez-Bl{\'a}zquez},
  {Peletier}, {Jim{\'e}nez-Vicente}, {Cardiel}, {Cenarro},
  {Falc{\'o}n-Barroso}, {Gorgas}, {Selam}, \& {Vazdekis}}]{2006MNRAS.371..703S}
{S{\'a}nchez-Bl{\'a}zquez}, P., {Peletier}, R.~F., {Jim{\'e}nez-Vicente}, J.,
  {et~al.} 2006, \mnras, 371, 703

\bibitem[{{S{\'a}nchez-Bl{\'a}zquez} {et~al.}(2014){S{\'a}nchez-Bl{\'a}zquez},
  {Rosales-Ortega}, {M{\'e}ndez-Abreu}, {P{\'e}rez}, {S{\'a}nchez}, {Zibetti},
  {Aguerri}, {Bland-Hawthorn}, {Catal{\'a}n-Torrecilla}, {Cid Fernandes}, {de
  Amorim}, {de Lorenzo-Caceres}, {Falc{\'o}n-Barroso}, {Galazzi}, {Garc{\'\i}a
  Benito}, {Gil de Paz}, {Gonz{\'a}lez Delgado}, {Husemann},
  {Iglesias-P{\'a}ramo}, {Jungwiert}, {Marino}, {M{\'a}rquez}, {Mast},
  {Mendoza}, {Moll{\'a}}, {Papaderos}, {Ruiz-Lara}, {van de Ven}, {Walcher}, \&
  {Wisotzki}}]{2014A&A...570A...6S}
{S{\'a}nchez-Bl{\'a}zquez}, P., {Rosales-Ortega}, F.~F., {M{\'e}ndez-Abreu},
  J., {et~al.} 2014, \aap, 570, A6

\bibitem[{{Sandage}(1986)}]{1986A&A...161...89S}
{Sandage}, A. 1986, \aap, 161, 89

\bibitem[{{Schaye} {et~al.}(2015){Schaye}, {Crain}, {Bower}, {Furlong},
  {Schaller}, {Theuns}, {Dalla Vecchia}, {Frenk}, {McCarthy}, {Helly},
  {Jenkins}, {Rosas-Guevara}, {White}, {Baes}, {Booth}, {Camps}, {Navarro},
  {Qu}, {Rahmati}, {Sawala}, {Thomas}, \& {Trayford}}]{2015MNRAS.446..521S}
{Schaye}, J., {Crain}, R.~A., {Bower}, R.~G., {et~al.} 2015, \mnras, 446, 521

\bibitem[{{Scholz-D{\'\i}az} {et~al.}(2022){Scholz-D{\'\i}az},
  {Mart{\'\i}n-Navarro}, \& {Falc{\'o}n-Barroso}}]{2022MNRAS.511.4900S}
{Scholz-D{\'\i}az}, L., {Mart{\'\i}n-Navarro}, I., \& {Falc{\'o}n-Barroso}, J.
  2022, \mnras, 511, 4900

\bibitem[{{Scholz-D{\'\i}az} {et~al.}(2023){Scholz-D{\'\i}az},
  {Mart{\'\i}n-Navarro}, \& {Falc{\'o}n-Barroso}}]{2023MNRAS.518.6325S}
{Scholz-D{\'\i}az}, L., {Mart{\'\i}n-Navarro}, I., \& {Falc{\'o}n-Barroso}, J.
  2023, \mnras, 518, 6325

\bibitem[{{Scholz-D{\'\i}az} {et~al.}(2024){Scholz-D{\'\i}az},
  {Mart{\'\i}n-Navarro}, {Falc{\'o}n-Barroso}, {Lyubenova}, \& {van de
  Ven}}]{2024NatAs...8..648S}
{Scholz-D{\'\i}az}, L., {Mart{\'\i}n-Navarro}, I., {Falc{\'o}n-Barroso}, J.,
  {Lyubenova}, M., \& {van de Ven}, G. 2024, Nature Astronomy, 8, 648

\bibitem[{{Schwarzschild}(1979)}]{1979ApJ...232..236S}
{Schwarzschild}, M. 1979, \apj, 232, 236

\bibitem[{{Shao} {et~al.}(2023){Shao}, {de Santi}, {Villaescusa-Navarro},
  {Teyssier}, {Ni}, {Angl{\'e}s-Alc{\'a}zar}, {Genel}, {Steinwandel},
  {Hern{\'a}ndez-Mart{\'\i}nez}, {Dolag}, {Lovell}, {Garrison}, {Visbal},
  {Kulkarni}, {Hernquist}, {Castro}, \& {Vogelsberger}}]{2023ApJ...956..149S}
{Shao}, H., {de Santi}, N. S.~M., {Villaescusa-Navarro}, F., {et~al.} 2023,
  \apj, 956, 149

\bibitem[{{Shao} {et~al.}(2022){Shao}, {Villaescusa-Navarro}, {Genel},
  {Spergel}, {Angl{\'e}s-Alc{\'a}zar}, {Hernquist}, {Dav{\'e}}, {Narayanan},
  {Contardo}, \& {Vogelsberger}}]{2022ApJ...927...85S}
{Shao}, H., {Villaescusa-Navarro}, F., {Genel}, S., {et~al.} 2022, \apj, 927,
  85

\bibitem[{{Shen} {et~al.}(2003){Shen}, {Mo}, {White}, {Blanton}, {Kauffmann},
  {Voges}, {Brinkmann}, \& {Csabai}}]{2003MNRAS.343..978S}
{Shen}, S., {Mo}, H.~J., {White}, S. D.~M., {et~al.} 2003, \mnras, 343, 978

\bibitem[{{Sobral} {et~al.}(2022){Sobral}, {van der Wel}, {Bezanson}, {Bell},
  {Muzzin}, {D'Eugenio}, {Darvish}, {Gallazzi}, {Wu}, {Maseda}, {Matthee},
  {Paulino-Afonso}, {Straatman}, \& {van Dokkum}}]{2022ApJ...926..117S}
{Sobral}, D., {van der Wel}, A., {Bezanson}, R., {et~al.} 2022, \apj, 926, 117

\bibitem[{{Spolaor} {et~al.}(2009){Spolaor}, {Proctor}, {Forbes}, \&
  {Couch}}]{Spolaor:2009}
{Spolaor}, M., {Proctor}, R.~N., {Forbes}, D.~A., \& {Couch}, W.~J. 2009, \apj,
  691, L138

\bibitem[{{Springel}(2005)}]{2005MNRAS.364.1105S}
{Springel}, V. 2005, \mnras, 364, 1105

\bibitem[{{Strateva} {et~al.}(2001){Strateva}, {Ivezi{\'c}}, {Knapp},
  {Narayanan}, {Strauss}, {Gunn}, {Lupton}, {Schlegel}, {Bahcall}, {Brinkmann},
  {Brunner}, {Budav{\'a}ri}, {Csabai}, {Castander}, {Doi}, {Fukugita},
  {Gy{\H{o}}ry}, {Hamabe}, {Hennessy}, {Ichikawa}, {Kunszt}, {Lamb}, {McKay},
  {Okamura}, {Racusin}, {Sekiguchi}, {Schneider}, {Shimasaku}, \&
  {York}}]{2001AJ....122.1861S}
{Strateva}, I., {Ivezi{\'c}}, {\v{Z}}., {Knapp}, G.~R., {et~al.} 2001, \aj,
  122, 1861

\bibitem[{{Taylor} {et~al.}(2020){Taylor}, {Cluver}, {Duffy}, {Gurri},
  {Hoekstra}, {Sonnenfeld}, {Bremer}, {Brouwer}, {Chisari}, {Dvornik}, {Erben},
  {Hildebrandt}, {Hopkins}, {Kelvin}, {Phillipps}, {Robotham}, {Sif{\'o}n},
  {Vakili}, \& {Wright}}]{2020MNRAS.499.2896T}
{Taylor}, E.~N., {Cluver}, M.~E., {Duffy}, A., {et~al.} 2020, \mnras, 499, 2896

\bibitem[{{Tempel} {et~al.}(2014){Tempel}, {Tamm}, {Gramann}, {Tuvikene},
  {Liivam{\"a}gi}, {Suhhonenko}, {Kipper}, {Einasto}, \&
  {Saar}}]{2014A&A...566A...1T}
{Tempel}, E., {Tamm}, A., {Gramann}, M., {et~al.} 2014, \aap, 566, A1

\bibitem[{{Thater} {et~al.}(2022){Thater}, {Jethwa}, {Tahmasebzadeh}, {Zhu},
  {den Brok}, {Santucci}, {Ding}, {Poci}, {Lilley}, {Tim de Zeeuw}, {Zocchi},
  {Maindl}, {Rigamonti}, {Yang}, {Fahrion}, \& {van de
  Ven}}]{2022A&A...667A..51T}
{Thater}, S., {Jethwa}, P., {Tahmasebzadeh}, B., {et~al.} 2022, \aap, 667, A51

\bibitem[{{Thomas} {et~al.}(2003){Thomas}, {Maraston}, \&
  {Bender}}]{2003MNRAS.339..897T}
{Thomas}, D., {Maraston}, C., \& {Bender}, R. 2003, \mnras, 339, 897

\bibitem[{{Thomas} {et~al.}(2005){Thomas}, {Maraston}, {Bender}, \& {Mendes de
  Oliveira}}]{2005ApJ...621..673T}
{Thomas}, D., {Maraston}, C., {Bender}, R., \& {Mendes de Oliveira}, C. 2005,
  \apj, 621, 673

\bibitem[{{Tinker}(2021)}]{2021ApJ...923..154T}
{Tinker}, J.~L. 2021, \apj, 923, 154

\bibitem[{{Tinker} {et~al.}(2013){Tinker}, {Leauthaud}, {Bundy}, {George},
  {Behroozi}, {Massey}, {Rhodes}, \& {Wechsler}}]{2013ApJ...778...93T}
{Tinker}, J.~L., {Leauthaud}, A., {Bundy}, K., {et~al.} 2013, \apj, 778, 93

\bibitem[{{Tojeiro} {et~al.}(2017){Tojeiro}, {Eardley}, {Peacock}, {Norberg},
  {Alpaslan}, {Driver}, {Henriques}, {Hopkins}, {Kafle}, {Robotham}, {Thomas},
  {Tonini}, \& {Wild}}]{2017MNRAS.470.3720T}
{Tojeiro}, R., {Eardley}, E., {Peacock}, J.~A., {et~al.} 2017, \mnras, 470,
  3720

\bibitem[{{Trager} \& {Somerville}(2009)}]{2009MNRAS.395..608T}
{Trager}, S.~C. \& {Somerville}, R.~S. 2009, \mnras, 395, 608

\bibitem[{{Tremonti} {et~al.}(2004){Tremonti}, {Heckman}, {Kauffmann},
  {Brinchmann}, {Charlot}, {White}, {Seibert}, {Peng}, {Schlegel}, {Uomoto},
  {Fukugita}, \& {Brinkmann}}]{2004ApJ...613..898T}
{Tremonti}, C.~A., {Heckman}, T.~M., {Kauffmann}, G., {et~al.} 2004, \apj, 613,
  898

\bibitem[{{Trujillo} {et~al.}(2020){Trujillo}, {Chamba}, \&
  {Knapen}}]{2020MNRAS.493...87T}
{Trujillo}, I., {Chamba}, N., \& {Knapen}, J.~H. 2020, \mnras, 493, 87

\bibitem[{{Trujillo} {et~al.}(2007){Trujillo}, {Conselice}, {Bundy}, {Cooper},
  {Eisenhardt}, \& {Ellis}}]{2007MNRAS.382..109T}
{Trujillo}, I., {Conselice}, C.~J., {Bundy}, K., {et~al.} 2007, \mnras, 382,
  109

\bibitem[{{Trujillo} {et~al.}(2004){Trujillo}, {Rudnick}, {Rix}, {Labb{\'e}},
  {Franx}, {Daddi}, {van Dokkum}, {F{\"o}rster Schreiber}, {Kuijken},
  {Moorwood}, {R{\"o}ttgering}, {van der Wel}, {van der Werf}, \& {van
  Starkenburg}}]{2004ApJ...604..521T}
{Trujillo}, I., {Rudnick}, G., {Rix}, H.-W., {et~al.} 2004, \apj, 604, 521

\bibitem[{{Trussler} {et~al.}(2020){Trussler}, {Maiolino}, {Maraston}, {Peng},
  {Thomas}, {Goddard}, \& {Lian}}]{2020MNRAS.491.5406T}
{Trussler}, J., {Maiolino}, R., {Maraston}, C., {et~al.} 2020, \mnras, 491,
  5406

\bibitem[{{Trussler} {et~al.}(2021){Trussler}, {Maiolino}, {Maraston}, {Peng},
  {Thomas}, {Goddard}, \& {Lian}}]{2021MNRAS.500.4469T}
{Trussler}, J., {Maiolino}, R., {Maraston}, C., {et~al.} 2021, \mnras, 500,
  4469

\bibitem[{Vallat(2018)}]{Vallat2018}
Vallat, R. 2018, Journal of Open Source Software, 3, 1026

\bibitem[{{van den Bosch} {et~al.}(2008){van den Bosch}, {van de Ven},
  {Verolme}, {Cappellari}, \& {de Zeeuw}}]{2008MNRAS.385..647V}
{van den Bosch}, R.~C.~E., {van de Ven}, G., {Verolme}, E.~K., {Cappellari},
  M., \& {de Zeeuw}, P.~T. 2008, \mnras, 385, 647

\bibitem[{{van der Wel} {et~al.}(2014){van der Wel}, {Franx}, {van Dokkum},
  {Skelton}, {Momcheva}, {Whitaker}, {Brammer}, {Bell}, {Rix}, {Wuyts},
  {Ferguson}, {Holden}, {Barro}, {Koekemoer}, {Chang}, {McGrath},
  {H{\"a}ussler}, {Dekel}, {Behroozi}, {Fumagalli}, {Leja}, {Lundgren},
  {Maseda}, {Nelson}, {Wake}, {Patel}, {Labb{\'e}}, {Faber}, {Grogin}, \&
  {Kocevski}}]{2014ApJ...788...28V}
{van der Wel}, A., {Franx}, M., {van Dokkum}, P.~G., {et~al.} 2014, \apj, 788,
  28

\bibitem[{{Verheijen} {et~al.}(2004){Verheijen}, {Bershady}, {Andersen},
  {Swaters}, {Westfall}, {Kelz}, \& {Roth}}]{Verheijen:2004aa}
{Verheijen}, M.~A.~W., {Bershady}, M.~A., {Andersen}, D.~R., {et~al.} 2004,
  Astronomische Nachrichten, 325, 151

\bibitem[{{Villaescusa-Navarro} {et~al.}(2021){Villaescusa-Navarro},
  {Angl{\'e}s-Alc{\'a}zar}, {Genel}, {Spergel}, {Somerville}, {Dave},
  {Pillepich}, {Hernquist}, {Nelson}, {Torrey}, {Narayanan}, {Li}, {Philcox},
  {La Torre}, {Maria Delgado}, {Ho}, {Hassan}, {Burkhart}, {Wadekar},
  {Battaglia}, {Contardo}, \& {Bryan}}]{2021ApJ...915...71V}
{Villaescusa-Navarro}, F., {Angl{\'e}s-Alc{\'a}zar}, D., {Genel}, S., {et~al.}
  2021, \apj, 915, 71

\bibitem[{Virtanen {et~al.}(2020)Virtanen, Gommers, Oliphant, Haberland, Reddy,
  Cournapeau, Burovski, Peterson, Weckesser, Bright, {van der Walt}, Brett,
  Wilson, Millman, Mayorov, Nelson, Jones, Kern, Larson, Carey, Polat, Feng,
  Moore, {VanderPlas}, Laxalde, Perktold, Cimrman, Henriksen, Quintero, Harris,
  Archibald, Ribeiro, Pedregosa, {van Mulbregt}, \& {SciPy 1.0
  Contributors}}]{2020SciPy-NMeth}
Virtanen, P., Gommers, R., Oliphant, T.~E., {et~al.} 2020, Nature Methods, 17,
  261

\bibitem[{{Wadekar} {et~al.}(2023{\natexlab{a}}){Wadekar}, {Thiele}, {Hill},
  {Pandey}, {Villaescusa-Navarro}, {Spergel}, {Cranmer}, {Nagai},
  {Angl{\'e}s-Alc{\'a}zar}, {Ho}, \& {Hernquist}}]{2023MNRAS.522.2628W}
{Wadekar}, D., {Thiele}, L., {Hill}, J.~C., {et~al.} 2023{\natexlab{a}},
  \mnras, 522, 2628

\bibitem[{{Wadekar} {et~al.}(2023{\natexlab{b}}){Wadekar}, {Thiele},
  {Villaescusa-Navarro}, {Hill}, {Cranmer}, {Spergel}, {Battaglia},
  {Angl{\'e}s-Alc{\'a}zar}, {Hernquist}, \& {Ho}}]{2023PNAS..12002074W}
{Wadekar}, D., {Thiele}, L., {Villaescusa-Navarro}, F., {et~al.}
  2023{\natexlab{b}}, Proceedings of the National Academy of Science, 120,
  e2202074120

\bibitem[{{Walcher} {et~al.}(2014){Walcher}, {Wisotzki}, {Bekerait{\'e}},
  {Husemann}, {Iglesias-P{\'a}ramo}, {Backsmann}, {Barrera Ballesteros},
  {Catal{\'a}n-Torrecilla}, {Cortijo}, {del Olmo}, {Garcia Lorenzo},
  {Falc{\'o}n-Barroso}, {Jilkova}, {Kalinova}, {Mast}, {Marino},
  {M{\'e}ndez-Abreu}, {Pasquali}, {S{\'a}nchez}, {Trager}, {Zibetti},
  {Aguerri}, {Alves}, {Bland-Hawthorn}, {Boselli}, {Castillo Morales}, {Cid
  Fernandes}, {Flores}, {Galbany}, {Gallazzi}, {Garc{\'\i}a-Benito}, {Gil de
  Paz}, {Gonz{\'a}lez-Delgado}, {Jahnke}, {Jungwiert}, {Kehrig}, {Lyubenova},
  {M{\'a}rquez Perez}, {Masegosa}, {Monreal Ibero}, {P{\'e}rez}, {Quirrenbach},
  {Rosales-Ortega}, {Roth}, {Sanchez-Blazquez}, {Spekkens}, {Tundo}, {van de
  Ven}, {Verheijen}, {Vilchez}, \& {Ziegler}}]{2014A&A...569A...1W}
{Walcher}, C.~J., {Wisotzki}, L., {Bekerait{\'e}}, S., {et~al.} 2014, \aap,
  569, A1

\bibitem[{{Wang} {et~al.}(2015){Wang}, {Dutton}, {Stinson}, {Macci{\`o}},
  {Penzo}, {Kang}, {Keller}, \& {Wadsley}}]{2015MNRAS.454...83W}
{Wang}, L., {Dutton}, A.~A., {Stinson}, G.~S., {et~al.} 2015, \mnras, 454, 83

\bibitem[{{Wang} {et~al.}(2006){Wang}, {Li}, {Kauffmann}, \& {De
  Lucia}}]{2006MNRAS.371..537W}
{Wang}, L., {Li}, C., {Kauffmann}, G., \& {De Lucia}, G. 2006, \mnras, 371, 537

\bibitem[{Waskom(2021)}]{Waskom2021}
Waskom, M.~L. 2021, Journal of Open Source Software, 6, 3021

\bibitem[{{Wechsler} {et~al.}(2002){Wechsler}, {Bullock}, {Primack},
  {Kravtsov}, \& {Dekel}}]{2002ApJ...568...52W}
{Wechsler}, R.~H., {Bullock}, J.~S., {Primack}, J.~R., {Kravtsov}, A.~V., \&
  {Dekel}, A. 2002, \apj, 568, 52

\bibitem[{{Wechsler} \& {Tinker}(2018)}]{2018ARA&A..56..435W}
{Wechsler}, R.~H. \& {Tinker}, J.~L. 2018, \araa, 56, 435

\bibitem[{{W}es {M}c{K}inney(2010)}]{mckinney-proc-scipy-2010}
{W}es {M}c{K}inney. 2010, in {P}roceedings of the 9th {P}ython in {S}cience
  {C}onference, ed. {S}t\'efan van~der {W}alt \& {J}arrod {M}illman, 56 -- 61

\bibitem[{{Whitaker} {et~al.}(2014){Whitaker}, {Franx}, {Leja}, {van Dokkum},
  {Henry}, {Skelton}, {Fumagalli}, {Momcheva}, {Brammer}, {Labb{\'e}},
  {Nelson}, \& {Rigby}}]{2014ApJ...795..104W}
{Whitaker}, K.~E., {Franx}, M., {Leja}, J., {et~al.} 2014, \apj, 795, 104

\bibitem[{{White} \& {Rees}(1978)}]{1978MNRAS.183..341W}
{White}, S.~D.~M. \& {Rees}, M.~J. 1978, \mnras, 183, 341

\bibitem[{{Wojtak} \& {Mamon}(2013)}]{2013MNRAS.428.2407W}
{Wojtak}, R. \& {Mamon}, G.~A. 2013, \mnras, 428, 2407

\bibitem[{{Wong} \& {Cranmer}(2022)}]{2022mla..confE..25W}
{Wong}, K. \& {Cranmer}, M. 2022, in Machine Learning for Astrophysics, 25

\bibitem[{{Worthey} {et~al.}(1994){Worthey}, {Faber}, {Gonzalez}, \&
  {Burstein}}]{1994ApJS...94..687W}
{Worthey}, G., {Faber}, S.~M., {Gonzalez}, J.~J., \& {Burstein}, D. 1994,
  \apjs, 94, 687

\bibitem[{{Worthey} \& {Ottaviani}(1997)}]{1997ApJS..111..377W}
{Worthey}, G. \& {Ottaviani}, D.~L. 1997, \apjs, 111, 377

\bibitem[{{Yang} {et~al.}(2007){Yang}, {Mo}, {van den Bosch}, {Pasquali}, {Li},
  \& {Barden}}]{2007ApJ...671..153Y}
{Yang}, X., {Mo}, H.~J., {van den Bosch}, F.~C., {et~al.} 2007, \apj, 671, 153

\bibitem[{{Zehavi} {et~al.}(2018){Zehavi}, {Contreras}, {Padilla}, {Smith},
  {Baugh}, \& {Norberg}}]{2018ApJ...853...84Z}
{Zehavi}, I., {Contreras}, S., {Padilla}, N., {et~al.} 2018, \apj, 853, 84

\bibitem[{{Zheng} {et~al.}(2007){Zheng}, {Coil}, \&
  {Zehavi}}]{2007ApJ...667..760Z}
{Zheng}, Z., {Coil}, A.~L., \& {Zehavi}, I. 2007, \apj, 667, 760

\bibitem[{{Zhou} {et~al.}(2024){Zhou}, {Arag{\'o}n-Salamanca}, \&
  {Merrifield}}]{2024MNRAS.530.4082Z}
{Zhou}, S., {Arag{\'o}n-Salamanca}, A., \& {Merrifield}, M. 2024, \mnras, 530,
  4082

\bibitem[{{Zhu} {et~al.}(2023){Zhu}, {Lu}, {Cappellari}, {Li}, {Mao}, \&
  {Gao}}]{2023MNRAS.522.6326Z}
{Zhu}, K., {Lu}, S., {Cappellari}, M., {et~al.} 2023, \mnras, 522, 6326

\bibitem[{{Zibetti}(2009)}]{2009arXiv0911.4956Z}
{Zibetti}, S. 2009, arXiv e-prints, arXiv:0911.4956

\bibitem[{{Zibetti} {et~al.}(2009){Zibetti}, {Charlot}, \&
  {Rix}}]{2009MNRAS.400.1181Z}
{Zibetti}, S., {Charlot}, S., \& {Rix}, H.-W. 2009, \mnras, 400, 1181

\bibitem[{{Zibetti} \& {Gallazzi}(2022)}]{2022MNRAS.512.1415Z}
{Zibetti}, S. \& {Gallazzi}, A.~R. 2022, \mnras, 512, 1415

\bibitem[{{Zibetti} {et~al.}(2017){Zibetti}, {Gallazzi}, {Ascasibar},
  {Charlot}, {Galbany}, {Garc{\'\i}a Benito}, {Kehrig}, {de
  Lorenzo-C{\'a}ceres}, {Lyubenova}, {Marino}, {M{\'a}rquez}, {S{\'a}nchez},
  {van de Ven}, {Walcher}, \& {Wisotzki}}]{2017MNRAS.468.1902Z}
{Zibetti}, S., {Gallazzi}, A.~R., {Ascasibar}, Y., {et~al.} 2017, \mnras, 468,
  1902

\bibitem[{{Zibetti} {et~al.}(2020){Zibetti}, {Gallazzi}, {Hirschmann},
  {Consolandi}, {Falc{\'o}n-Barroso}, {van de Ven}, \&
  {Lyubenova}}]{2020MNRAS.491.3562Z}
{Zibetti}, S., {Gallazzi}, A.~R., {Hirschmann}, M., {et~al.} 2020, \mnras, 491,
  3562

\end{thebibliography}

\appendix

\section{Data and adaptative smoothing illustration}
\label{ap:smoothing}

\begin{figure*}
\centering
\includegraphics[width=\hsize]{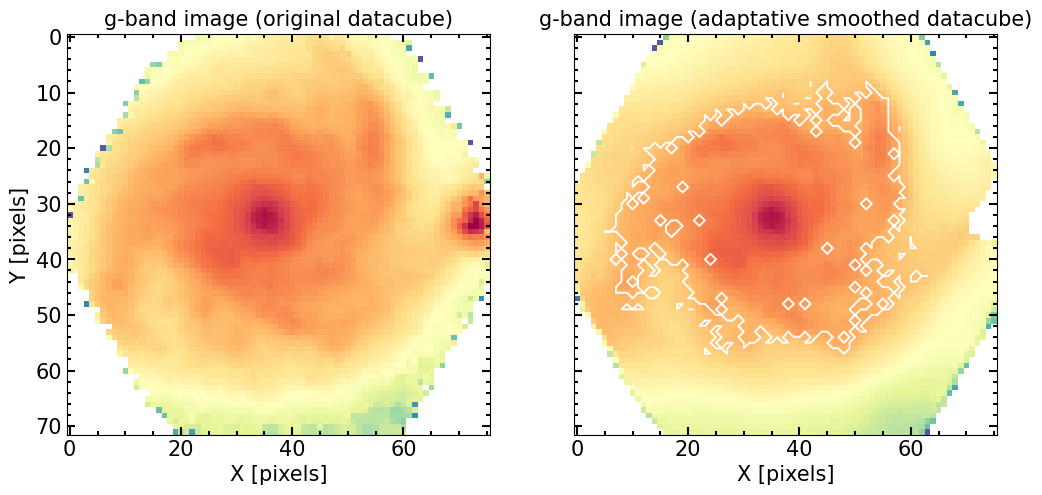}
  \caption{Synthesized g-band images of NGC0234 derived from both the original datacube (left panel) and after applying the adaptive smoothing to the datacube (right panel). The white contours in the left panel indicate the region in which no smoothing is applied due to the higher SNR, i.e., effectively the central region.
  }
     \label{fig:smooth}
\end{figure*}

In this section, we illustrate the adaptive smoothing scheme for an example galaxy of our CALIFA sample. In Fig. \ref{fig:smooth} we show synthesized g-band images of NGC0234 derived from both the original datacube (left panel) and after applying the adaptive smoothing (right panel). The contours indicate the region of the galaxy where no smoothing is effectively applied. We highlight how the spatial resolution is remarkably well preserved. Moreover, we note that the smoothing typically is applied in the outer parts of the galaxies due to the higher SNR in their central regions (within 1 $R_e$) which are effectively not smoothed.

\section{Model-free alternative}
\label{ap:break}

In order to test the robustness of our bayesian fitting framework, we measure the D4000$\rm_n$ break, which is mainly sensitive to age, to have a model-independent first-order estimation of age.

\begin{figure*}
\centering
\includegraphics[width=\hsize]{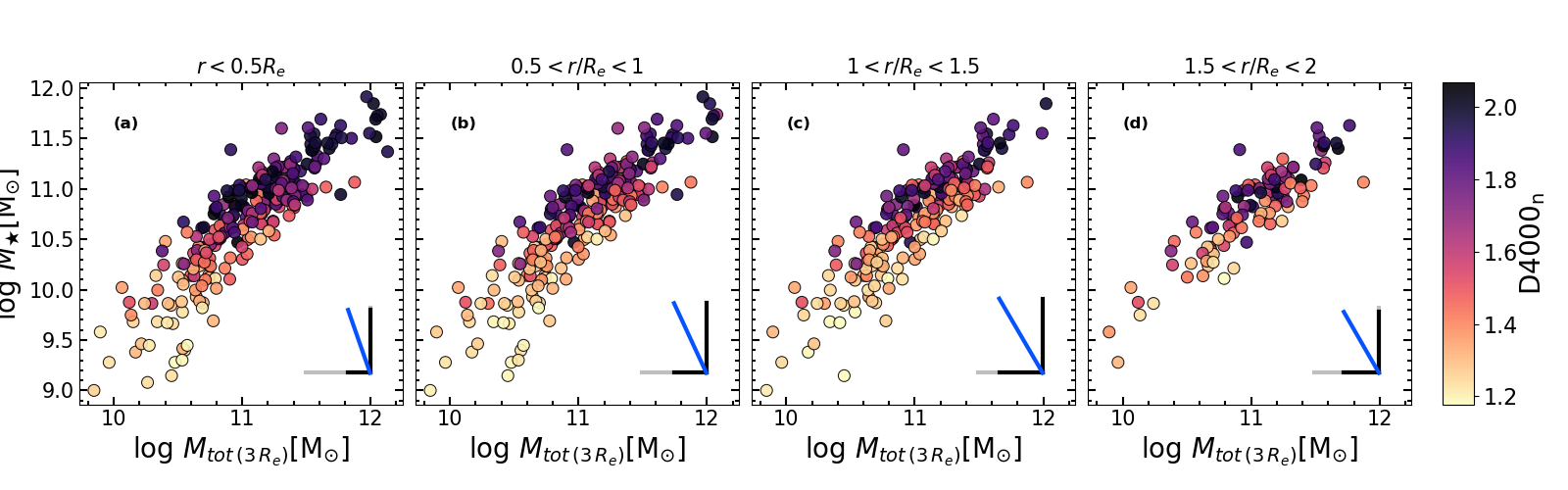}
  \caption{Stellar-to-total dynamical mass relation for our CALIFA galaxies in terms of the age-sensitive D4000$\rm_n$ break measured at different annuli. Each panel corresponds to a different annuli with increasing galactocentric distance from left to right (see text). Galaxies are shown as circles colored-coded by the median D4000$\rm_n$  measured within their corresponding annuli. Partial correlation coefficient strengths are shown in the bottom right corner (solid black lines) between the D4000$\rm_n$ and $M_{\star}$ (vertical) and $M_{tot}$ (horizontal). Grey solid lines have a length which corresponds to a correlation coefficient of 0.6 for reference. The direction of maximal increase of the stellar population parameters (see text) is indicated as a blue solid line.}
     \label{fig:indices}
\end{figure*}

Fig. \ref{fig:indices}, similarly to Fig. \ref{fig:stel_pops_annuli}, shows the STDMR color-coded with median D4000$\rm_n$. Each panel corresponds to the different radial annulus defined in section \ref{sec:STDMR:radialregions} (with increasing galactocentric distance from left to right), where the D4000$\rm_n$ is measured. The trends are in remarkably good agreement with the ones derived for age with our bayesian framework (Fig. \ref{fig:stel_pops_annuli}). Moreover, the partial correlation coefficient strengths also indicate a primary dependence on stellar mass and a second one on total dynamical mass. Given that this method is model independent and the very good agreement between these two approaches, we are confident in our results obtained with our stellar population analysis.

\section{Mean STDMR through symbolic regression}
\label{sec:symbreg}

To compute the mean STDMR, we use open-source and publicly available PySR symbolic regression tool \citep{2020arXiv200611287C} based on genetic programming\footnote{This approach builds on a set of operators (e.g., $ \rm +, -, \times, / $), which are combined with the input variables to form initial equations. These expressions are evaluated, and those with the highest accuracy are retained for subsequent generations. Then, mutations and crossovers are applied to discover and refine optimal equations}. In a nutshell, PySR is a machine learning task that aims to find an interpretable symbolic expression, approximating the relation between input and output parameters through analytical mathematical formulas. In this sense, the advantage of employing symbolic regression instead other machine learning regression models is that it yields analytical formulas that facilitate the interpretation of the underlying physics and can be easily generalized. We note that symbolic regression has already been used in several astrophysical studies \citep[e.g.,][]{2012arXiv1208.2480G, 2020PhRvD.101l3525A,2020arXiv200611287C,2021PhRvL.127m1102B,2021ApJ...915...71V, 2022MNRAS.515.2733D,2022arXiv221111461B,2022arXiv221106393B,2022ApJ...927...85S,2023ApJ...956..149S,2022mla..confE..25W}, including some using PySR \citep[e.g.,][]{2022ApJ...930...33M, 2022ApJ...927...85S, 2023MNRAS.522.2628W, 2023PNAS..12002074W}. 

By default PySR selects the best-fitting models with compromise between accuracy and complexity to avoid data over-fitting and complex non interpretable equations. After testing, we selected the default least-squares loss and employed basic mathematical operators ($ \rm +, -, \times, /, pow $) together with low-order polynomials ($ \rm x^2,x^3,x^4$) and $\rm 1/x$ (we do not include exponentials and logarithms to avoid overly complex equations). 

Genetic algorithms do not tend to converge, and we are applying PySR to real noisy data. Yet, after several iterations we observed convergence on similar functional forms. Therefore we performed 10 runs of with a larger number of iterations (5000), out of which we selected most repeated functional form with less complexity of the ones with similar low losses.

\section{Gradients as a function of stellar mass}
\label{ap:gradients}

In other to facilitate comparisons with other works in the literature, in this section we also look at gradients across projections of the STDMR, and also as a function of morphology.

\begin{figure*}
\centering
\includegraphics[width=\hsize]{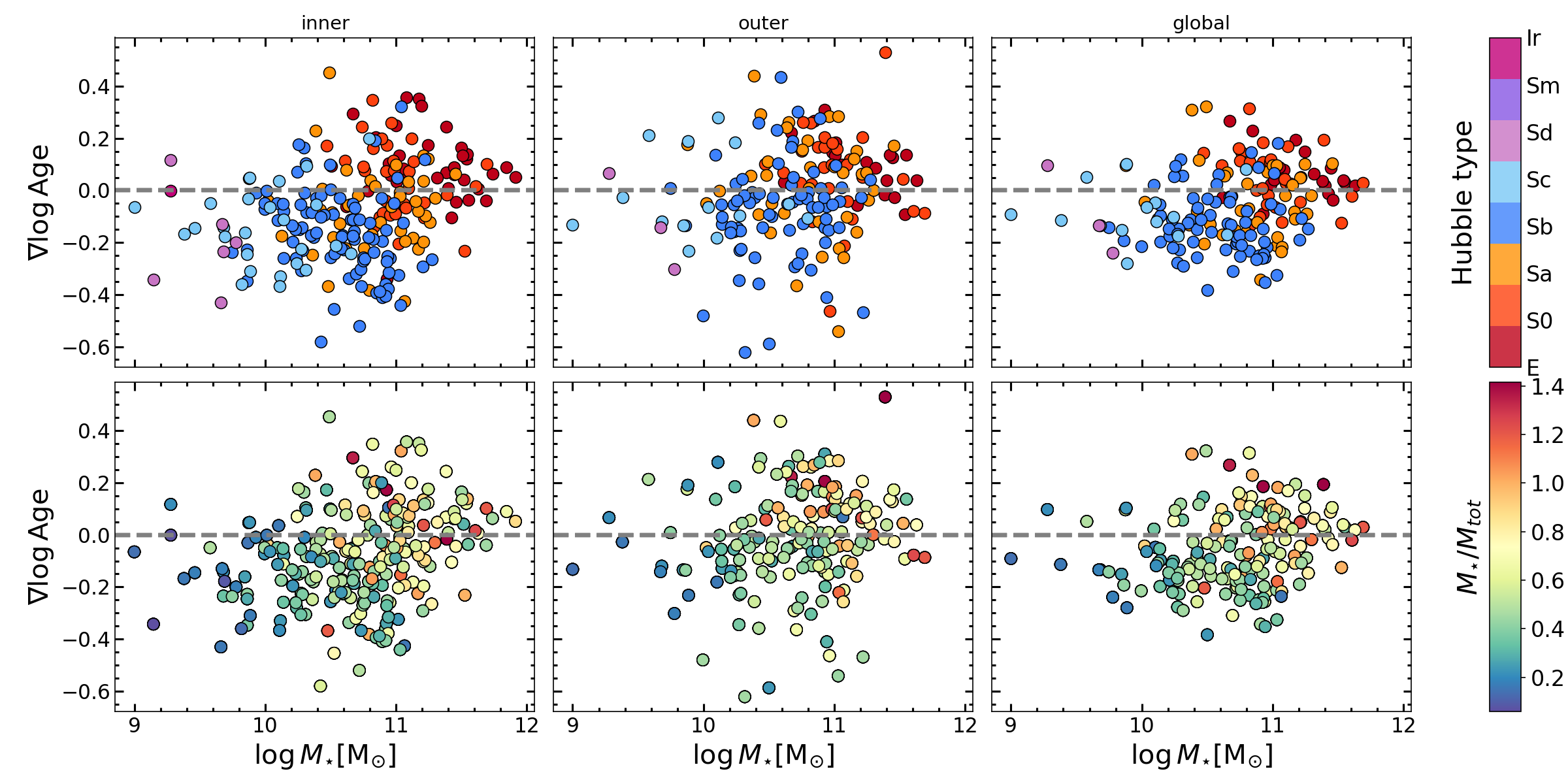}
  \caption{Age gradients as a function of stellar mass. Inner (left column), outer (middle column) and global (right column) gradients (as defined in section \ref{sec:STDMR:gradients}) for individual galaxies (circles) are color-coded with morphology (top row) and the $M_{\star}/M_{tot}$ ratio (bottom row). For reference, the grey horizontal dashed line corresponds to zero.}
     \label{fig:proj_gradients_age}
\end{figure*}

\begin{figure*}
\centering
\includegraphics[width=\hsize]{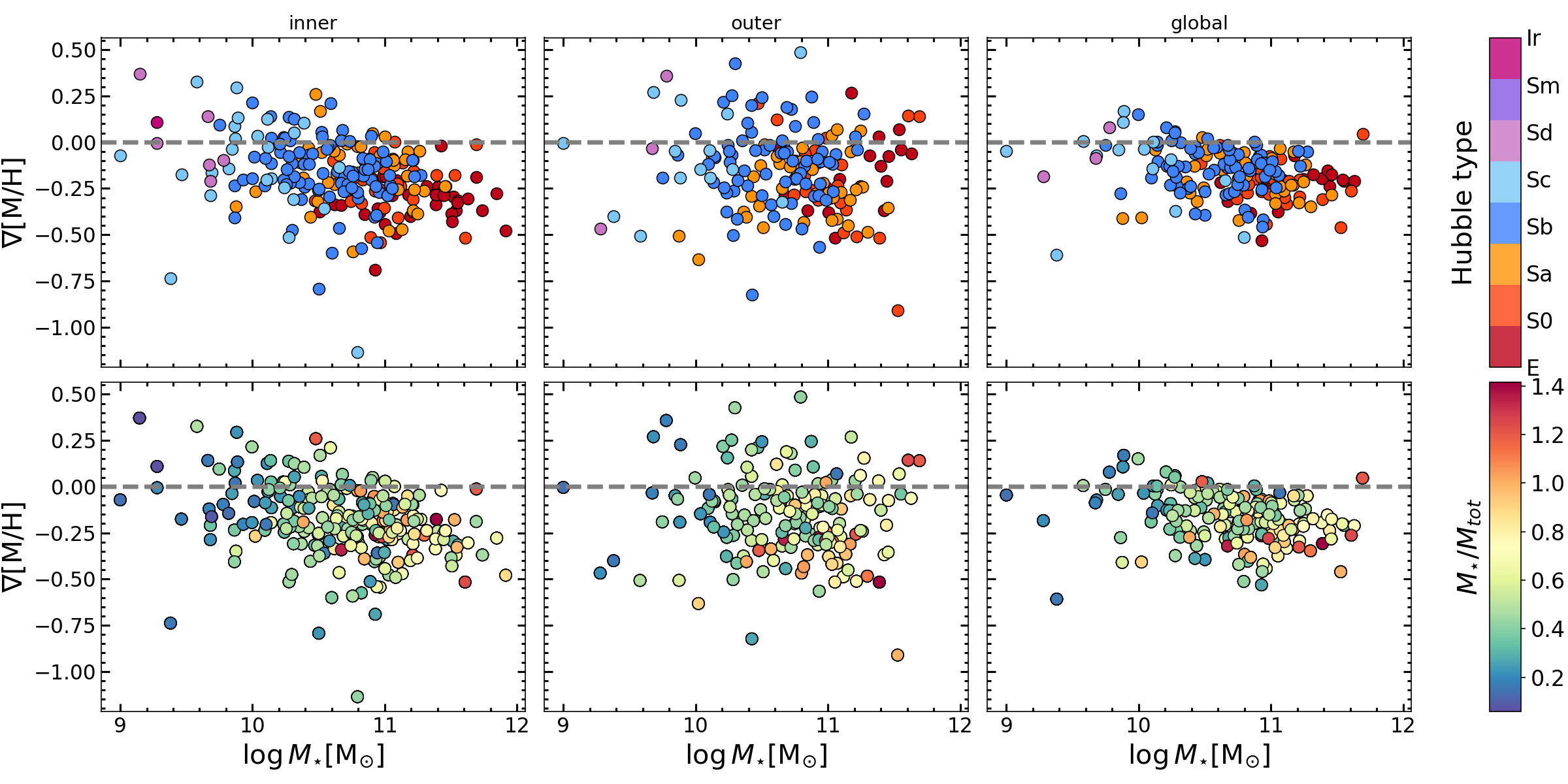}
  \caption{Metallicity gradients as a function of stellar mass. Inner (left column), outer (middle column) and global (right column) gradients (as defined in section \ref{sec:STDMR:gradients}) for individual galaxies (circles) are color-coded with morphology (top row) and the $M_{\star}/M_{tot}$ ratio (bottom row). For reference, the grey horizontal dashed line corresponds to zero.}
     \label{fig:proj_gradients_met}
\end{figure*}

Fig. \ref{fig:proj_gradients_age} and \ref{fig:proj_gradients_met} show age and [M/H] gradients, respectively, as a function of stellar mass. Inner (left column), outer (middle column) and global (right column) gradients are color-coded with morphology (top row) and the $M_{\star}/M_{tot}$ ratio (bottom row). 

\paragraph{\emph{Age} --} In Fig. \ref{fig:proj_gradients_age}, we clearly observe that the inner gradients of early- and late-type galaxies follow different sequences with stellar mass. While late types tend to follow negative gradients, that become steeper with increasing stellar mass, ETGs present positive, flat gradients or negative age gradients. For the outer gradients, we see a similar behavior, but with more scatter and less difference between these morphological types, as there are more LTGs with positive gradients. In contrast, global gradients show again a clear difference between the groups, although with less scatter.

Furthermore, we see here again that morphology is connected to the $M_{\star}/M_{tot}$ ratio of the galaxies, with ETGs having higher stellar-to-total mass ratios than LTGs. We observe that intermediate-mass galaxies with positive gradients tend to have lower total masses at fixed stellar mass, while galaxies with negative gradients have higher total masses. 

\paragraph{\emph{[M/H]} --} In  Fig. \ref{fig:proj_gradients_met} we see a different behavior than for age gradients. Now, the gradients of ETGs and LTGs generally follow the same regions in the $\rm \nabla[M/H]-M_{\star}$ plane. 

Inner gradients of LTGs and ETGs follow the same scaling relation with stellar mass, being generally negative although ETGs tend to have steeper gradients. Although some less massive ($ \rm M_{\star}<10^{10}M_{\odot}$) LTGs show mildly positive ones. The behavior for global gradients is quite similar, although with less scatter. Yet, the outer gradients exhibit a greater scatter and the relation is less clear.

Consistently with Fig. \ref{fig:gradients}, we do not observe any strong trend with the $M_{\star}/M_{tot}$ ratio, as it mainly correlates with stellar mass. Yet, the outer gradients of intermediate-mass and massive galaxies tend to be more negative if they have lower total masses at fixed stellar mass.

\end{document}